\documentclass[preprint,amsmath,amssymb,superscriptaddress]{revtex4}
\usepackage{graphicx}
\usepackage{indentfirst}
\usepackage{epsfig}
\usepackage{subfigure}
\usepackage{amsmath}
\usepackage{amssymb}
\def\be{\begin{eqnarray}}
\def\ee{\end{eqnarray}}

\def\k{{\bf k}}

\begin{document}
\title{The Quantum Vacuum of Complex Media.
A Unified Approach to the Dielectric Constant,
the Spontaneous Emission and the Zero-Temperature Electromagnetic Pressure}
\author{M. Donaire}
\email{mdonaire@fc.up.pt}
\address{Centro de F\'{\i}sica do Porto, Faculdade de Ci\^{e}ncias da Universidade do Porto, Rua do Campo Alegre 687, 4169-007 Porto, Portugal.}
\address{Departamento de F\'{\i}sica de la Materia Condensada, Universidad Aut\'{o}noma de
Madrid, E-28049 Madrid, Spain.}
\begin{abstract}
We study from a critical perspective several quantum-electrodynamic phenomena commonly related to vacuum electromagnetic (EM) fluctuations in complex media. We compute the resonance-shift, the spontaneous emission rate, the local density of states and the van-der-Waals-Casimir pressure in a dielectric medium using a microscopic diagrammatic approach. We find, in agreement  with some recent works, that these effects cannot be attributed to variations on the energy of the EM vacuum but to variations of the dielectric self-energy. This energy is the result of the interaction of the bare polarizability of the dielectric constituents with the EM fluctuations of an \emph{actually polarized vacuum}. We have found an exact expression for the spectrum of these fluctuations in a statistically homogeneous dielectric. Those fluctuations turn out to be different to the ones of normal radiative modes. It is the latter that carry the zero-point-energy (ZPE). Concerning spontaneous emission, we clarify the nature of the radiation and the origin of the so-called local field factors. Essential discrepancies are found with respect to previous works. We perform a detailed analysis of the phenomenon of radiative and non-radiative energy transfer. Analytical formulae are given for the decay rate of an interstitial impurity in a Maxwell-Garnett dielectric and for the decay rate of a substantial impurity sited in a large cavity. The construction of the effective dielectric constant is found to be a self-consistency problem. The van-der-Waals pressure in a complex medium is computed in terms of variations of the dielectric self-energy at zero-temperature. An additional radiative pressure appears associated to variations of the EM vacuum energy.
\end{abstract}
%
%
\maketitle

\tableofcontents

\newpage
\section{Introduction}\label{Sect1}
\indent The  concepts of radiative fluctuations and virtual particles are inherent to the nature of quantum mechanical processes in the framework of Quantum Field Theory (QFT). The actual existence of fluctuations is however questionable as the manner they enter quantum phenomena depends on the formalism employed. For instance, one can say that radiative corrections renormalize the 'bare' mass of a free electron as long as one can postulate the existence of a 'bare' electron in absence of coupling to the electromagnetic (EM) field. Likewise, one can say that zero-point quantum fluctuations of the EM field give rise to a shift on the energy of stationary atomic states provided that the actual existence of those states could be postulated in the absence of coupling to radiation. However, quantum fields couple to each other and such a coupling cannot be switched off. Bare masses and stationary energy states are actually unobservable. In addition, if the electromagnetic field appears coupled to a stochastic system, eg. a thermal reservoir or a gas of (actual) charged particles randomly distributed, additional fluctuations appear and add up to the once previously  considered. The crucial difference of these with respect to the zero-point fluctuations is that they can be modified or even switched off by tuning some external parameter. It is implicit in this picture that the spectrum of fluctuations can be expanded as a power series in the free-space electromagnetic propagator and the coupling to actual charges, in application of the Lippman-Schwinger equation. Many electromagnetic phenomena are explained appealing to the actual existence of  zero-point fluctuations using quantum or semiclassical approaches \cite{Milonnibook}. That is the case of the Casimir-Polder effect, the Lamb shift, the van der Waals forces and the spontaneous decay of excited atoms. Recently \cite{Jaffetpaper}, it has been pointed out that the Casimir effect as formulated in Quantum Electrodynamics (QED) is not actually a manifestation of the zero-point vacuum fluctuations of quantum fields but the result of the interaction between actual charges and currents --see also \cite{Bordag}.\\
\indent In this paper, we are concerned with all the above phenomena in the context of random media. We will show that they all can be addressed employing a unified formalism. Our study includes the computation of  the modified decay rate of an excited emitter and the calculation of the dielectric constant of a homogeneous medium. Closely related, we will compute the local density of states and the EM energy density of a dielectric. Van-der-Waals-Casimir forces and radiative pressure manifest as a response to virtual variations on the total EM energy density due to changes in the spacial distribution of dipoles. To explain all these phenomena we will make use of both Classical Optics and QFT formalisms.\\
\indent To what Optics concerns, it is a general issue the characterization of a
material system which couples to radiation by means of its coherent transport properties and the study of the decay of unstable local states. With respect to the latter, it is known since the work of Purcell \cite{Purcell} that the spontaneous emission rate of an atom, $\Gamma$,
in a dielectric host medium depends on the interaction of the emitter with the material environment. This is so because  the medium determines the spectrum of the EM fluctuations which mediate the atom self-interactions and hence, its self-energy.  In the first place the host medium modifies the density of channels into which the atom can radiate --i.e. the Local Density of States (LDOS)-- and hence the value of $\Gamma$. Second, the integration of the associated self-energy gives rise to a shift in the resonance frequency of the atom. Third, the  dipole-transition-amplitude gets also modified. As a result, the medium is said to renormalize the polarizability of the emitter.\\
\indent In addition and complementarily, the medium polarizes the EM vacuum. This reflects in the fact that the dispersion relations that EM normal modes satisfy are determined by the effective susceptibility of the medium. Complementarity can be seen in that LDOS and $\Gamma$
depend as well on those parameters which determine the
coherent transport features of the medium. That is, on the refractive index and the mean free path. As a matter of fact, both null transmittance
and inhibition of spontaneous emission are expected to occur in photonic band gap materials \cite{JohnI}.\\
\indent For practical purposes, the
understanding of life-times in random media is relevant in the context of fluorescence biological imaging \cite{Suhling}  and nano-antennas \cite{Antenas}. Also, understanding of unconventional coherent transport properties is essential in engineering metamaterials for electromagnetic  and acoustic waves \cite{PendryLiu}.\\
\indent To what QFT concerns, we will postulate the existence of two distinguishable EM vacua attending to the the existence of two different spectra of fluctuations. These are, a \emph{sourceless vacuum} in which normal modes propagate and a \emph{self-polarization vacuum} in which both the radiative emission and the photons mediating the self-polarization of the emitter propagate. While the fluctuations which carry the divergent ZPE live on the former vacuum, those which gives rise to the van-der-Waals-Casimir forces live in the latter. In addition, a \emph{coherent-vacuum} in which coherent emission propagates will be identified.\\
\indent The paper is organized as follows,\\
\indent In Section \ref{Sect2} we first analyze the role of electromagnetic fluctuations in the paradigmatic quantum process of spontaneous dipole emission. That section serves also to describe the features of the framework in which our approach fits.\\
\indent  Next, we develop our approach in several steps. It is based on a microscopical diagrammatic treatment which explodes the seminal works carried out by Foldy \cite{Foldy}, Lax \cite{Lax}, Frisch \cite{Frisch}, Bullough and Hynne \cite{Bullough,BulloughHynne,BulloughHynne2} and Felderhof \emph{et al.} \cite{Feldoher,Feldoher2}. Our main goal will be  to develop a general formalism suitable for treating in the same footing any emission-related process in any particular scenario  within the framework of linear random media. Thus, the next sections are organized in a manner that allows us to address systematically and separately each of the features which characterize a given scenario and a given process. These features are,\\
\indent A) The nature of the emission. That is, it can be spontaneous emission from an isolated atom; stimulated emission by an external exciting field on a polarizable particle; a combination of both the spontaneous emission of an atom and the emission induced on the polarizable molecule which hosts the atom. This matter is addressed in Section \ref{Sect3}.\\
\indent B) The nature of the embedding of the emitter in the host medium. When the emitter is itself a host particle or it resides in a host particle we talk of virtual cavity as the medium is in fact unaltered by the presence of the emitter. In any other case, the emitter sites within a real cavity of radius $R$. This matter is addressed in Sections \ref{Sect4} and \ref{Sect4B}.\\
\indent C) The nature of the host medium. For the sake of simplicity, we will restrict ourselves to statistically homogeneous and isotropic three dimensional host media. Further classifications are made attending to the relation between the wavelength of the radiation, $\lambda$, the correlation length between the host medium constituents, $\xi$, and, if it applies, the radius of the emitter cavity, $R$. In the virtual cavity scenario, $\xi=R$ and the topology is refereed to as cermet topology as the medium is composed by disconnected point dipoles. An effective dielectric medium can be defined as long as $\lambda\gg\xi$. In the latter case the propagation of coherent radiation is fully characterized by an effective dielectric constant, $\epsilon_{eff}$. On the contrary, any real cavity breaks manifestly translation invariance and no effective medium can be defined generally. Nevertheless, if $R\gg\xi$, the cavity is said large and the medium is seen by the emitter as a continuum. Further, if $\lambda\gg\xi$, the medium behaves as an almost-effective one w.r.t. the propagation of the coherent radiation emitted from the center of the cavity. It is in that case that the topology is that of a simply-connected-non-contractible manifold in which the emitter is placed at the center of a cavity of radius $R$ surrounded by a continuous effective medium. All these matters are addressed in Sections \ref{Sect4B} and \ref{Sect5B2}.\\
\indent D) The nature of the radiation. In the first place, radiation can be either transverse or longitudinal attending to its direction with respect to the propagation wave vector. Second, radiation can be coherent and incoherent. The coherent part has two contributions. The first one is that emitted directly by the source dipole into the host medium. The second one comes from the coherent interference of the latter with the radiation induced in the surrounding dipoles. Coherent radiation propagates through a coherent-vacuum. Incoherent radiation is that which gets dispersed or absorbed. These matters are addressed in Section \ref{Sect5} together with a study of the process of radiative/non-radiative energy transfer.\\
\indent Equipped with the above detailed analysis, we address explicit computations which can be performed analytically. In Section \ref{Sect6} we concentrate on the virtual cavity scenario. First, in Section \ref{Sect6A} we compute the spontaneous decay rate of a weakly-polarizable
interstitial impurity within a Maxwell-Garnett dielectric. In Section \ref{Sect6B} we compute formally the dielectric constant of a Maxwell-Garnett atomic gas for the case that the single-atom-polarizabilities are known in free space and non-radiative effects can be ignored. We analyze old experiments and propose suitable modifications in order to test our approach.\\
\indent We address the real cavity scenario in Section \ref{Sect7P}. There, we compute the spontaneous decay rate of a weakly-polarizable substantial impurity placed in a large real cavity.\\
\indent In Section \ref{Sect7} we compare our results with previous ones. Several arguments are given to explain why those fail. In particular, we concentrate on explaining why of the erroneous use of the so-called local field factors.\\
\indent Finally, we explain in Section \ref{Sect9} the computation of the zero-temperature EM pressure in a gas. Both a contribution coming from variations on the matter self-energy and another one coming from normal mode fluctuations are found. Only the former is identified with the van-der-Waals pressure in agreement with recent interpretations of van-der-Waals-Casimir forces.\\
\indent We summarize our results in the Conclusions, Section \ref{TheEnd}.
\section{The spontaneous emission process in a dielectric medium. The role of the electromagnetic vacuum  fluctuations}\label{Sect2}
\indent This section has a double purpose. On the one hand, it offers a comprehensive overview of the features which characterize the physical scenarios to which the approach developed in the next sections applies. We will clarify to which extend our approach is really \emph{microscopical} and to which extend quantization is taken into account. On the other hand, it describes qualitatively some of the subtle points related to the role of the EM vacuum fluctuations. To these aims, we analyze qualitatively the phenomenon of spontaneous emission which, since the pioneering work by Einstein  \cite{Einstein}, has been considered paradigmatic in the understanding of the EM quantum vacuum. Elements of both Quantum Field Theory and Classical Optics will be used. There is little new in this section and our arguments base on well-known results of standard textbooks --eg. \cite{Sakurai}. Nevertheless, we feel it is necessary in order to appreciate the subtle points of the next sections. Reference will be made, with no further explanation at this stage, to concepts and results which will appear later.
\subsection{Fermi's golden rule}\label{Sect2A}
The formula for $\Gamma$ is given by Fermi's golden rule. According to it, in second order of perturbation theory an excited state of matter localized at $\vec{r}$ decays via electric dipole transition with a rate given by
\begin{eqnarray}
\Gamma&=&\frac{2\pi}{\epsilon_{0}\hbar^{2}}\sum_{I,\gamma}|\langle A|\hat{H}_{int}(\vec{r},t)|I,\gamma_{\omega_{IA}}\rangle\langle I,\gamma_{\omega_{IA}}|\hat{H}_{int}(\vec{r},t)|A\rangle|\delta\Bigl(\omega_{IA}-(\omega_{A}-\omega_{I})\Bigr),\label{jg}\\
\Gamma&\simeq&\frac{2\pi}{\epsilon_{0}\hbar^{2}}\sum_{I,\gamma_{IA}}|\langle A|\hat{\vec{\mu}}\cdot\hat{\vec{E}}(\vec{r},t)|I,\gamma_{\omega_{IA}}\rangle\langle I,\gamma_{\omega_{IA}}|\hat{\vec{\mu}}\cdot\hat{\vec{E}}(\vec{r},t)|A\rangle|\delta\Bigl(\omega_{IA}-(\omega_{A}-\omega_{I})\Bigr),\label{jj}
\end{eqnarray}
where the dipole approximation has been taken in Eq.(\ref{jj}). $|A\rangle$ is the initial state in which the emitter is in an excited atomic stationary state of energy $\hbar\omega_{A}$ with wave function $\psi_{A}(x)$  and the electromagnetic field is in its ground state in which there are no actual photons. $\{|I,\gamma_{\omega_{IA}}\rangle\}$ is the set of intermediate states in which the emitter is in a lower energy stationary state of wave function $\psi_{I}(x)$ after the release of a quantum of EM energy, which equals $\hbar\omega_{IA}=\hbar(\omega_{A}-\omega_{I})$ according to the delta function of energy conservation. In free space, $|\gamma_{\omega_{IA}}\rangle$ is a transverse photon.  $\hat{\vec{\mu}}$ is the transition dipole operator and $\hat{\vec{E}}(\vec{r},t)$ is the electric field operator in the Schr\"{o}dinger picture. Following the introduction by Barnett et al. in \cite{LaudonJPB}, by Fourier-transforming  the time dependence of $\hat{\vec{E}}(\vec{r},t)$ into $\hat{\vec{E}}_{\omega}(\vec{r})$, decomposing the latter in terms of creation/annihilation operators and using the completeness of the space of EM states, we can eliminate the time dependence in Eq.(\ref{jj}) in favor of $\omega_{IA}$ and arrive at
\begin{equation}
\Gamma=\frac{2\pi}{\epsilon_{0}\hbar}\sum_{I}|\vec{\mu}_{AI}|^{2}\:^{s.p.}\langle \Omega|\hat{\vec{E}}_{\omega_{IA}}(\vec{r})\cdot\hat{\vec{E}}^{\dagger}_{\omega_{IA}}(\vec{r})|\Omega\rangle^{s.p.},\label{hh}
\end{equation}
where $\dagger$ denotes hermitian conjugate and $|\Omega\rangle^{s.p.}$ is the \emph{self-polarization vacuum state} which is left for a more precise definition. From now on we will simplify matters assuming the emitter is a two-level atom --with state labels $A,B$-- with a unique transition dipole matrix element $\vec{\mu}_{AB}=\langle A|\hat{\vec{\mu}}|B\rangle=\int\textrm{d}^{3}x\:\psi_{A}(\vec{x})\vec{\mu}(\vec{x})\psi^{*}_{B}(\vec{x})$. It is important to bear in mind that the stationary atomic states so far considered are bound states of actual charges. Those are, the atomic nuclei and the atomic electrons. Therefore, they are eigenstates of some effective Hamiltonian which already contain electromagnetic interactions as a result of the partial integration of 'virtual' photons. No actual photons (i.e. propagating photons) are involved in those atomic states and the interaction integrated is essentially electrostatic. Hence, the resultant $\vec{\mu}_{AB}$ is directly related to the so-called electrostatic polarizability --see Section \ref{Sect3C}. Therefore, once knowing $\mu_{AB}$ the problem becomes one of computing the radiative vacuum fluctuations of energy $\hbar\omega_{AB}$ which couple to the otherwise stationary dipole.\\
\indent Applying next the fluctuation-dissipation theorem \cite{Huang} we can express the electric field fluctuations of frequency $\omega_{IA}$ in terms of the imaginary part of a Green's tensor,
\begin{equation}\label{Disip}
^{s.p.}\langle \Omega|\hat{\vec{E}}_{\omega_{AB}}(\vec{r})\hat{\vec{E}^{\dagger}}_{\omega_{AB}}(\vec{r}')|\Omega\rangle^{s.p.}=
-\hbar\frac{\omega_{AB}^{2}}{\epsilon_{0}c^{2}\pi}\Im{\{\bar{\mathcal{G}}(\vec{r},\vec{r}';\omega_{AB})\}}.
\end{equation}
$\bar{\mathcal{G}}(\vec{r},\vec{r};\omega_{AB})$ is the Green's tensor which propagates the electric field from the point dipole source  which oscillates with frequency $\omega_{AB}$ at $\vec{r}$ back to itself. That field mediates the interaction of the dipole with itself. That is, it is the self-polarization field which propagates in  $|\Omega\rangle^{s.p.}$. Alternatively, $\bar{\mathcal{G}}$ is interpreted as the propagator of virtual photons which are emitted and reabsorbed at the dipole source.\\
\indent Up to know, we have followed a quite standard procedure. Alternatively, other microscopical approaches base on  generalized optical Bloch equations in the context of second-quantization formalism  \cite{CrenBow,Crenshaw,BerMil}. Throughout this paper we will stick to a Green's function based formalism.
\begin{figure}[h]
\includegraphics[height=7.7cm,width=13.2cm,clip]{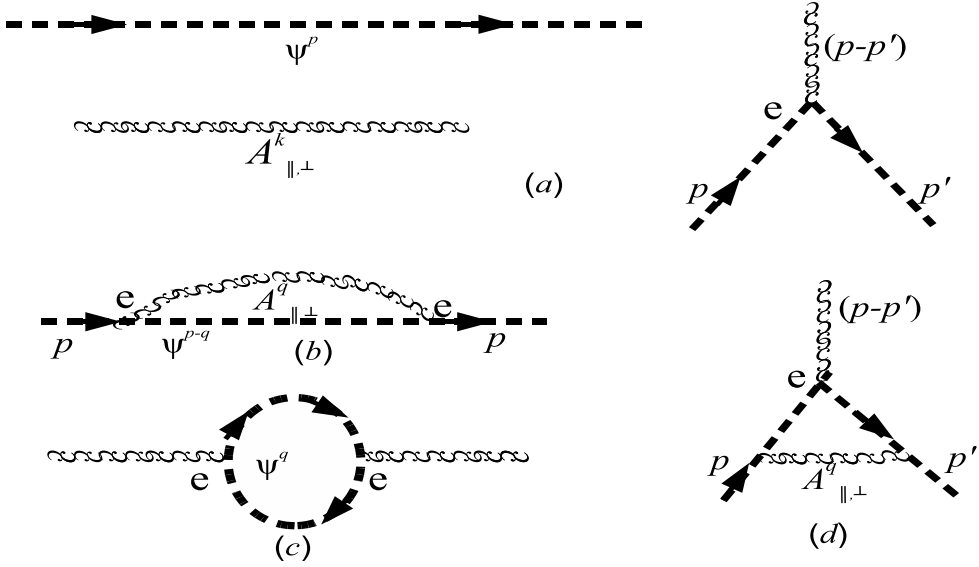}
\caption{($a$) Feynman rules of the QED field theory. The wavy line denotes the propagator of transverse ($A_{\perp}$) and longitudinal ($A_{\parallel}$) photons. The thick dashed line denotes the propagator of charged particles. On the right hand side, tree level interaction diagram from Dirac's lagrangian \cite{Peskin}. ($b$) One-loop mass renormalization diagram for charged particles. ($c$) One-loop vertex renormalization diagram. ($d$) One-loop polarized photon propagator.}\label{FIGnew1}
\end{figure}
\subsection{Zero-point and in-free-space vacuum fluctuations}\label{Sect2B}
\begin{figure}[h]
\includegraphics[height=7.7cm,width=13.2cm,clip]{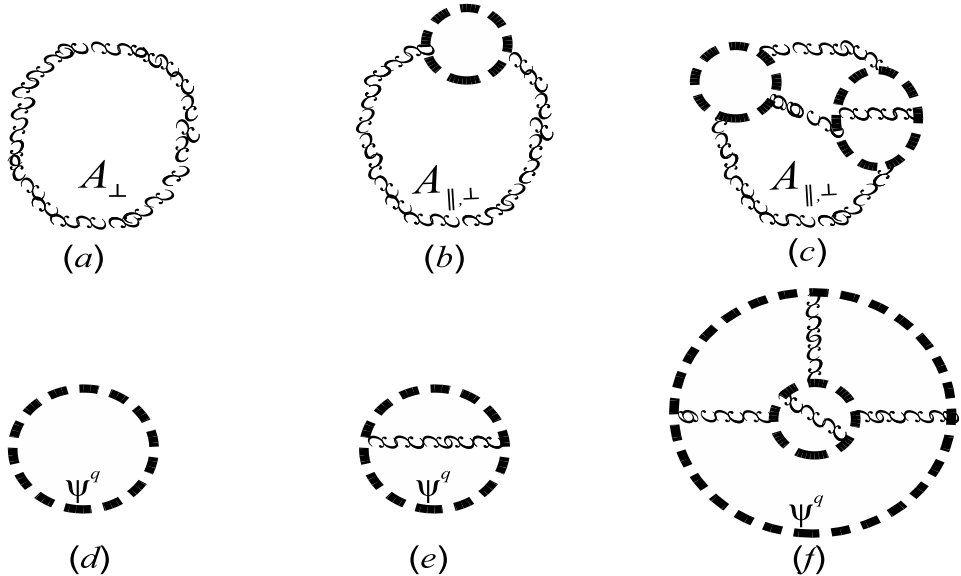}
\caption{QED vacuum diagrams contributing to the ZPE. Diagrams $(a)$ and $($d$)$ refer to the non-interacting theory. The diagrams $(b)$ and $(e)$, $(c)$ and $(f)$ are topologically identical. They give rise to a polarization of the EM vacuum as seen by photons (upper row) or to a renormalization of the electric charge $e^{2}$ as seen by charged fields (lower row).}\label{FIGnew1p}
\end{figure}

\begin{figure}[h]
\includegraphics[height=5.cm,width=13.5cm,clip]{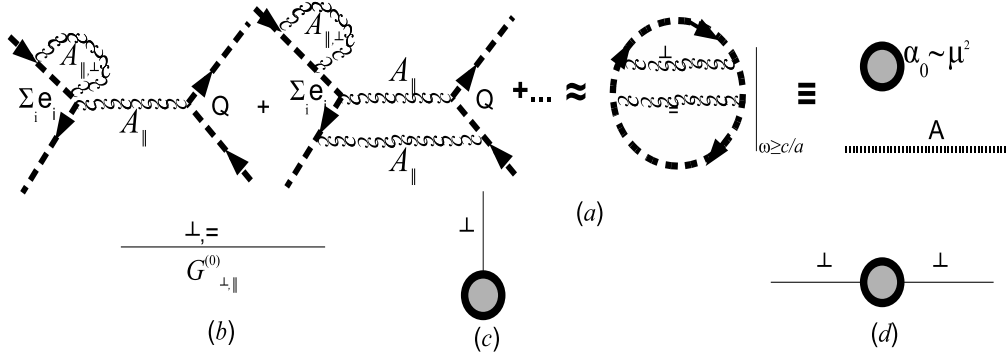}
\caption{($a$) Formal integration of electrostatic and rest of high frequency modes in the stationary atomic state A and bare electrostatic polarizability $\alpha_{0}$. ($b$) Effective interaction vertex in the dipole approximation after integration of high frequency modes. ($c$) Vacuum polarization diagram in presence of a unique dipole.}\label{FIGnew4}
\end{figure}

\begin{figure}[h]
\includegraphics[height=2.7cm,width=9.cm,clip]{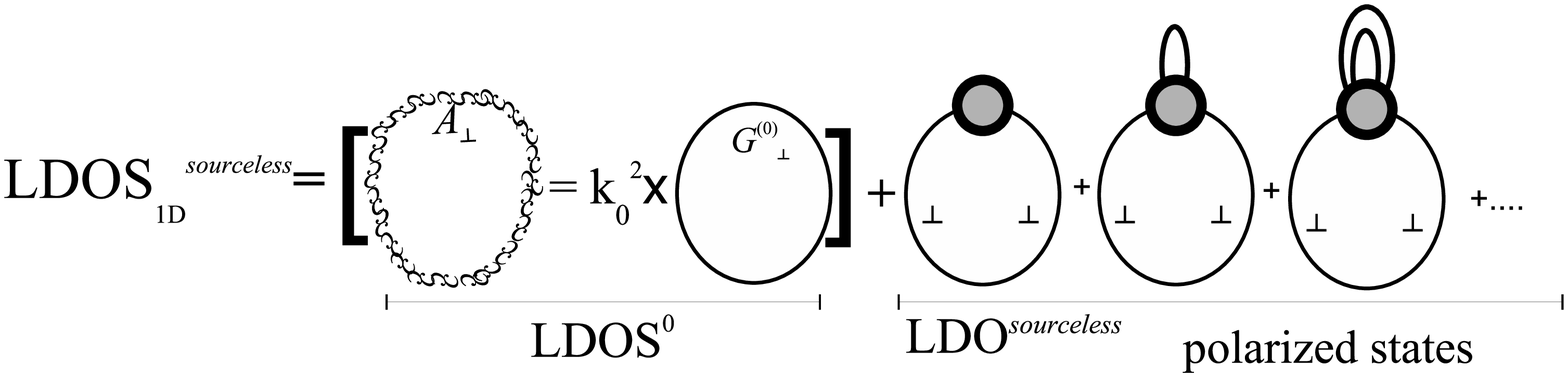}
\caption{Diagrammatical representation of $LDOS^{sourceless}_{1D}$. The two terms on the \emph{r.h.s.} correspond to the integrands of the two terms on the \emph{r.h.s.} of Eq.(\ref{uma}) respectively. The geometrical series of the second term stand for the states multiply-polarized by the presence of the unique dipole.}\label{FIGnew6}
\end{figure}

\begin{figure}[h]
\includegraphics[height=11.7cm,width=14.5cm,clip]{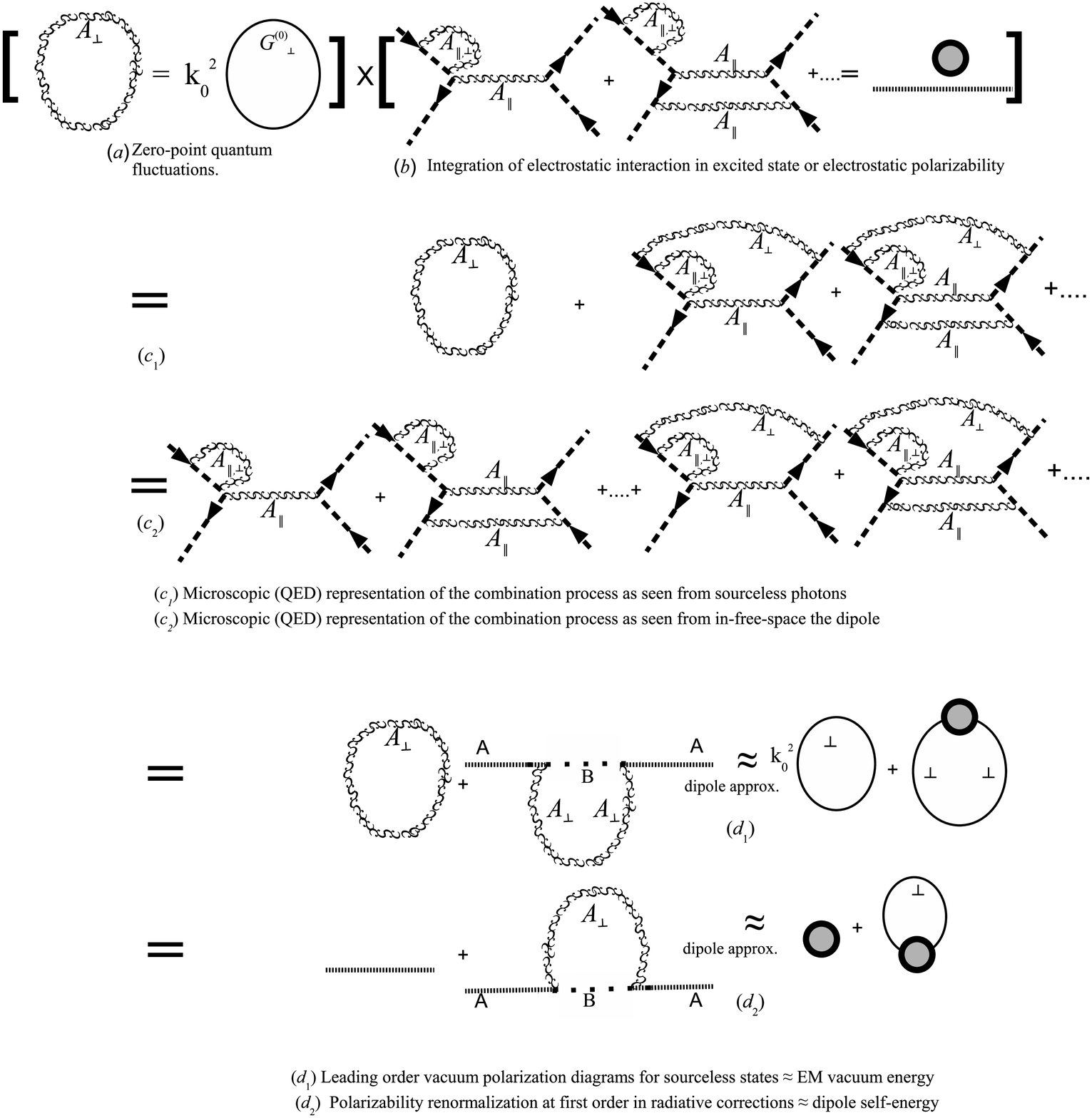}
\caption{Interpretation of the 'recombination' process between a point dipole and zero-point EM vacuum fluctuations. ($a$) Zero-point vacuum fluctuation loop of the electromagnetic field. The equivalence in Eq.(\ref{EA}) has been depicted. ($b$) Feynman diagrams contributing to a stationary atomic state and its equivalence after integration of high frequency modes as in Fig.\ref{FIGnew4}. ($c_{1}$) Microscopic representation of the recombination process leading to polarized EM vacuum fluctuations in the presence of an actual dipole. ($c_{2}$) Microscopic representation of the recombination process leading to an actual fluctuating dipole. Equivalent interpretation: renormalization  of the dipole polarizability through radiative corrections. In ($d_{1,2}$), same phenomena as in ($d_{1,2}$) using the effective nomenclature of Fig\ref{FIGnew4}.}\label{FIGnew2}
\end{figure}

\begin{figure}[h]
\includegraphics[height=6.5cm,width=14.0cm,clip]{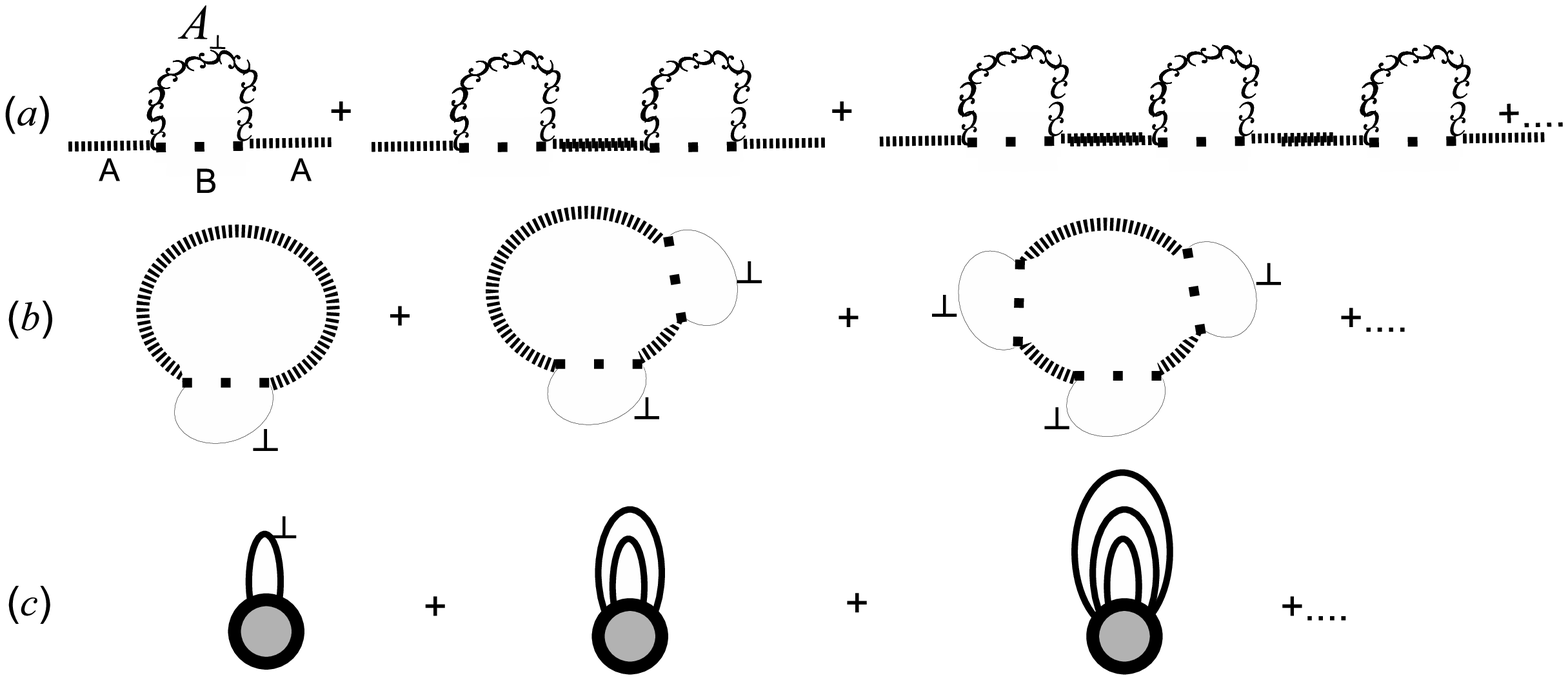}
\caption{Diagrammatical representation of the geometrical series which give rise to the radiative renormalization of the  electrostatic polarizability of a two-level atom in free space. In ($a,b$) it is written as a succession of coupled virtual states. In ($c$), the effective rules in Fig.\ref{FIGnew4} are used. In ($b,c$), the dipole approximation is implicit.}\label{FIGnew5}
\end{figure}

\indent We give here a non-rigorous overview on the derivation of Eqs.(\ref{jg}-\ref{hh}) in free space in terms of the interaction between actual charges and EM fluctuations.\\
\indent At zero temperature, with no actual charges at all and disregarding virtual ones, photons are non-interacting species. They are radiative and hence transverse. Virtual photons are depicted as closed wavy lines which stand for frequency modes of the propagator of $\vec{A}_{\perp}$ with origin and end at the same point. The 'bare' EM vacuum of sourceless modes at zero temperature will be denoted by $|0\rangle$. As noted in \cite{Agerwal}, the propagator of $\vec{A}^{\omega}$ equals that for $\vec{E}^{\omega}$ modulo a prefactor $(\omega/c)^{-2}$. Hence, $\vec{E}=\frac{\partial}{\partial t}\vec{A}$. In application of the fluctuation-dissipation theorem we can write,
\begin{equation}\label{EA}
\langle0|\hat{\vec{E}}_{\perp}^{\omega}(\vec{r})\hat{\vec{E}}_{\perp}^{\omega\dag}(\vec{r})|0\rangle=\frac{\omega^{2}}{c^{2}}
\langle0|\hat{\vec{A}}_{\perp}^{\omega}(\vec{r})\hat{\vec{A}}_{\perp}^{\omega\dag}(\vec{r})|0\rangle\textrm{ and}
\end{equation}
\begin{eqnarray}
\langle0|\hat{\vec{E}}(\vec{r})\hat{\vec{E}}^{\dag}(\vec{r})|0\rangle&=&
\int\langle0|\hat{\vec{E}}_{\perp}^{\omega}(\vec{r})\hat{\vec{E}}_{\perp}^{\omega\dag}(\vec{r})|0\rangle\:\textrm{d}\omega
\nonumber\\
&=&-\frac{\hbar}{\epsilon_{0}\pi c^{2}}\int\omega^{2}\Im{\{\bar{G}_{\perp}^{(0)}(\vec{r},\vec{r},\omega)\}}\textrm{d}\omega=           \frac{\hbar}{6\epsilon_{0}\pi^{2}c^{3}}\bar{\mathbb{I}}\:\int\textrm{d}\omega\:\omega^{3}.\label{disi2}
\end{eqnarray}
Only sourceless fluctuations satisfying the transversality condition $\nabla\cdot\vec{A}=0$ contribute in the above equations. Thus, the Coulomb gauge $\nabla\cdot\vec{A}=0$ is a natural choice in this context. Longitudinal photons cannot propagate energy as such propagation needs of charged matter support. The Green's function in Eq.(\ref{disi2}) is that which propagates the electric field of an isolated oscillating stationary dipole,
\begin{equation}\label{Maxwellb}
\Bigr[\vec{\nabla}\times\vec{\nabla}\times-\frac{\omega^{2}}{c^{2}}\mathbb{I}\Bigl]\bar{G}^{(0)}(\vec{r},\vec{r}',\omega)
=-\delta^{(3)}(\vec{r}-\vec{r}').
\end{equation}
The left hand side of Eq.(\ref{Maxwellb}) is also the Maxwell's equation for the $\omega$-mode of the electric field in free space.
Although the real part of $\bar{G}^{(0)}(\vec{r},\vec{r}',\omega)$ diverges for $\vec{r}\rightarrow\vec{r}'$, its imaginary part yields the desired result in Eq.(\ref{disi2}) \cite{BulloughHynne}.\\
\indent Further,  $\Im{\{\bar{G}^{(0)}_{\perp}(\vec{r},\vec{r},\omega)\}}$ relates to the LDOS of the electric field which propagate in the bare vacuum through \cite{Bloch,Economou}
\begin{equation}\label{disort3}
\textrm{LDOS}_{\omega}^{0}]^{E}=-\frac{\omega}{2\pi c}\textrm{Tr}\Bigl\{\Im{\{\bar{G}^{(0)}_{\perp}(\vec{r},\vec{r},\omega)\}}\Bigr\}.
\end{equation}
Same expression holds for the LDOS of the electromagnetic field $\vec{A}^{\omega}$, LDOS$_{\omega}^{0}$ \cite{Agerwal}.
It follows that the spectrum of fluctuations of $\vec{E}^{\omega}$ and LDOS$_{\omega}^{0}$ are proportional,
\begin{equation}\label{diso3}
\textrm{LDOS}_{\omega}^{0}=\frac{\epsilon_{0}c}{2\hbar\omega}
\textrm{Tr}\{\langle0|\hat{\vec{E}}_{\perp}^{\omega}(\vec{r})\hat{\vec{E}}_{\perp}^{\omega\dag}(\vec{r})|0\rangle\},
\end{equation}
and the total EM energy of the bare vacuum per unit volume reads
\begin{eqnarray}
\mathcal{E}^{0}&=&c^{-1}\int\hbar\omega\:\textrm{LDOS}_{\omega}^{0}\:\textrm{d}\omega\nonumber\\&=&\frac{\epsilon_{0}}{2}
\int\textrm{Tr}\{\langle0|\hat{\vec{E}}_{\perp}^{\omega}(\vec{r})\hat{\vec{E}}_{\perp}^{\omega\dag}(\vec{r})
|0\rangle\}\:\textrm{d}\omega=\frac{\hbar}{4\pi^{2}c^{3}}\int\textrm{d}\omega\:\omega^{3}.
\label{disi3}
\end{eqnarray}
The above integral has no cut-off and diverges as $\sim\omega^{4}$. That is the so-called zero-point-energy (ZPE) and the corresponding fluctuations are referred to as zero-point vacuum fluctuations.\\
\indent When the electromagnetic interaction turns on (i.e. $e\neq0$), the theory is that of QED. The Feynman rules are the tree-level diagrams in Fig.\ref{FIGnew1}($a$).  Loop corrections give rise to fermion mass renormalization, interaction vertex renormalization and electric charge renormalization -- Figs.\ref{FIGnew1}($b,c,d$) respectively. The latter is also responsible for the (virtual) polarization of the EM vacuum due to charged-field fluctuations. In Fig.\ref{FIGnew1p} we depict a number of diagrams contributing to the ZPE of QED. The diagrams in Fig.\ref{FIGnew1p}($a$) and  Fig.\ref{FIGnew1p}($d$) refer to the zero-point fluctuations of the non-interacting fields. In the rest, vacuum fluctuations of both fields combine. Diagrams $(b)$ and ($c$) refer to fluctuations in a polarized EM vacuum.  The fermionic loop of Fig.\ref{FIGnew1}$(d)$ enters the diagrams of the upper row as a one-loop polarization function. The pairs of diagrams ($b$) and ($e$), ($c$) and ($f$) are actually identical. That is, from the point of view of the EM field, fermionic loops polarize the EM vacuum. From the point of view of charged fields, radiative corrections renormalize their charge. In any case, the inclusion of these fluctuations does not make any better the convergence of the ZPE. In addition, at finite temperature, the spectrum of thermal fluctuations is Planck's and additional photon interaction vertices show up --see \cite{Ranvdal}.\\
\indent Next, let us introduce actual charges. Let us consider an isolated atomic dipole source made of a positively charged heavy nucleus and a number of negatively charged electrons in motion around it. In this scenario longitudinal photons are also relevant for they mediate the electrostatic interaction between charges. The primitive stationary atomic states incorporate actual charges and longitudinal photons. They are eigenstates of the Hamiltonian
\begin{equation}
\hat{H}_{0}=\sum_{i=1}^{Z}\frac{1}{2m_{e}}\:|\hat{\vec{p}}_{i}|^{2}+\sum_{i>j}\frac{\tilde{e}^{2}}{\widehat{|\vec{r}_{i}-\vec{r}_{j}|}}
+\sum_{i=1}^{Z}\frac{\tilde{e}Q}{\widehat{|\vec{r}_{i}|}},
\label{Hmat}
\end{equation}
where $Z$ is the atomic number; $Q$ is the total charge of the nucleus placed at the center of the atom; the subscripts $i,j$ label each electron with renormalized mass and charge $m_{e}$ and $\tilde{e}$ respectively; and $\hat{\vec{p_{i}}}(x_{i},t)$ is the ordinary linear momentum operator of the $i^{th}$ electron. The radiative interactions which are not yet integrated in stationary states (i.e. transverse photons) enter the (non-relativistic) Hamiltonian of interaction \cite{Sakurai},
\begin{equation}\label{hint}
\hat{H}_{int}=\sum_{i}-\frac{\tilde{e}}{m_{e}c}\hat{\vec{A}}_{\perp}(r_{i},t)\cdot\hat{\vec{p_{i}}}(r_{i},t)
+\frac{\tilde{e}^{2}}{2m_{e}c^{2}}\hat{\vec{A}}_{\perp}(r_{i},t)\cdot\hat{\vec{A}}_{\perp}(r_{i},t),
\end{equation}
where the magnetic interaction due to spin coupling has been neglected. The dominant interaction is given by the first term on the \emph{r.h.s.}  At leading order in $\tilde{e}$, the  representation of the self-energy of state A is  the second diagram on the \emph{l.h.s.} of the approx. symbol in Fig.\ref{FIGnew2}($d_{2}$). The dotted line between the points of emission and reabsorbtion there depicts the virtual transition state B. For typical frequency transition energies, the wavelength of the emitted photons is much longer than the radius of the atoms. In such a case, the first term in Eq.(\ref{hint}) reduces to
\begin{equation}\label{dipole}
-\frac{\tilde{e}}{m_{e}c}\hat{\vec{A}}_{\perp}(x_{i},t)\cdot\hat{\vec{p_{i}}}(x_{i},t)\simeq-\tilde{e}\hat{\vec{x}}\cdot\hat{\vec{E}}_{\perp}(x_{i},t)=
-\hat{\vec{\mu}}_{i}\cdot\hat{\vec{E}}_{\perp}(x_{i},t),
\end{equation}
where $\hat{\vec{\mu}}_{i}$ is the electrostatic dipole moment operator associated to the $i^{th}$ electron. This is the electric dipole approximation, restricted to transverse modes in free space, that is used in Eq.(\ref{jj}). This way, the propagator of virtual photons in the self-energy diagrams can be reduced to the electric field propagator. That propagator is also the scattering amplitude, computed at second-order of perturbation theory in QED, of the process through which an excited two-level point dipole transfers a quantum of energy (not necessarily by means of radiation) to a non-excited dipole. One can verify for instance in
\cite{Andrews} that such a scattering amplitude equals the Green's function of Eq.(\ref{Maxwellb}),
\begin{eqnarray}
\bar{G}^{(0)}(R,\omega_{AB})&=&[\vec{\mu}_{AB}\vec{\mu}_{BA}]^{-1}\sum_{\gamma_{\vec{q},\kappa}}\Bigl[\frac{\langle A,B|\otimes \langle0|\:\hat{H}_{int}\:|\gamma_{\vec{q},\kappa}\rangle\otimes|A,A\rangle\langle A,A|\otimes\langle\gamma_{\vec{q},\kappa}|\:\hat{H}_{int}\:|0\rangle\otimes|A,B\rangle}{\hbar\omega_{AB}-\hbar cq}\nonumber\\&+&\frac{\langle A,B|\otimes \langle0|\:\hat{H}_{int}\:|\gamma_{\vec{q},\kappa}\rangle\otimes|B,B\rangle\langle B,B|\otimes\langle\gamma_{\vec{q},\kappa}|\:\hat{H}_{int}\:|0\rangle\otimes|A,B\rangle}{-\hbar\omega_{AB}-\hbar cq}\Bigr]\label{Gqm},
\end{eqnarray}
where $\hat{H}_{int}$ is given by Eq.(\ref{dipole}), $R$ is the distance between the dipoles and the sum runs over all possible values of the momentum $\vec{q}$ and polarization states $\kappa$ of the only photon in the intermediate states, $|\gamma_{\vec{q},\kappa}\rangle\otimes|A,A\rangle$ and $|\gamma_{\vec{q},\kappa}\rangle\otimes|B,B\rangle$.\\
\indent In application of the fluctuation-dissipation theorem in-free-space we can write
\begin{eqnarray}\label{dison3}
\Gamma_{0}&=&
\frac{2\pi}{\epsilon_{0}\hbar}|\vec{\mu}_{AB}|^{2}\:_{1D}^{s.p.}\langle \Omega|\hat{\vec{E}}_{\perp}^{\omega_{AB}}(\vec{r})\cdot\hat{\vec{E}}_{\perp}^{\dagger\omega_{AB}}(\vec{r}')
|\Omega\rangle_{1D}^{s.p.}\nonumber\\&=&
-\frac{2\omega_{AB}^{2}}{3\epsilon_{0}\hbar c^{2}}|\mu_{AB}|^{2}\Im{\Bigl\{\textrm{Tr}_{-}\{\bar{G}^{(0)}_{\perp}(\vec{r},\vec{r}';\omega_{AB})\}\Bigr\}},
\qquad\quad|\vec{r}-\vec{r}'|\ll a\nonumber,
\end{eqnarray}
where $|\Omega\rangle^{s.p.}_{1D}$ stands for the EM self-polarization vacuum seen from the unique dipole in-free-space. Because $\bar{G}^{(0)}_{\perp}(\vec{r},\vec{r}';\omega_{AB})$ above equals that in Eq.(\ref{Maxwellb}) for photons propagating through the bare zero-point vacuum, we conclude that the spectrum of fluctuations of $|\Omega\rangle^{s.p.}_{1D}$ forms part of the spectrum of fluctuations of $|0\rangle$.\\
\indent In a sense, it can be interpreted that zero-point EM fluctuations interact with the dipole to make it emit a photon. However, there is no 'borrowing' of vacuum energy as suggested in the literature (eg. \cite{Andrews}). The zero-point EM fluctuations and hence the associated ZPE remain as depicted in Figs.\ref{FIGnew2}($c_{1},d_{1}$). Polarized fluctuations add up on top of those which yield the ZPE. On the one hand, fluctuations of the self-polarization vacuum 'interact' with the actual dipole making it fluctuate and acquire an additional self-energy. Reciprocally, the dipole randomly localized yields an additional effective vertex of interaction which enters polarizing the EM fluctuations of the sourceless vacuum $|\Omega\rangle^{s.l.}_{1D}$. In Fourier space, the vertex read $-(\omega/c)^{2}\mathcal{V}^{-1}\alpha_{0}$, where $\alpha_{0}$ is the electrostatic bare polarizability of the dipole and $\mathcal{V}$ the total volume of the sample (eventually infinite). Its diagrammatic representation is that in Fig.\ref{FIGnew4}($a$). The reason why we depict the bare dipole there as a closed loop is that it is a bound state. Momentum and energy can only leak out of it through the coupling to EM modes. It behaves w.r.t. the EM fluctuations in $|\Omega\rangle^{s.l.}_{1D}$ as one of the loops of fermionic fluctuations. In the process of joining the external fermionic legs of the QED diagrams of Fig.\ref{FIGnew4}($a$) in order to get an $\alpha_{0}$ loop there is an implicit integration of high frequency degrees of freedom. Those are, all the electrostatic interactions which yield the stationary atomic state plus high frequency transverse modes. This implies that there is a frequency cut-off in all the processes in which $\alpha_{0}$ plays the role of an effective vertex. If the radius of the dipole is $a$, the cut-off must be of the order of $c/a$ to preserve consistency with the dipole approximation. An analogous reasoning may apply to restrict the interval of frequencies which are integrated in the Casimir energy of a system formed by two perfectly conducting parallel plates. In that case, the restriction is imposed to preserve consistency with the perfect conductivity approximation \cite{Miltonbook,Greiner}.\\
\indent So far, we have distinguished between the EM vacuum  seen by sourceless photons (i.e. normal EM modes), $|\Omega\rangle^{s.l.}_{1D}$, and that seen by the photons whose source is the dipole, $|\Omega\rangle^{s.p.}_{1D}$. This distinction is based on the double role played by loops of actual charged-fields and loops of virtual photons. Those are, the renormalization of the bare polarizability and the polarization of the EM vacuum. The same as shown in Fig.\ref{FIGnew1p} for the ZP QED vacuum, both effects are complementary. The EM vacuum seen from the emitter was identified above, at least partially, with the bare zero-point vacuum. That is, the radiation emitted by the dipole sees empty space in its way. Likewise, self-polarization photons see empty space from and back to the dipole. However, the restrictions $\omega<c/a$ must apply in order to consider the emitter as a point dipole. Hence, the equivalence $\vec{r}=\vec{r}'$ implicit in Eqs.(\ref{EA}-\ref{disi3}) has been substituted by the limit $|\vec{r}-\vec{r}'|\ll a$ in Eq.(\ref{dison3}) to set explicitly the limitation in spatial resolution. Therefore we will write $|\Omega\rangle^{s.p.}_{1D}=|0\rangle|_{\omega<c/a}$. The associated LDOS is
\begin{equation}
\textrm{LDOS}^{emission}_{1D}\simeq\frac{\omega^{2}}{4\pi^{2}c^{2}}\Bigr|_{\omega<c/a}.
\end{equation}
In an analogous manner that fermionic loops renormalize the fine-structure constant $\sim e^{2}$ in QED through its coupling to radiation, the $\alpha_{0}$-loop renormalizes the value of $\alpha_{0}$ itself. This self-polarization effect is diagrammatically given by the geometric series in Fig.\ref{FIGnew5}. It yields the renormalized polarizability $\alpha$ in free space. Note that the photon loops in Fig.\ref{FIGnew5}($c$) denote $\mathcal{\bar{G}}^{\omega}(\vec{r},\vec{r})$ which equals $\bar{G}^{(0)}(\vec{r},\vec{r};\omega)$ in free space, where $\vec{r}$ is the position vector of the dipole.\\
\indent On the contrary, the vacuum that sourceless normal modes see, $|\Omega\rangle^{s.l.}_{1D}$, is polarized by the presence of the dipole. In other words, radiation traveling from infinity and normal modes can meet a randomly distributed scatterer in their way. Its EM fluctuations are made of two additive contributions. The first one  yields the divergent ZPE. The second one is depicted by those diagrams in which virtual photons of frequency less than $c/a$ scatter with the dipole of 'dressed' polarizability $\alpha$ --see Fig.\ref{FIGnew6}. Note the difference with respect to those in Fig.\ref{FIGnew5}($c$). In the diagrams of the polarized sourceless EM vacuum two propagators $\bar{G}^{(0)}_{\perp}$ attach to the dipole from below. That is so because the whole series in Fig.\ref{FIGnew6} denote the bulk loop-propagator, $\Im{\{\bar{G}_{\perp}(\vec{r},\vec{r})\}}$, where $\vec{r}$ is any point. In contrast, the loops in Fig.\ref{FIGnew5}($c$) stand for $\Im{\{\bar{\mathcal{G}}(\vec{r},\vec{r})\}}=\Im{\{\bar{G}^{(0)}(\vec{r},\vec{r})\}}$ for $\vec{r}$ being the position vector of the dipole.
The physical distinction between $\bar{G}$ and $\mathcal{\bar{G}}$  will be explained in due course. The LDOS for sourceless-normal modes reads
\begin{equation}\label{uma}
\textrm{LDOS}^{sourceless}_{1D}\simeq\textrm{LDOS}^{0}_{\omega}+\frac{\Re{\{\alpha\}}}{2\mathcal{V}}
\frac{\omega^{2}}{4\pi^{2}c^{2}}\Bigr|_{\omega<c/a}.
\end{equation}
Note that, because $\alpha$ depends on the fluctuations of $|\Omega\rangle^{s.p.}_{1D}$, LDOS$^{emission}_{1D}$ is implicitly accounted for in the second term on the \emph{r.h.s} of Eq.(\ref{uma}).\\
\indent At leading order, the counterpart of the addition of the diagrams of Figs.\ref{FIGnew1p}$(a),(b)$ of the virtually polarized vacuum is the sum on the \emph{r.h.s} of the approx. symbol in Fig.\ref{FIGnew2}$(d_{1})$ of the actually polarized vacuum. Likewise, the counterpart of the addition of the diagrams in Figs.\ref{FIGnew1p}$(d),(e)$ for the renormalization of the fine structure constant, is the sum of the diagrams on the \emph{r.h.s} of the approx. symbol in Fig.\ref{FIGnew2}$(d_{2})$ for the renormalization of the polarizability. However, the equivalence between the last diagram on Fig.\ref{FIGnew2}$(d_{1})$ and the last diagram on Fig.\ref{FIGnew2}$(d_{2})$ does not hold because the 'fermionic loop' in Figs.\ref{FIGnew1p}$(d),(e)$ turn into an actual dipole in Fig.\ref{FIGnew2}. Although both sourceless and sourced
photons see statistically homogeneous vacua, $|\Omega\rangle^{s.p.}$ is attached to actual dipoles while $|\Omega\rangle^{s.l.}$ is translation invariant.
As a matter of fact, the series in Fig.\ref{FIGnew2}($d_{2}$) stands for the self-energy of the actual dipole.\\
\indent We next overview the renormalization of the polarizability in-free-space.
 As mentioned above, the imaginary part of $\bar{G}^{(0)}(\vec{r},\vec{r})$ yields the right result consistent with the quantum radiative corrections which dress up the electrostatic
polarizability \cite{BulloughHynne}. However, it contains both longitudinal and real transverse parts which diverge. The phenomenological regularization scheme that we use in Section \ref{Sect3C} follows that of \cite{RMPdeVries}. It allows
us to attribute a physical interpretation to the divergences.
The longitudinal divergence is associated to the longitudinal modes which are integrated out in the original
stationary state. Therefore, it relates to $\mu_{AB}$ and $\omega_{AB}$. The associated electrostatic
polarizability $\alpha_{0}$ is proportional to $\mu_{AB}^{2}$ in a two-level atom. The real
transverse divergence is interpreted in terms of a shift in the transition energy with respect to the
 stationary original state. That gives rise to a static shifted polarizability,
$\alpha_{stat.}^{0}(\omega)=\alpha_{0}\frac{\omega_{res}^{2}}{\omega_{res}^{2}-\omega^{2}}$, $\alpha_{0}$
 being real. The shift in energy
 is the analog to the Lamb shift in the Hydrogen atom. That scheme of renormalization yields the
 Lorentzian-type polarizability in free space
$\alpha(\omega)\approx\frac{\alpha_{0}\omega^{2}_{res}}{\omega_{res}^{2}-\omega^{2}+i\Gamma_{0}}$.
There have been proposed however several schemes to accomplish the computation of $\alpha$ in-free-space
for  idealized two and three-level atoms. We refer to the reader to the works of Berman, Boyd and Milonni
\cite{BerBoMil} for a computation based on Heisenberg and Schr\"{o}dinger's pictures; to that of Barnett and Loudon \cite{BarLou} for a Green function approach; and to that of Bialynicki-Birula and Sowinski for an approach based on Feynman diagrams \cite{Birula}. Even in the latter work, the authors begin with \emph{a priori} modeled electronic bound states and perform a dimensional reduction from three to zero spatial dimensions.
It has been pointed out in \cite{BerBoMil} that a Lorentzian profile is an approximation not fully justified
and it has been found in \cite{MilLouBerBar} that other more complicated form can be more suitable.
We will not attempt here to analyze those approaches and we will stick to the phenomenological Lorentzian profile. Also, we will adapt in Section \ref{Sect3}  a phenomenological treatment appealing to the existence of a classical effective dielectric constant within the scatterers. That procedure is in all equivalent to the actual QED treatment once divergences in $\bar{G}^{(0)}(\vec{r},\vec{r},\omega)$ are conveniently regularized.
\subsection{In-random-medium vacuum fluctuations}\label{Sect2C}
\indent  In a random medium, on top of the  in-free-space fluctuations, additional fluctuations are induced by the multiple scattering of light with the host scatterers and the stochasticity of the medium itself. The additional fluctuations affect both the polarization of the sourceless EM vacuum and the self-energy of matter.\\
\indent The ondulatory nature of light implies that fluctuations appear as considering the interference between all the different paths that a virtual photon can follow in its way. The particularity of a random medium with respect to the free space is that the different paths are not statistically equivalent as a result of the spatial correlations between host scatterers. Therefore, each path has a statistical weight and fluctuations are affected by the randomness of the ensemble of scatterer configurations. In turn, this gives rise to a stochastic equation whose Green function incorporates the EM fluctuations we seek for. Purely thermal fluctuations of the electromagnetic field will not be considered in this paper. However, it can be intended that the fluctuations induced by spatial correlations have an indirect thermal origin. That is, it is the classical degrees of freedom which characterize the configuration of host scatterers, vector positions, velocities and angular momenta, that obey a thermal distribution. In our approach, atomic orbitals do not overlap and the separation between the emitter and the host scatterers and the separation between host scatterers themselves are determined by the range and strength of the cohesive forces --but for the dipole interaction- as well as the temperature. In fluids such  atomic gases, one can think of that force as the short-ranged repulsive force between the external electrons of the atoms. In experiments, the range of this force can be tuned by ionazing the atoms or molecules \cite{PRLRojasFranck}. For practical uses, throughout this paper we will consider an effective rigid exclusion volume around both the emitter and the host scatterers. In the construction of the dielectric constant in Section \ref{Sect6B} we will only assume knowledge of the in-free-space polarizabilities of the atomic spaces considered as isolated. In crystals and tight-binding models in general, atomic orbitals naturally overlap and the electronic band structure of the system is completely different to that of the atomic species. The assignment of a 'new' single polarizability to each atomic site might be still possible  provided that the only remaining interaction between individual atoms is effectively dipole-like. Nevertheless, for the case that the emitter is a distinguishable dipole, the only relevant information for studying the influence of the medium in its decay rate is that of the spatial embedding of the emitter into the medium provided the medium is correctly described by an effective dielectric constant --see Section \ref{Sect4A}.\\
\indent We summarize here the basis of our approach. In the process of dressing up the electrostatic polarizability $\alpha_{0}$, all the virtual photons running inside each dipole are assumed integrated out. The interaction between atoms is dipole-like provided atomic orbitals do not overlap. The 'bare' propagator of such an interaction is the one given by Eq.(\ref{Gqm}) and the 'bare' interaction vertex is given by the strength coupling $-k_{0}^{2}\rho\alpha_{0}$. This way, in-free-space radiative fluctuations and those induced by the host medium can be incorporated in the same footing. Quantization enters at the level of atomic polarizabilities, dipole-dipole interaction and EM fluctuations. In our approach, matter fluctuations enter classically through the implementation of the correlation functions between scatterers. These correlation functions are affected neither by the dipole-dipole nor by the radiation-dipole interaction. Thus, the coupling between matter and radiation is assumed weak unless specified otherwise. Explicit treatment of quantum long wavelength excitons, typical of tight-binding models, is disregarded. The reader is referred to the pioneer works by Fano and Hopfield \cite{Fano,Hopfield} for more convenient approaches to highly ordered systems. As long as collective quantum  excitons can be neglected, and in the weak-coupling regime between medium and radiation, there is no need to develop any further macroscopic quantization scheme. In particular, there is no need to postulate any additional noise polarization operator --compare with \cite{KnollBarnett}. Only for the case that the emitter is a substantial impurity we will assume any other fluctuation incorporated in the given dielectric function of the host medium without further explicit computations --see Sections \ref{Sect4B},\ref{Sect7P}. Additional non-radiative effects, such as collisional effects, can be incorporated through convenient renormalization of the vertex $-k_{0}^{2}\rho\alpha_{0}$.\\
\indent Next, let us consider formally the host medium as an stochastic configuration of dipoles, where stochasticity  concerns the classical variables which determine the configuration.
In order to treat the  EM fluctuations induced by the classical matter fluctuations in the same footing as the rest, it
is necessary for them to fulfill some conditions. The Green's function which appears in the
fluctuation-dissipation theorem of Eq.(\ref{Disip}) refers to an effective and stationary configuration. That implies
 conditions over the dynamics of the ensemble of scatterers with respect to that of the emission process
and photon propagation. In the first place the host medium must pass through all the accessible
configurations in a time scale much less than $\Gamma^{-1}$. That is, the relaxation time of the ensemble,
 $\tau$, satisfies $\tau\ll\Gamma^{-1}$. This guarantees that the degrees of freedom of the host medium are
the fast variables which can be integrated out. Second, stationarity demands that $\tau\gg\omega^{-1}$.
In addition, it is implicit in the derivation of Fermi's Golden rule that $\Gamma^{-1}\gg\omega^{-1}$, which is a weak-coupling condition on the coupling radiation-emitter. That way, the spontaneous emission process in such a random medium adjusts to the Born-Markov approximation \cite{Breuer}. We can also appeal to the ergodic character of the electric field so that ensemble averages are equivalent to time averages \cite{Wolf,Glauber}.\\
\indent We follow next a similar treatment to that of \cite{Glauber} in the introduction of the matrix density operator. $|\Omega\rangle^{s.p.}$ is not $|0\rangle|_{\omega<c/a}$ any more as the surrounding dipoles polarize  $|\Omega\rangle^{s.p.}$ too. There exists a one-to-one correspondence between the spatial configurations of scatterers and the corresponding polarized vacuum states. Let us denote those vacuum states by  $\{|\phi_{m}\rangle^{s.p.}\}$. The density matrix operator $\hat{M}$ is a diagonal matrix whose entries $M_{mn}$ are the statistical weights of the vacuum states,
\begin{equation}
\hat{M}\equiv\sum_{m,n}M_{mn}|\phi_{m}\rangle^{s.p.\:s.p.}\langle\phi_{n}|, \qquad M_{mn}\propto\delta_{mn}.
\end{equation}
In the simplest case, $M_{mn}$ is a Boltzmann weight.
Application of the fluctuation-dissipation theorem over each EM vacuum yields
\begin{equation}
^{s.p.}\langle\phi_{m}|\hat{M}\hat{\vec{E}}^{\omega_{AB}}(\vec{r})\hat{\vec{E}}^{\omega_{AB}\dag}(\vec{r})|\phi_{n}\rangle^{s.p.}=
-\frac{\hbar\omega_{AB}^{2}}{\epsilon_{0}\pi c^{2}}\Im{\{\bar{\mathfrak{G}}_{mn}(\vec{r},\vec{r};\omega_{AB})\}M_{mn}},
\end{equation}
where the Green function $\bar{\mathfrak{G}}_{mn}(\vec{r},\vec{r};\omega_{AB})$ corresponds to that of the self-polarization field of a point dipole sited at $\vec{r}$ surrounded by the $m^{th}$ configuration of scatterers. As the medium is stochastic and statistically homogeneous and isotropic, the final expression for the  decay rate so obtained reads
\begin{equation}\label{laprimeras}
\Gamma=-\frac{2\omega_{AB}^{2}}{3\epsilon_{0}\hbar c^{2}}|\mu_{AB}|^{2}\Im{\Bigl\{\textrm{Tr}\{M_{mn}\bar{\mathfrak{G}}_{mn}(\vec{r},\vec{r};\omega_{AB})\}\Bigr\}},
\end{equation}
where Tr denotes the trace operator. The action of Tr must be intended not only over the spatial indices of the tensor propagator, but also over those indices corresponding to the space of EM vacuum configurations. The EM vacuum $|\Omega\rangle^{s.p.}$ is this way  a superposition of states. Following \cite{Glauber}, each of those vacuum states $|\phi_{m}\rangle^{s.p.}$ is weighted by the
root-squared of the coefficients of the density matrix, $|\Omega\rangle^{s.p.}=\sum_{m}\sqrt{M_{mm}}|\phi_{m}\rangle^{s.p.}$. In terms of it, we can write
\begin{equation}\label{Gcal0}
\sum_{m,n}\:^{s.p.}\langle\phi_{m}|\hat{M}\hat{\vec{E}}^{\omega_{AB}}(\vec{r})\hat{\vec{E}}^{\omega_{AB}\dag}(\vec{r})|\phi_{n}\rangle^{s.p.}
=\:^{s.p.}\langle\Omega|\hat{\vec{E}}^{\omega_{AB}}(\vec{r})\hat{\vec{E}}^{\omega_{AB}\dag}(\vec{r})|\Omega\rangle^{s.p.},
\end{equation}
Summing over the indices denoting the vacuum space configurations we obtain
\begin{equation}\label{Gcal}
\bar{\mathcal{G}}(\vec{r},\vec{r};\omega_{AB})=\sum_{m,n}\bar{\mathfrak{G}}_{mn}(\vec{r},\vec{r};\omega_{AB})M_{mn}.
\end{equation}
Note that in the above development we are implicitly employing a picture in which the electric field operator remains unperturbed under changes in the host medium while it is the vacuum state that changes.\\
\indent Alternatively, $\bar{\mathcal{G}}(\vec{r},\vec{r};\omega_{AB})$ is the Green function of the stochastic equation for the self-polarization field of a point dipole surrounded by a stochastic configuration of host scatterers. In turn, $\bar{\mathcal{G}}$ is determined by a stochastic kernel that relates to the electric susceptibility tensor, $\bar{\chi}$, which characterizes the medium. Finally, we can write $\vec{\mu}\cdot\bar{\mathcal{G}}(\vec{r},\vec{r};\omega_{AB})\cdot\vec{\mu}=
\frac{1}{3}|\mu|^{2}\textrm{Tr}_{-}\{\bar{\mathcal{G}}(\vec{r},\vec{r};\omega_{AB})\}$, where now the trace operator Tr$_{-}$ only acts over spatial indices and the factor $1/3$ amounts to averaging over the three equivalent directions so that
\begin{eqnarray}
\Gamma&=&-\frac{2\omega_{AB}^{2}}{3\epsilon_{0}\hbar c^{2}}|\mu_{AB}|^{2}\Im{\Bigl\{\textrm{Tr}_{-}\{\bar{\mathcal{G}}(\vec{r},\vec{r};\omega_{AB})\}\Bigr\}},\label{GammaF}\\
\textrm{and correspondingly, }\qquad\textrm{LDOS}^{emission}_{\omega}&=&-\frac{\omega}{2\pi c}\Im{\Bigl\{\textrm{Tr}_{-}\{\bar{\mathcal{G}}(\vec{r},\vec{r};\omega)\}\Bigr\}}\label{LDOSF}.
\end{eqnarray}
\indent In summary, during a time of the order of $\Gamma^{-1}$, self-polarizing photons of frequency $\omega_{AB}$ explore all the accessible states compatible with both energy conservation requirements and environmental constraints. To each configuration of the environment corresponds a different set of possible photonic paths. They all add up, interfere, and give rise to a spectrum of fluctuations $\sim\Im{\Bigl\{\textrm{Tr}_{-}\{\bar{\mathcal{G}}(\vec{r},\vec{r};\omega)\}\Bigr\}}$ which obey certain stochastic equation.\\
\indent The propagator $\bar{\mathcal{G}}(\vec{r},\vec{r};\omega)$ must not be confused with the Green's function of the normal modes which propagate in $|\Omega\rangle^{s.l.}$. The latter is denoted throughout this paper by $\bar{G}$ and is referred to as Dyson or bulk propagator. The LDOS associated to $\bar{G}$ is  LDOS$_{\omega}^{sourceless}$. These matters will be addressed in detail in the next Section.\\
\subsection{Longitudinal vs. transverse emission}\label{Sect3A}
As mentioned in Subsection \ref{Sect2B}, it is only in free-space that $\bar{\mathcal{G}}(\vec{r},\vec{r};\omega)$
equals the propagator of Maxwell's equation for the electric field in free space,
$\bar{G}^{(0)}(\vec{r},\vec{r};\omega)$.
$\bar{G}^{(0)}(\vec{r};\omega)$ consists of an electrostatic (Coulombian) dipole field propagator,
\begin{equation}
\bar{G}_{stat.}^{(0)}(r)=\Bigl[\frac{1}{k_{0}^{2}}
\vec{\nabla}\otimes\vec{\nabla}\Bigr]\Bigl(\frac{-1}{4\pi\:r}\Bigr)
\end{equation}
plus a radiation field propagator,
\begin{equation}
\bar{G}_{rad.}^{(0)}(r)=\frac{e^{i\:k_{0}r}}{-4\pi
r}\mathbb{I}+\bigl[\frac{1}{k_{0}^{2}}\vec{\nabla}\otimes\vec{\nabla}\bigr]\frac{e^{i\:k_{0}r}-1}{-4\pi r},
\end{equation}
with $k_{0}=\omega/c$. In reciprocal space and for isotropic systems, any tensor can be decomposed into a longitudinal and a transverse part with respect to the propagation direction of the wave vector $\vec{k}$. That is, $\bar{\mathcal{G}}(\vec{k})=\mathcal{G}_{\perp}(k)(\bar{\mathbb{I}}-\hat{k}\otimes\hat{k})+\mathcal{G}_{\parallel}(k)
\hat{k}\otimes\hat{k}$, where $\hat{k}$ is the unitary vector parallel to $\vec{k}$.
In free space,
\begin{eqnarray}
\bar{G}^{(0)}(\vec{k})=\int\textrm{d}^{3}r\:e^{i\vec{k}\cdot\vec{r}}\bar{G}^{(0)}(r)&=&
\int\textrm{d}^{3}r\:e^{i\vec{k}\cdot\vec{r}}\Bigl[\bar{G}^{(0)}_{rad.}(r)\cdot(\bar{\mathbb{I}}-\hat{k}\otimes\hat{k})
+\bar{G}^{(0)}_{stat.}(r)\cdot\hat{k}\otimes\hat{k}\Bigr]\nonumber\\&=&G_{\perp}^{(0)}(k)(\bar{\mathbb{I}}-\hat{k}\otimes\hat{k})+
G_{\parallel}^{(0)}(k)\hat{k}\otimes\hat{k},
\end{eqnarray}
with
\begin{equation}\label{losfreek}
G_{\perp}^{(0)}(k)=\frac{1}{k_{0}^{2}-k^{2}},\qquad G_{\parallel}^{(0)}(k)=\frac{1}{k_{0}^{2}}.
\end{equation}
While the radiative component is fully transverse, the electrostatic one is fully longitudinal.
In most of this work we will deal with the Fourier transform of $\bar{G}^{(0)}$. The reason being that it is easier to keep track of the longitudinal and transverse contributions to the emission by direct computation of
\begin{eqnarray}\label{b5}
\bar{\mathcal{G}}_{\omega}(\vec{r},\vec{r})&=&\frac{1}{3}\textrm{Tr}_{-}\{\bar{\mathcal{G}}_{\omega}(\vec{r},\vec{r})\}\bar{\mathbb{I}}
=\frac{1}{3}\textrm{Tr}_{-}\Bigl\{\int\frac{\textrm{d}^{3}k}{(2\pi)^{3}}\bar{\mathcal{G}}(\vec{k})\Bigr\}\bar{\mathbb{I}}\nonumber\\&=&
\frac{1}{3}\Bigl[\int\frac{\textrm{d}^{3}k}{(2\pi)^{3}}2\mathcal{G}_{\perp}(k)\:
+\:\int\frac{\textrm{d}^{3}k}{(2\pi)^{3}}\mathcal{G}_{\parallel}(k)\Bigr]\bar{\mathbb{I}},
\end{eqnarray}
where the factor $2$ in front of $\mathcal{G}_{\perp}(k)$ comes from Tr$_{-}\{\bar{\mathbb{I}}-\hat{k}\otimes\hat{k}\}$ and stands for the two transverse modes.\\
\indent In a generic medium, there does not exist direct identification between radiative and transverse emission and neither does it between electrostatic and longitudinal emission, but for in free space. In free space the whole radiation consists of transverse modes which are all coherent propagating modes as they satisfy the free-space (on-shell) dispersion relation $k^{2}=k_{0}^{2}$. That is, they are poles of $G^{(0)}_{\perp}(k)$. Because $G^{(0)}_{\parallel}$ does not contain poles and longitudinal modes do not couple to transverse radiative ones in free space, the above identifications are possible. On the contrary, the presence of a polarizable environment gives rise to non-coherent radiation as well as absorbtion which contain both longitudinal and transverse modes. Unless a very specific configuration of scatterers holds --eg. a dipole chain \cite{Citrus}-- the coherent propagation is entirely transverse and corresponds to the far-field coherent signal which would be received by a distant antenna in the medium. The incoherent radiation is the noise signal received by that antenna. It has its origin in the coupling of 'bare' radiative-transverse modes to 'bare' electrostatic-longitudinal modes and contains both transverse and longitudinal effective modes. The coupling takes place both at the host dipoles and the cavity surface. Note also that, differently to the self-polarization field in free space that appears in Eq.(\ref{dison3}) which is purely transverse, the one in Eq.(\ref{Gcal0}) and thereafter contains both transverse and longitudinal modes. The reason being that while all the longitudinal modes have been integrated out in the bare polarizability of the unique dipole in free space, still long wavelength longitudinal modes mediate the interaction (i.e. the induction) between distant dipoles in a random medium. Nevertheless, the fact that the emission be longitudinal does not imply that it is not radiative as erroneously interpreted in some works.\\
\indent In free space,
\begin{eqnarray}
\Im{\Bigl\{\textrm{Tr}_{-}\{\bar{\mathcal{G}}_{\omega}(\vec{r},\vec{r})\}\Bigr\}}|_{free}&=&\Im{\Bigl\{\textrm{Tr}_{-}
\{\bar{G}_{rad.}^{(0)}(\vec{r},\vec{r})\}\Bigr\}}
\nonumber\\&=&2\Im{\Bigl\{\int\frac{\textrm{d}^{3}k}{(2\pi)^{3}}\frac{1}{k_{0}^{2}-k^{2}}\Bigr\}}=-k_{0}/2\pi,
\end{eqnarray}
the emission is entirely radiative and, in application of Eq.(\ref{dison3}), we can write $\Gamma_{0}=\frac{\omega_{AB}^{3}}{3\pi\epsilon_{0}\hbar c^{3}}|\mu_{AB}|^{2}$, where the $0$ script refers to the use of in-free-space parameters. The real parts of both
$\gamma^{(0)}_{\perp}\equiv\int\frac{\textrm{d}^{3}k}{(2\pi)^{3}}2\:G^{(0)}_{\perp}(k)$ and
$\gamma^{(0)}_{\parallel}\equiv\int\frac{\textrm{d}^{3}k}{(2\pi)^{3}}G_{\parallel}(k)$ diverge and are needed of regularization. Because they are the zero-order terms of the perturbative expansion of $\textrm{Tr}\{\bar{\mathcal{G}}_{\omega}(\vec{r},\vec{r})\}$, we can write
\begin{eqnarray}
\textrm{Tr}\{\bar{\mathcal{G}}_{\omega}(\vec{r},\vec{r})\}&=&\int\frac{\textrm{d}^{3}k}{(2\pi)^{3}}2\mathcal{G}_{\perp}(k)\:
+\:\int\frac{\textrm{d}^{3}k}{(2\pi)^{3}}\mathcal{G}_{\parallel}(k)\label{yo}\\&\equiv&2\gamma_{\perp}^{Tot.}+\gamma_{\parallel}^{Tot.}
\:=\:[2\gamma_{\perp}^{(0)}+\gamma_{\parallel}^{(0)}]+2\gamma_{\perp}+\gamma_{\parallel},
\end{eqnarray}
where $2\gamma_{\perp}$ and $\gamma_{\parallel}$ are the divergenceless pieces. Physically, the $\gamma$-factors account for the dipole self-energy associated to intermediate EM states of frequency $\omega$. The self-polarization field propagated with $\bar{\mathcal{G}}_{\omega}$ is also referred to as \emph{local field}. The imaginary part of the $\gamma$-factors relates  to $LDOS^{emission}_{\omega}$ through
\begin{equation}\label{gammaLDOS}
\textrm{LDOS}^{emission}_{\omega}=-\frac{\omega}{2\pi c}\Im{\{2(\gamma^{(0)}_{\perp}+\gamma_{\perp})+\gamma_{\parallel}\}}.
\end{equation}
\indent It is also possible to relate $\Gamma$ with the power emitted in the decay process. That is, because such power is due to the self-interaction of the dipole with its own field, we can write
\begin{eqnarray}\label{laprimeraW}
W_{\mu}&=&\frac{\omega_{res}}{2}\Im{\{\vec{\mu}\cdot\vec{E}^{*}_{exc}\}}=
\frac{\omega_{res}^{3}}{2c^{2}\epsilon_{0}}\Im{\{\vec{\mu}\cdot\bar{\mathcal{G}}^{*}_{\omega_{res}}(\vec{r},\vec{r})\cdot\vec{\mu}^{*}\}}
\nonumber\\&=&
-\frac{\omega_{res}^{3}}{6c^{2}\epsilon_{0}}|\mu|^{2}\Im{\Bigl\{\textrm{Tr}\{\bar{\mathcal{G}}_{\omega_{res}}(\vec{r},\vec{r})\}\Bigr\}},
\end{eqnarray}
where in the second equality the exciting field $\vec{E}^{*}_{exc}$ has been identified with the self-polarization field, $\frac{\omega_{res}^{2}}{c^{2}\epsilon_{0}}\bar{\mathcal{G}}^{*}_{\omega_{res}}(\vec{r},\vec{r})\cdot\vec{\mu}^{*}$. The values of $\omega_{res}$ and $\vec{\mu}$ are those of the resonance frequency and the transition dipole in the medium (which may differ from the in-free-space values, $\omega_{AB}$, $\vec{\mu}_{AB}$). Comparing Eq.(\ref{LDOSF}) with Eq.(\ref{laprimeraW}) we can write $\Gamma_{\mu}=\frac{4}{\omega_{res}\hbar}W_{\mu}$.
\section{Spontaneous vs. Stimulated Emission}\label{Sect3}
\subsection{Stimulated emission, $W_{\omega}$}\label{Sect3B}
\begin{figure}[h]
\includegraphics[height=6.5cm,width=14.9cm,clip]{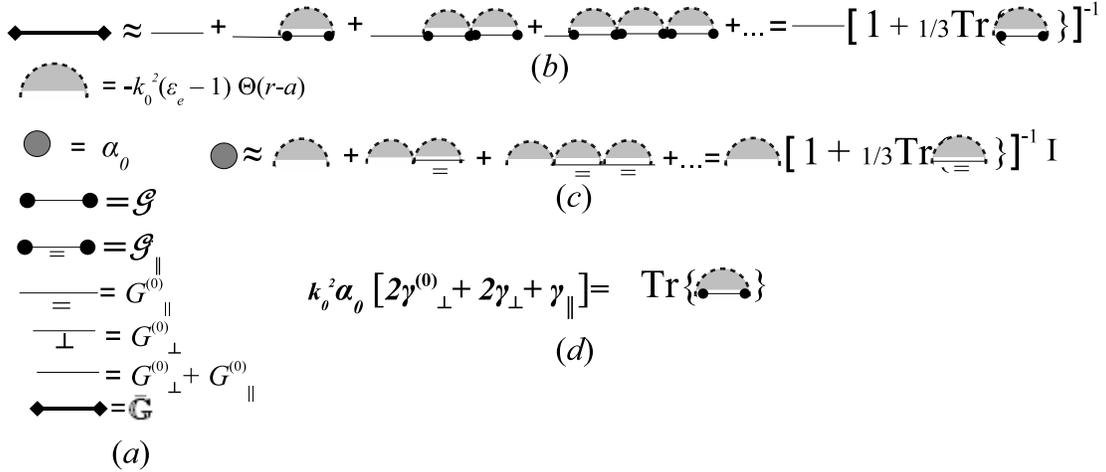}
\caption{($a$) Feynman's rules for the classical regularization scheme of Section \ref{Sect3B}. ($b$) Diagrammatic representation
of Eq.(\ref{b3}). ($c$) Diagrammatic representation of the
dressing up of $\chi_{e}$ leading to $\alpha_{0}$.
Approximation symbols denote that the field within the emitter is
taken uniform. ($d$) Diagrammatic representation of
 a the action of a self-polarization cycle on the electrostatic polarizability.}\label{fig21}
\end{figure}

\indent Let us considered first that the emitter is a dipole stimulated by an stationary external field. The
exciting field,
$\vec{E}_{exc}(\vec{r})=\vec{E}_{0}^{\omega}(\vec{r})$, oscillates in time with  frequency $\omega$ far from any resonance frequency of the dipole. In this case, the processes of absorbtion and emission of radiation by the emitter become stationary after a relaxation time of the order of $\Gamma^{-1}$, which is supposed much greater than the relaxation time of the host medium. Therefore, the process through which the absorbtion and emission become stationary is Markovian with respect to the dynamics of the host medium. The stochastic computation of $\mathcal{G}$ is so justified as in the spontaneous emission process. In order to regularize the divergences in $\bar{G}^{(0)}$ we consider the dipole as a spherical scatterer of radius $a$ and relative dielectric constant
$\epsilon_{e}$. This model implies a classical regularization scheme.
Again, the
dipole approximation requires  $a\ll k_{0}^{-1}$, with $k_{0}=\omega/c$,
$\lambda=2\pi/k_{0}$.
Formally, the emitted power reads,
\begin{equation}\label{lasimple}
W_{\omega}=\frac{\omega}{2}\Im{\{\int\textrm{d}^{3}r\:\Theta(r-a)
\vec{p}^{\omega}(\vec{r})\cdot\vec{E}_{0}^{\omega*}(\vec{r})\}},
\end{equation}
where $\vec{p}^{\omega}(\vec{r})$ is the density of dipole moment induced, which is proportional to $\vec{E}^{\omega}_{0}(\vec{r})$ in our linear and small particle approximation and is affected by self-polarization effects. We introduce the self-polarizing field through the insertion of appropriate Green's functions in the above expression,
\begin{equation}
\vec{p}^{\omega}(\vec{r})=\int\:\textrm{d}^{3}r''
\Theta(r-a)\epsilon_{0}\chi_{e}\int\textrm{d}^{3}r'
\bar{\mathbb{G}}_{\omega}(\vec{r},\vec{r}')\cdot[\bar{G}^{(0)}]^{-1}(\vec{r}',\vec{r}'')
\cdot\vec{E}^{\omega}_{0}(\vec{r}''),
\end{equation}
\begin{eqnarray}\label{lacomplex}
W_{\omega}&=&
\frac{\omega}{2}\Im{}\Bigl\{\int\textrm{d}^{3}r\:\chi_{e}\:\Theta(r-a)\int\textrm{d}^{3}r'
\textrm{d}^{3}r''
\bar{\mathbb{G}}_{\omega}(\vec{r},\vec{r}')\nonumber\\&\cdot&[\bar{G}^{(0)}]^{-1}(\vec{r}',\vec{r}'')
\cdot\vec{E}^{\omega}_{0}(\vec{r}'')\cdot
\vec{E}^{\omega*}_{0}(\vec{r})\Bigr\}.
\end{eqnarray}
In these expressions, $\chi_{e}=(\epsilon_{e}-1)$ is the relative electrostatic susceptibility of the emitter --not to be
confused with the susceptibility of the random medium-- and
\begin{equation}\label{b3}
\bar{\mathbb{G}}_{\omega}(\vec{r})\approx\bar{G}^{(0)}(\vec{r})
\sum_{m=0}^{\infty}\Bigl[-k_{0}^{2}\chi_{e}\int\Theta(v-a)\bar{\mathcal{G}}_{\omega}(v)\textrm{d}^{3}v\Bigr]^{m}
\end{equation}
is the propagator which makes account of the infinite number of self-polarization cycles which give rise to radiative corrections. $\bar{\mathbb{G}}_{\omega}(\vec{r}-\vec{r}')$ propagates virtual photons from a point $\vec{r}'$ inside the emitter back to another point $\vec{r}$ also within the emitter.  All the equations above become simple in the small particle limit, $a\ll k_{0}^{-1}$, for the electric field
is nearly uniform within the emitter and so are the density of dipole moment and the propagator
$\bar{\mathbb{G}}_{\omega}(\vec{r},\vec{r}')$. The $n$-points irreducible diagrams which enter the computation of $\bar{\mathbb{G}}_{\omega}(\vec{r})$
can be approximated by the series of Fig.\ref{fig21}($b$) in which the two-point correlation functions $\Theta(r-a)$ appear consecutively as factors of a product. That way the corresponding integrals appear disentangled and the corresponding series becomes  geometrical. The underlying approximation is
$\int\textrm{d}^{3}r\:\Theta(r-a)\bar{\mathcal{G}}_{\omega}(r)\simeq\frac{4\pi}{3}a^{3}\bar{\mathcal{G}}_{\omega}(0)=
\frac{4\pi}{3}a^{3}\Bigl[2\gamma_{\perp}^{(0)}+\gamma_{\parallel}^{(0)}+
2\gamma_{\perp}+\gamma_{\parallel}\Bigr]\frac{1}{3}\bar{\mathbb{I}}$. We already mentioned that  $2\gamma_{\perp}^{(0)}$ and $\gamma_{\parallel}^{(0)}$ are divergent. Those divergences are cured in our classical model by the presence of the finite radius $a$. Since the limit of the integral  Lim$\{\int\textrm{d}^{3}r\:\Theta(r-a)\bar{G}_{stat.}^{(0)}(r)\}=\frac{1}{3k_{0}^{2}}\mathbb{I}$ as $k_{0}a\rightarrow0$ is conditionally convergent, it is the Heviside function $\Theta(r-a)$ of the integrand which we used to model the spherical shape of the dipole which yields the finite value $\frac{1}{3k_{0}^{2}}\mathbb{I}$ \cite{Japa}. Any other geometry would give a different numerical value. By equating that result with  $\frac{4\pi}{3}a^{3}\gamma^{(0)}_{\parallel}\frac{1}{3}\bar{\mathbb{I}}$ (leaving $\Re{\{2\gamma^{(0)}_{\perp}\}}$ still free)
we obtain $\gamma^{(0)}_{\parallel}=(\frac{4\pi}{3}a^{3}k_{0}^{2})^{-1}$. The net effect of this regularization procedure is the dressing up of the single particle
susceptibility with all the in-free-space electrostatic corrections. This procedure is depicted in Fig.\ref{fig21}($c$). Its quantum counterpart is the integration of the electrostatic interactions which give rise to an atomic bound state in a two-level atom.  That way we can define
$\tilde{\chi}_{e}\equiv\frac{3}{\epsilon_{e}+2}\chi_{e}$ and obtain the bare electrostatic polarizability $\alpha_{0}\equiv 4\pi
a^{3}\frac{\epsilon_{e}-1}{\epsilon_{e}+2}$. We emphasize that it is the electrostatic polarizability
$\alpha_{0}$ which really has physical meaning regardless of the regularization scheme applied in its computation.
We refer to section III of \cite{RMPdeVries} for a comprehensive summary of regularization methods.
With the above
definitions we can rewrite Eq.(\ref{lacomplex}) in terms of
electrostatically renormalized operators,
\begin{eqnarray}\label{laleche2}
W_{\omega}&=&
\frac{\omega\epsilon_{0}}{2}\Im{}\Bigl\{\int\textrm{d}^{3}r\:\tilde{\chi}_{e}\:\Theta(r-a)\int\textrm{d}^{3}r'
\textrm{d}^{3}r''
\bar{\tilde{\mathbb{G}}}_{\omega}(\vec{r},\vec{r}')\nonumber\\&\cdot&[\bar{G}^{(0)}]^{-1}(\vec{r}',\vec{r}'')
\cdot\vec{E}^{\omega}_{0}(\vec{r}'')\cdot
\vec{E}^{\omega*}_{0}(\vec{r})\Bigr\},
\end{eqnarray}
where
\begin{equation}\label{b4}
\bar{\tilde{\mathbb{G}}}_{\omega}(\vec{r},\vec{r}')\equiv\bar{G}^{(0)}(\vec{r},\vec{r}')
\sum_{m=0}^{\infty}(-k_{0}^{2}\alpha_{0})^{m}3^{-m}\Bigl(2\gamma^{(0)}_{\perp}+2\gamma_{\perp}+\gamma_{\parallel}\Bigr)^{m}.\nonumber
\end{equation}
Note that longitudinal and
transverse modes couple to each other in the series of Eq.(\ref{b4})
as scattering takes place off the dipole surface. This is the classic phenomenological analog to the quantum coupling of longitudinal and transverse modes of the diagrams of Fig.\ref{FIGnew2}($c_{2}$).
Finally, in function of the $\gamma$-factors the power emitted and absorbed by the induced dipole reads
\begin{eqnarray}
W_{\omega}&=&
\frac{\omega\epsilon_{0}}{2}\Im{}\Bigl\{\frac{\alpha_{0}}{1+\frac{1}{3}k_{0}^{2}\alpha_{0}[2\gamma^{(0)}_{\perp}+2\gamma_{\perp}+
\gamma_{\parallel}]}
\Bigr\}|E^{\omega}_{0}|^{2}\label{la1}\\&=&\frac{-\omega^{3}\epsilon_{0}}{6c^{2}}\Bigl\{\frac{|\alpha_{0}|^{2}}
{|1+\frac{1}{3}k_{0}^{2}\alpha_{0}[2\gamma^{(0)}_{\perp}+2\gamma_{\perp}+\gamma_{\parallel}]|^{2}}
\Im{\{2\gamma^{Tot.}_{\perp}
+\gamma^{Tot.}_{\parallel}\}}\label{la2}\\
&-&\frac{3}{k_{0}^{2}}
\frac{\Im{\{\alpha_{0}\}}}{|1+\frac{1}{3}k_{0}^{2}\alpha_{0}[2\gamma^{(0)}_{\perp}+2\gamma_{\perp}+\gamma_{\parallel}]|^{2}}\Bigr\}
|E^{\omega}_{0}|^{2}\label{la3}.
\end{eqnarray}
The term in Eq.(\ref{la3}) corresponds to the power
 absorbed within the emitter.  The term in Eq.(\ref{la2}) corresponds to the power radiated into the medium. The former is non-zero only if $\epsilon_{e}$ contains an imaginary part \footnote{In metals,  $\epsilon_{e}$  rather contains an imaginary part in the
denominator which accounts for dissipative damping in agreement with Drude's model.}.
We can write Eq.(\ref{la1}) in terms of a renormalized polarizability $\tilde{\alpha}$ as $W_{\omega}=\frac{\omega\epsilon_{0}}{2}\Im{\{\tilde{\alpha}\vec{E}_{0}^{\omega}\cdot(\vec{E}_{0}^{\omega})^{*}\}}=
\frac{\omega\epsilon_{0}}{2}|E_{0}^{\omega}|^{2}\Im{\{\tilde{\alpha}\}}$ with
\begin{equation}\label{alpha1}
\tilde{\alpha}=\frac{\alpha_{0}}{1+\frac{2}{3}k_{0}^{2}\alpha_{0}\gamma^{(0)}_{\perp}+\frac{1}{3}k_{0}^{2}\alpha_{0}
[2\gamma_{\perp}+\gamma_{\parallel}]},
\end{equation}
so that Eqs.(\ref{la2},\ref{la3}) can be written also as
\begin{eqnarray}\label{laWshort}
W_{\omega}=
-\frac{\omega^{3}\epsilon_{0}}{6c^{2}}|\tilde{\alpha}E_{0}^{\omega}|^{2}\Bigl[\Im{\{2\gamma^{Tot.}_{\perp}
+\gamma^{Tot.}_{\parallel}\}}-\frac{3}{k_{0}^{2}}\frac{\Im{\{\alpha_{0}\}}}{|\alpha_{0}|^{2}}\Bigr].
\end{eqnarray}
\subsection{Relation between induced and spontaneous emission for the Lorentz dipole model, $\Gamma_{\alpha}$}\label{Sect3C}
\indent The expression for $\tilde{\alpha}$ still contains a divergent term in the denominator,  $\frac{2}{3}k_{0}^{2}\alpha_{0}\gamma^{(0)}_{\perp}$. While its imaginary part is convergent and gives rise to the usual radiative corrections for an off-resonance point dipole in vacuum, $\alpha\simeq\alpha_{0}[1-\frac{i}{6\pi}k_{0}^{3}\alpha_{0}]^{-1}$, its real part diverges and needs of regularization. Following the regularization scheme of \cite{RMPdeVries} we can take advantage of this and incorporate the divergence into a resonance in $\alpha$. The procedure  must be compatible with the regularization scheme used for $\alpha_{0}$ and it is our choice that it reproduces the phenomenological Lorentzian model (L) for the polarizability of a two-level atom,
\begin{equation}\label{alphaLorentz}
\alpha_{L}(\tilde{k})=\widetilde{\alpha_{0}}k_{res}^{2}[k_{res}^{2}-\tilde{k}^{2}-i\Gamma_{\alpha}\tilde{k}^{3}/(c k_{res}^{2})]^{-1},
\end{equation}
where $k_{res}$ is the resonance wave number and $\widetilde{\alpha_{0}}$ is the 'renormalized' value of the electrostatic polarizability of the emitter, both evaluated within the medium. We proceed by equating Eq.(\ref{alpha1}) and Eq.(\ref{alphaLorentz}) in absence of absorbtion within the emitter and with all the parameters evaluated in free-space -- which is denoted with script 0,
\begin{equation}\label{tup}
\alpha_{0}[1-\frac{i}{6\pi}\alpha_{0}\tilde{k}^{3}+\frac{1}{3}\alpha_{0}\tilde{k}^{2}\Re{\{2\gamma_{\perp}^{(0)}\}}]^{-1}=
\alpha_{0}[1-\tilde{k}^{2}/k_{0}^{2}-i\Gamma_{0}\frac{\tilde{k}^{3}}{ck_{0}^{4}}]^{-1}.
\end{equation}
From Eq.(\ref{tup}) we identify $\Re{\{2\gamma_{\perp}^{(0)}\}}=\frac{-3}{k_{0}^{2}\alpha_{0}}$ and
$\Gamma_{0}=c\alpha_{0}k_{0}^{4}/6\pi$. We refer to $\alpha_{stat}^{0}=\alpha_{0}[1-\tilde{k}^{2}/k_{0}^{2}]^{-1}$ as the 'shifted' electrostatic polarizability at frequency $\omega=c\tilde{k}$. By consistency with the  decay rate of a dipole with transition amplitude $\mu_{0}$ in free-space according to Fermi's Golden rule, $\Gamma_{0}=\frac{k_{0}^{3}}{3\epsilon_{0}\pi\hbar}|\mu_{0}|^{2}$, we find the relation $\alpha_{0}=\frac{2|\mu_{0}|^{2}}{\epsilon_{0}\hbar c k_{0}}$. The latter can be found in textbooks (eg. \cite{Davydov,Band}). It is important to emphasize that, although the parametrization of $\alpha$ in Eq.(\ref{alphaLorentz}) is phenomenological and questionable on the  basis of some two-level atom quantum models \cite{MilLouBerBar}, the renormalization procedure after regularization of divergences is independent of such parametrization. Our renormalization scheme bases on the renormalization of the photon propagator in which, after regularization,  $-\tilde{k}^{2}\rho\alpha^{0}_{stat}$ enters as a point vertex. Hence, although the computed values of all the  renormalized parameters depend on the functional form attributed to the in-free-space polarizability, the local density of states  computed through Eq.(\ref{LDOSF}) is independent of such arbitrary choice provided that the value of $\alpha^{0}_{stat}$ is the correct one.\\
\indent In the host medium, the values of $\Gamma$, $k_{res}$ and $\widetilde{\alpha_{0}}$ get renormalized according to the equations,
\begin{eqnarray}\label{Gamares}
\Gamma_{\alpha}&=&-\frac{c}{3}\widetilde{\alpha_{0}}\tilde{k}^{3}\Im{\{2\gamma^{Tot.}_{\perp}+\gamma^{Tot.}_{\parallel}\}}|_{\tilde{k}=k_{res}}
\nonumber\\&=&-\Gamma_{0}\frac{2\pi}{k_{0}^{2}}\tilde{k}\Im{\{2\gamma^{Tot.}_{\perp}+\gamma^{Tot.}_{\parallel}\}}|_{\tilde{k}=k_{res}},
\end{eqnarray}
where $k_{res}$ is a real non-negative root of the equation
\begin{equation}\label{kres}
(\tilde{k}/k_{0})^{2}-1=\frac{1}{3}\alpha_{0}\tilde{k}^{2}\Re{\{2\gamma_{\perp}+\gamma_{\parallel}\}}|_{\tilde{k}=k_{res}},
\end{equation}
\begin{equation}\label{alpha0reg}
\textrm{ and }\quad\widetilde{\alpha_{0}}=\alpha_{0}(k_{0}/k_{res})^{2}.
\end{equation}
Eq.(\ref{kres}) is the Lamb-shift of the resonance frequency, which is as a result of the variation of the real part of the dipole self-energy \cite{Wylie}. In addition, Eqs.(\ref{Gamares}-\ref{alpha0reg}) together with the requirement of consistency with Fermi's Golden rule imply that  $\mu$ gets renormalized with respect to $\mu_{0}$.
\subsection{Combination of spontaneous and induced emission, $\Gamma_{\mu}^{\alpha}$}\label{Sect3D}
\begin{figure}[h]
\includegraphics[height=3.6cm,width=13.8cm,clip]{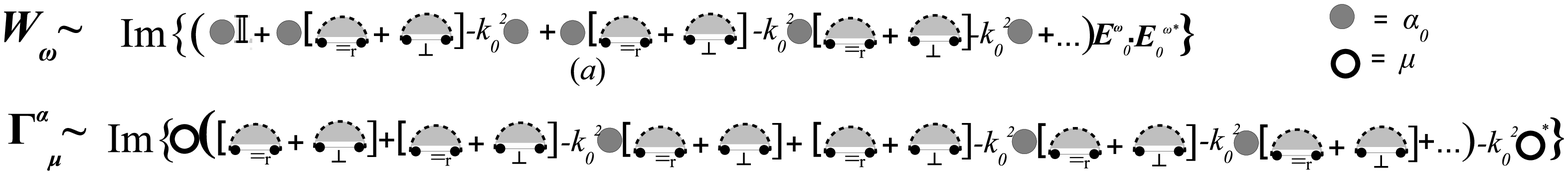}
\caption{($a$) Diagrammatic representation of Eq.(\ref{la1}).
 ($b$) Diagrammatic representation of Eqs.(\ref{latercera1},\ref{latercera2}).}\label{fig22}
\end{figure}

\indent Finally, consider the spontaneous emission of a point
dipole like that in Eq.(\ref{GammaF}), but now with an
additional bare electrostatic polarizability $\alpha_{0}$. The situation is analogous to
that of a fluorescent atom with transition dipole amplitude $\mu$ within a complex molecule. Let us assume that the value of $\mu$ is fixed so that $\mu$ gets effectively regularized only by the self-polarization cycles due to the presence of polarizable atoms within the host molecule, being $\alpha_{0}$ the total bare polarizability of the molecule. The net effect is that the spontaneous field emitted by the atom in the decay process gives rise to an induced dipole moment in the molecule which
modifies the decay rate. If there existed other kinds of interactions between the fluorescent atom and the host particle, $\mu$ and $k_{res}$ might be modified by additional non-radiative effects. In the following, we will refer to the atom as emitter, to the molecule of polarizability $\alpha_{0}$ as host particle and to the surrounding medium as host-medium.
The dipole moment of the system emitter-host-particle reads
$\vec{p}=\vec{\mu}+\frac{\omega^{2}}{c^{2}}\tilde{\alpha}\bar{\mathcal{G}}_{\omega}(\vec{r},\vec{r})\vec{\mu}$,
$\tilde{\alpha}$ being the renormalized polarizbility of the host
particle.  The
perturbative series in Fig.\ref{fig22} allows us to build up the analytical expression for $\Gamma^{\alpha}_{\mu}$ straight away from that of $W_{\omega}$. The integration of the electrostatic part which yields the term proportional to $\alpha_{0}\mathbb{I}$ must be removed with respect to the series in Fig.\ref{fig22}$(a)$ for in the present case the source is the spontaneous emitter with transition amplitude $\mu$. The rest of the terms remain unaltered but for the substitution of the induced non-radiative dipole moments $\epsilon_{0}\alpha_{0}\vec{E}^{\omega}_{0}$  and $\epsilon_{0}\alpha_{0}\vec{E^{*}}^{\omega}_{0}$ by the fixed dipole moments $\vec{\mu}$ and $\vec{\mu}^{*}$ at the emission and reception sites respectively,
\begin{eqnarray}
\Gamma^{\alpha}_{\mu}&=&
\frac{2\epsilon_{0}}{\hbar}|\mu|^{2}\Im{}\Bigl\{(\epsilon_{0}\alpha_{0})^{-2}\Bigl[\frac{\alpha_{0}}{1-\frac{i}{6\pi}k_{0}^{3}\alpha_{0}+\frac{1}{3}k_{0}^{2}\alpha_{0}[2\gamma_{\perp}+\gamma_{\parallel}]}
-\alpha_{0}\Bigr]\Bigr\}\label{latercera}\nonumber\\&=&\frac{-2\omega^{2}}{3\epsilon_{0}c^{2}\hbar}\frac{|\mu|^{2}}
{|1-\frac{i}{6\pi}k_{0}^{3}\alpha_{0}+\frac{1}{3}k_{0}^{2}\alpha_{0}[2\gamma_{\perp}+\gamma_{\parallel}]|^{2}}
\Bigl[\Im{\{2\gamma^{Tot.}_{\perp}+\gamma^{Tot.}_{\parallel}\}}\label{latercera1}\\
&-&\frac{k_{0}^{2}}{3}\Im{\{\alpha_{0}\}}|-i\frac{k_{0}}{2\pi}+2\gamma_{\perp}+\gamma_{\parallel}|^{2}\Bigr]\label{latercera2},
\end{eqnarray}
We recognize again  the power absorbed within the
host particle in the last term and the power radiated into the
medium in the remaining. Because any possible resonance of the host particle is assumed far from that of the emitter, we have ignored the regularization of $\Re{\{2\gamma^{(0)}_{\perp}\}}$.
\section{The virtual cavity scenario}\label{Sect4}
\subsection{Virtual cavity vs. real cavity}\label{Sect4I}
\indent The $\gamma$-factors are the traces of the solution of the Lippmann-Schwinger stochastic equation for the propagator of the self-polarization field of a point source, $\mathcal{G}(\vec{r},\vec{r'})$,
\begin{equation}\label{sthocastic}
\mathcal{G}_{ij}(\vec{r},\vec{r}')=G^{(0)}_{ij}(\vec{r},\vec{r}')-k_{0}^{2}\int\textrm{d}^{3}r_{1}
G^{(0)}_{ik}(\vec{r},\vec{r}_{1})\Bigl[\Bigl<\epsilon_{km}(\vec{r}_{1}-\vec{r})-\delta_{mj}\Bigr>\Bigr|_{\vec{r},\vec{r}'}\Bigr]
\mathcal{G}_{mj}(\vec{r}_{1},\vec{r}').
\end{equation}
See \footnote{In all the calculations of this Section we will evaluate the $\gamma$-factors at frequency $\omega=ck_{0}$. It must be intended as the frequency of the exciting field for the case of an induced dipole like that of Eqs.(\ref{la1}-\ref{la3}). It must be intended as the resonance frequency for the case of spontaneous emission by a  dipole with given transition amplitude $\mu$ like those of Eqs.(\ref{GammaF},\ref{latercera}-\ref{latercera2}). It must be intended as the frequency $c\tilde{k}$ of intermediate photons for the self-energy of a Lorentzian-type dipole like that of Eq.(\ref{Gamares}).} for an explanation about the notation. The brackets of the stochastic kernel $\Bigl[\Bigl<\epsilon_{km}(\vec{r}_{1}-\vec{r})-\delta_{mj}\Bigr>\Bigr|_{\vec{r},\vec{r}'}\Bigr]$ denote average taking over all possible configurations of the surrounding host scatterers and both $\vec{r}$ and $\vec{r}'$ are inside the emitter such that $|\vec{r}'-\vec{r}|<a\ll k_{0}^{-1}$. In turn, the distinction between $\vec{r}$ and $\vec{r}'$ is just formal as it cannot be resolved. The explicit dependence of $\epsilon(\vec{r}_{1}-\vec{r})$ on the emitter position $\vec{r}$ and the restriction symbolized by $|_{\vec{r},\vec{r}'}$ signal the fact that $\vec{r},\vec{r}'$ are kept fixed inside the emitter as performing the average over all possible configurations of host scatterers.  This way, any scattering process of the virtual photons in their way from and towards the emitter correlates to the emitter position. In case the emitter is in all equivalent to the rest of host scatterers within the host medium, such a correlation is the same as that among the host scatterers themselves. It is in this sense that translation invariance is only virtually broken and the exclusion volume around the dipole emitter is referred to as \emph{virtual cavity}. In case the emitter is distinguishable w.r.t. any other scatterer, that correlation differs. Hence, translation invariance is actually broken and the cavity is not
virtual but a \emph{real cavity}. The break of translation invariance, either virtually or actually, makes the computation of $\bar{\mathcal{G}}(\vec{r},\vec{r}')$ different to that of the ordinary Dyson propagator. This is the reason why the propagator appearing in Eq.(\ref{GammaF}) for Fermi's Golden rule is not Dyson propagator as erroneously suggested in \cite{RMPdeVries,RemiMole}. In any case, it is important to emphasize that a translation invariant Dyson propagator does exist for normal modes in the virtual cavity scenario. In contrast, a strictly translation invariant propagator for normal modes does not exist in the real cavity scenario.\\
\indent This section and the next one are devoted to the computation of $\bar{\mathcal{G}}$ and the $\gamma$-factors in both the virtual cavity (VC) and the real cavity (RC) scenarios. Notice that not only the equivalence/inequivalence of the spatial correlation of the emitter matters in order to define each scenario but also the equivalence/inequivalence of the polarizability of the emitter w.r.t. that of the host scatterers. That is so because the $\gamma$-factors also depend on the polarizability of the emitter. In particular, the VC scenario requires that the dipole transition amplitude of the emitter satisfies the relation,  $\alpha_{0}=\frac{2|\mu_{0}|^{2}}{\epsilon_{0}\hbar c k_{0}}$. $\alpha_{0}$ being the electrostatic bare polarizability of the host particles. As pointed out in \cite{Toptygin}, the bare polarizability of a two-level excited emitter has the opposite sign with respect to that of the same particle in its ground state \cite{Davydov}. More importantly, the polarizability during the decay process is undefined. Therefore, \emph{stricto sensu} only in the case of induced emission with all the scatterers of the medium in their ground state and for the case of a fixed dipole on top of a ground state host particle, emission in a virtual cavity scenario holds. Hence, it is only in these scenarios that a translation-invariant actually polarized sourceless vacuum can be defined.\\
\indent The case of emission from a real cavity is much more generic and its treatment, at least formally, more complicated. There are nevertheless two situations where calculations simplify. These are, in the first case the excited emitter is weakly-polarizable and occupies an interstitial position in the host medium. In the second case, the interaction emitter-host medium is such that the emitter expels the surrounding host particles at a distance $R$ much greater than the the typical distance between scatterers.
\subsection{Computation of the virtual-cavity $\gamma$-factors}\label{Sect4A}
\begin{figure}[h]
\includegraphics[height=4.7cm,width=12.5cm,clip]{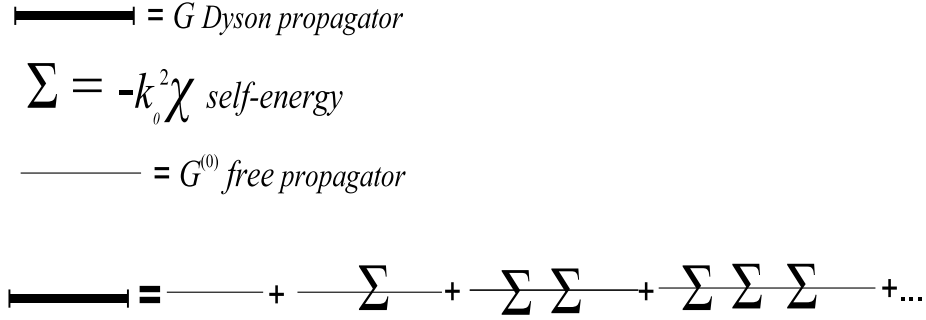}
\caption{Diagrammatic representation
of the Dyson propagator $\bar{G}$.}\label{FigI03}
\end{figure}

\begin{figure}[h]
\includegraphics[height=7.9cm,width=13.8cm,clip]{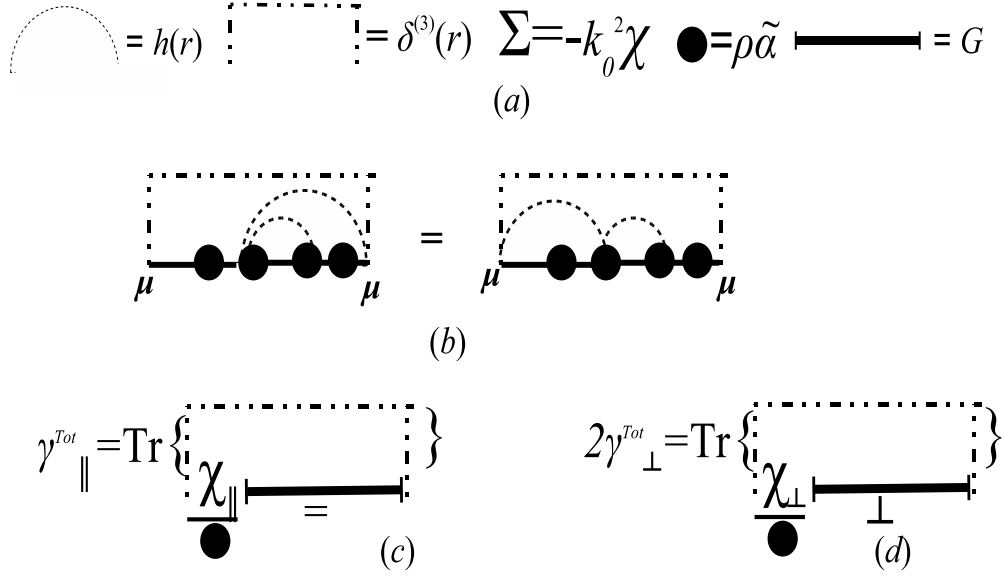}
\caption{($a$) Feynman rules. Only two-point irreducible correlation functions have been used for the sake of simplicity. ($b$) Diagrammatic representation of the
equivalence between multiple-scattering processes amounting to
$\bar{\mathcal{G}}$. ($c$),($d$) Diagrammatic representations of Eq.(\ref{LDOSIperg}) and Eq.(\ref{LDOSIparalg}) respectively.}\label{FigI04}
\end{figure}

\indent Rather than solving the stochastic integral equation Eq.(\ref{sthocastic}) directly, we will compute first the translation invariant bulk propagator $\bar{G}(\vec{r}-\vec{r}')$ which obey the Dyson equation,
\begin{equation}\label{bulk}
G_{ij}(\vec{r}-\vec{r}')=G^{(0)}_{ij}(\vec{r},\vec{r}')-k_{0}^{2}\int\textrm{d}^{3}r_{1}
G^{(0)}_{ik}(\vec{r},\vec{r}_{1})\Bigl[\Bigl<\epsilon_{km}(\vec{r}_{1}-\vec{r})-\delta_{mj}\Bigr>\Bigr]
G_{mj}(\vec{r}_{1},\vec{r}').
\end{equation}
In contrast to Eq.(\ref{sthocastic}), the average has no-constraint in this case.
Hence, $\bar{G}(\vec{r}-\vec{r}')$ is a function of $\vec{r}-\vec{r}'$ for any two given points in the bulk.
The field so propagated is the propagating coherent field or Dyson field, $\vec{E}_{D}(\vec{r})$. That is, the
field obtained by averaging point-wise the value of the classical Maxwell electric field $\vec{E}_{Max.}(\vec{r})$
in the ensemble of spatial host scatterers configurations,
$\vec{E}_{D}(\vec{r})=\Bigl\langle\{\vec{E}^{m}_{Max.}(\vec{r})\}\Bigr\rangle_{ensemble}$ --see Section \ref{Sect5B1} too.
The photons which propagate with $\bar{G}(\vec{r},\vec{r}')$ are said on-shell or normal modes as they obey certain dispersion relations --see below. Because the set of fields $\{\vec{E}^{m}_{Max.}(\vec{r})\}$ are in one-to-one correspondence with the set of scatterer configurations and since each one obeys a linear Maxwell equation with certain dielectric function $\bar{\epsilon}_{m}(r)$ specific of each particular configuration, $\vec{E}_{D}(\vec{r})$ obeys the same functional equation but with an averaged dielectric function $\Bigl\langle\{\bar{\epsilon}_{m}(r)\}\Bigr\rangle_{ensemble}$. Hence, Eq.(\ref{bulk}) \cite{Frisch}. There exists a relation between the propagators $\bar{G}$ of Eq.(\ref{bulk}) and $\bar{\mathcal{G}}$ of Eq.(\ref{sthocastic}) which we proceed to find out.\\
\indent In Fourier space, isotropy allows to split the Dyson
equation for $\bar{G}(k)$ in two uncoupled and mutually orthogonal algebraic equations,
\begin{eqnarray}
G_{\perp}(k)&=&G_{\perp}^{(0)}(k)\:-\:k_{0}^{2}\:G_{\perp}^{(0)}(k)\:\chi_{\perp}(k)\:G_{\perp}(k),\label{DysonI}\\
G_{\parallel}(k)&=&G_{\parallel}^{(0)}(k)\:-\:k_{0}^{2}\:G_{\parallel}^{(0)}(k)\:\chi_{\parallel}(k)\:G_{\parallel}(k),
\label{DysonII}
\end{eqnarray}
where $\bar{G}_{\perp}^{(0)}(k)$ and
$\bar{G}_{\parallel}^{(0)}(k)$ are given in Eq.(\ref{losfreek}) and $\bar{\chi}(\vec{k})$ is the electric susceptibility tensor such that $\bar{\epsilon}(\vec{k})=\bar{\mathbb{I}}+\bar{\chi}(\vec{k})$ is the effective dielectric tensor. Statistical isotropy allows to decompose $\bar{\epsilon}$ and $\bar{\chi}$ in longitudinal and transverse components, $\epsilon_{\perp,\parallel}(k)$, $\chi_{\perp,\parallel}(k)$.  For a medium made of a collection of disconnected dipoles, $\chi_{\perp,\parallel}(k)$ can be expanded as a series of one-particle-irreducible (1PI) multiple-scattering terms of order $n$,
\begin{equation}\label{laXenAes}
\chi_{\perp,\parallel}(k)=\sum_{n=1}^{\infty}X^{(n)}_{\perp,\parallel}(k)\rho^{n}
\tilde{\alpha}^{n}.
\end{equation}
The functions $\chi_{\perp,\parallel}$ are named 1PI as they contain integrals over undefined photon momenta which  cannot be disentangled \cite{Frisch,Peskin}.
In the above formula, $\rho$ is the average numerical volume density of scatterers and $\tilde{\alpha}$ is the renormalized polarizability  of single scatterers. The functions
 $X^{(n)}_{\perp,\parallel}(k)$  incorporate the spatial dispersion due to the spatial correlation within clusters of $n$ scatterers. In particular,  $X^{(1)}_{\perp}=X^{(1)}_{\parallel}=1$ incorporates all the self-correlation factors. In field theory terminology, $\bar{\chi}_{\perp,\parallel}$ are proportional to the photon self-energy functions, $\Sigma_{\perp,\parallel}(\vec{k})=-k_{0}^{2}\chi_{\perp,\parallel}(\vec{k})$ --not to be confused with the dipole self-energy. Alternatively, Eqs.(\ref{DysonI},\ref{DysonII}) can be written in terms of the so-called $T$-matrix, $T_{\perp,\parallel}(k)\equiv\Sigma_{\perp,\parallel}(k)+\Sigma_{\perp,\parallel}(k)G_{\perp,\parallel}(k)\Sigma_{\perp,\parallel}(k)$, as
\begin{equation}
G_{\perp,\parallel}(k)=G_{\perp,\parallel}^{(0)}(k)\:+\:G_{\perp,\parallel}^{(0)}(k)\:T_{\perp,\parallel}(k)\:
G^{(0)}_{\perp,\parallel}(k).\label{Dysonalter}
\end{equation}
\indent Two remarks are in order at this point. The first one concerns the topology of the host medium. Because the emitter is treated as a point emitter and it is equivalent to the host scatterers, a cermet topology is inherent to the virtual cavity scenario. In the expression of Eq.(\ref{laXenAes}), spatial dispersion in $\bar{\chi}$ cannot be disregarded as there must be at least a minimum distance $\xi$ between scatterers which determines their exclusion volume. Hence, this distance is the virtual cavity radius. For the sake of consistency with the linear medium approximation, $\xi$ must be greater than the dipole radius to avoid overlapping between orbitals and so that the dipole approximation be valid. Second, in view of
Eqs.(\ref{DysonI},\ref{DysonII}), longitudinal and transverse
modes of the Dyson field do not couple to each other as traveling
throughout a random medium the same as it occurs with photons in free
space. Eqs.(\ref{DysonI},\ref{DysonII}) can be solved
independently yielding the Dyson propagator components,
\begin{eqnarray}\label{effectivG}
G_{\perp}(k)&=&\frac{1}{k_{0}^{2}[1+\chi_{\perp}(k)]-k^{2}},\nonumber\\
G_{\parallel}(k)&=&\frac{1}{k_{0}^{2}[1+\chi_{\parallel}(k)]}.
\end{eqnarray}
In terms of free propagators and self-energy functions they can be depicted perturbatively as in Fig.\ref{FigI03}.
A more detailed examen in Section \ref{Sect5} will show that longitudinal and
transverse bare photons --i.e., the ones with propagator  $\bar{G}^{(0)}$--
do couple necessarily when they experience multiple scattering processes. In
Eqs.(\ref{DysonI},\ref{DysonII}), longitudinal and transverse bare
photons enter both $\chi_{\perp}(k)$ and $\chi_{\parallel}(k)$ by
means of the spatial correlations among scatterers.\\
\indent Getting back to the computation of $\bar{\mathcal{G}}$ and the $\gamma$-factors, we first observe that $2\gamma_{\perp}$ and $\gamma_{\parallel}$ are made of 1PI diagrams in which the starting and ending points coincide at the emitter location. Those diagrams belong to $\bar{\chi}(k)$ and amount to the so-called recurrent scattering. As a matter of fact, the emitter itself enters the bulk propagator of Eq.(\ref{effectivG}) as an ordinary scatterer in the VC scenario. The second observation is that, as the averaging process in Eq.(\ref{sthocastic}) is subject to the fixed location of the emitter, every scattering event in the 1PI diagrams contributing to $\gamma$ is correlated to the emitter either at the near end or at the rare end of each diagram, indistinguishably. This is a consequence of reciprocity. Taking advantage of this feature in every multiple-scattering diagram like that in Fig.\ref{FigI04}($b$), we can attribute all the irreducible correlations of the intermediate scattering events to the emitter on the
left. By so proceeding, we end up with an effective separation of all those pieces irreducibly correlated to the emitter on the left completely disentangled from those non irreducibly correlated which form non-1PI pieces on the right. The sum of the 1PI pieces on the left amounts to  $\bar{\chi}/\rho\tilde{\alpha}$, where the factor $1/\rho\tilde{\alpha}$ stands for amputation of the the first random scatter which enters the diagrams of $\bar{\chi}$ in favor of the 'virtually' fixed emitter location.
The sum of the non-1PI pieces on the right amounts to the bulk propagator $\bar{G}$. Therefore, we end up with the formulae \cite{MeI} --Figs.\ref{FigI04}($c,d$),
\begin{eqnarray}
\mathcal{G}^{VC}_{\perp}(k)&=&\frac{1}{\rho\tilde{\alpha}}\:\chi_{\perp}(k)\:G_{\perp}(k)=\frac{1}{\rho\tilde{\alpha}}\:
\frac{\chi_{\perp}(k)}{k_{0}^{2}[1+\chi_{\perp}(k)]-k^{2}},\label{LDOSIper}\\
\mathcal{G}^{VC}_{\parallel}(k)&=&\frac{1}{\rho\tilde{\alpha}}\:\chi_{\parallel}(k)\:G_{\parallel}(k)
=\frac{1}{\rho\tilde{\alpha}}\:\frac{\chi_{\parallel}(k)}{k_{0}^{2}[1+\chi_{\parallel}(k)]}.\label{LDOSIparal}
\end{eqnarray}
The above expressions contain both $G_{\perp}^{(0)}$ and $G_{\parallel}^{(0)}$ which carry divergences. The expressions
\begin{eqnarray}
2\gamma^{VC}_{\perp}&=&
\int\frac{\textrm{d}^{3}k}{(2\pi)^{3}}\Bigl[\frac{2\chi_{\perp}(k)/(\rho\tilde{\alpha})}{k_{0}^{2}
[1+\chi_{\perp}(k)]-k^{2}}-2G_{\perp}^{(0)}(k)\Bigr],
\label{LDOSIperg}\\
\gamma^{VC}_{\parallel}&=&\int\frac{\textrm{d}^{3}k}{(2\pi)^{3}}\:\Bigl[\frac{1}{\rho\tilde{\alpha}}
\frac{\chi_{\parallel}(k)}{k_{0}^{2}[1+\chi_{\parallel}(k)]}-G_{\parallel}^{(0)}(k)\Bigr],\label{LDOSIparalg}
\end{eqnarray}
are however fully convergent. We emphasize that the above expressions for the virtual-cavity $\gamma$-factors are exact under the assumption that the emitter can be treated as a point dipole equivalent
in all to the rest of host scatterers.
The explicit inclusion of $\tilde{\alpha}$ --as given by Eq.(\ref{alpha1})-- in the computation of  $\mathcal{G}_{\perp,\parallel}^{VC}$ through Eqs.(\ref{LDOSIper},\ref{LDOSIparal}) makes it depend implicitly on their integral quantities $\gamma_{\perp,\parallel}$ given in  Eqs.(\ref{LDOSIperg},\ref{LDOSIparalg}).
Later on we will see that the form of the above expressions for $2\gamma^{VC}_{\perp}$ and $\gamma^{VC}_{\parallel}$ will allow us to classify the nature of the emission in a transparent manner.\\
\indent Alternatively, we can write $\bar{\mathcal{G}}^{VC}(k)$ in other forms making use of  Dyson's equation. In function of the bulk propagator of Eqs.(\ref{DysonI},\ref{DysonII}) it reads
\begin{equation}\label{infunctionofG}
\mathcal{G}^{VC}_{\perp,\parallel}(k)=\frac{1}{k_{0}^{2}\rho\tilde{\alpha}}
\Bigl[1-\frac{G_{\perp,\parallel}}{G_{\perp,\parallel}^{(0)}}\Bigr].
\end{equation}
In function of the $T$-matrix of Eq.(\ref{Dysonalter}) it reads
\begin{equation}\label{infunctionofT}
\mathcal{G}^{VC}_{\perp,\parallel}(k)=\frac{-1}{k_{0}^{2}\rho\tilde{\alpha}}\:
T_{\perp,\parallel}(k)\:G^{(0)}_{\perp,\parallel}(k).
\end{equation}
\indent Finally, we write the Lippmann-Schwinger equation of Eq.(\ref{sthocastic}) for the self-polarization local field propagator in Fourier space,
\begin{eqnarray}
\mathcal{G}^{VC}_{\perp}(k)&=&G_{\perp}^{(0)}(k)\:+\:G_{\perp}^{(0)}(k)\:\Xi^{VC}_{\perp}(k)\:\mathcal{G}^{VC}_{\perp}(k),\label{DysonImathG}\\
\mathcal{G}^{VC}_{\parallel}(k)&=&G_{\parallel}^{(0)}(k)\:+\:G_{\parallel}^{(0)}(k)\:\Xi^{VC}_{\parallel}(k)\:\mathcal{G}^{VC}_{\parallel}(k),
\label{DysonIImathG}
\end{eqnarray}
where
\begin{equation}\label{Xi}
\Xi^{VC}_{\perp,\parallel}(k)=-\frac{\rho\tilde{\alpha}}{\chi_{\perp,\parallel}(k)\:G^{(0)}_{\perp,\parallel}(k)}
\Bigl[1\:-\:\frac{\chi_{\perp,\parallel}(k)}{\rho\tilde{\alpha}}\:+\:k_{0}^{2}\chi_{\perp,\parallel}(k)\:G^{(0)}_{\perp,\parallel}(k)\Bigr]
\end{equation}
is the stochastic kernel expressed as a function of $\bar{\chi}$, $\bar{G}^{(0)}$ and $\rho\tilde{\alpha}$.\\
\indent Because the \emph{transference matrix} ($\bar{t}$-matrix in brief) formalism is profusely used in numerical simulations which involve  discrete configurations of point dipoles (see eg.\cite{Latakia,Draine}), we will next show that analogous formulae to the ones above are obtained for a fixed configuration of point dipoles. Let us take the $m^{th}$ configuration of the statistical ensemble described in Section \ref{Sect2C}. Its corresponding self-polarization vacuum state there was denoted by $|\phi_{m}\rangle^{s.p.}$. Let us assume the configuration consists of $N+1$ host scatterers with fixed position vectors denoted by $\{\vec{r}_{i}\}$ with $i=0,...,N$. We then decide to excite the dipole at $\vec{r}_{0}$ with an external monochromatic field of frequency $\omega=k_{0}c$ such that $\vec{E}_{exc}(\vec{r})=\vec{E}^{\omega}_{0}$ iff $\vec{r}=\vec{r}_{0}$ and $\vec{E}_{exc}(\vec{r})=0$ otherwise, as in Section \ref{Sect3B}. All the dipoles are equivalent and have renormalized polarizability $\tilde{\alpha}$. Therefore, the dipole moment of the emitter reads
\begin{equation}\label{pemuto}
\vec{p}(\vec{r}_{0})=\epsilon_{0}\tilde{\alpha}\vec{E}^{\omega}_{0}.
\end{equation}
The total field at the emitter location is the sum of the exciting field plus the self-polarization field which the dipole moment $\vec{p}(\vec{r}_{0})$ itself creates at $\vec{r}_{0}$. That is,
\begin{eqnarray}\label{Eemit}
\vec{E}(\vec{r}_{0})&=&\vec{E}^{\omega}_{0}+\frac{k_{0}^{2}}{\epsilon_{0}}\bar{\mathcal{G}}(\vec{r}_{0},\vec{r}_{0})\cdot\vec{p}(\vec{r}_{0})
\nonumber\\&=&\vec{E}^{\omega}_{0}+k_{0}^{2}\tilde{\alpha}\bar{\mathcal{G}}(\vec{r}_{0},\vec{r}_{0})\cdot\vec{E}^{\omega}_{0}.
\end{eqnarray}
The second term on the \emph{r.h.s.} of Eq.(\ref{Eemit}) is the field which propagates freely in space from all the induced dipoles to $\vec{r}_{0}$,
\begin{equation}\label{pemit}
k_{0}^{2}\tilde{\alpha}\bar{\mathcal{G}}(\vec{r}_{0},\vec{r}_{0})\cdot\vec{E}^{\omega}_{0}=\frac{k_{0}^{2}}{\epsilon_{0}}
\sum_{\{\vec{r}_{i}\}}\bar{G}^{(0)}(\vec{r}_{0},\vec{r}_{i})\cdot\vec{p}(\vec{r}_{i}).
\end{equation}
Except for the emitter, all the rest dipoles are only induced by their mutual interactions. The $\bar{t}$-matrix, with components $\bar{t}(\vec{r}_{i},\vec{r}_{j})$, in the same spirit as that in Eq.(\ref{Dysonalter}), yields the dipole induced at some point $\vec{r}_{i}$ as a result of its interaction with the dipole excited by an external source at some other point $\vec{r}_{j}$. That is,
\begin{equation}
\vec{p}(\vec{r}_{i})=\Bigl(\frac{k_{0}^{2}}{\epsilon_{0}}\Bigr)^{-1}\sum_{\{\vec{r}_{j}\}}-\bar{t}(\vec{r}_{i},\vec{r}_{j})\cdot\vec{E}_{exc}(\vec{r}_{j}).
\end{equation}
Analogously to the susceptibility function in Eq.(\ref{laXenAes}), $\bar{t}^{(n)}(\vec{r}_{i},\vec{r}_{j})$ can be expanded in powers of $-k_{0}^{2}\tilde{\alpha}$, $\bar{t}(\vec{r}_{i},\vec{r}_{j})=\sum_{n=1} \bar{t}^{(n)}(\vec{r}_{i},\vec{r}_{j})$. Each term in the sum contains $n$ factors $-k_{0}^{2}\tilde{\alpha}$ and $n-1$ tensors $\bar{G}^{(0)}(\vec{r}_{m},\vec{r}_{l})$ which propagate the field though all possible paths connecting the points $\vec{r}_{i}$ and $\vec{r}_{j}$. As an example,
\begin{equation}
\bar{t}^{(4)}(\vec{r}_{i},\vec{r}_{j})=(-k_{0}^{2}\tilde{\alpha})^{4}\sum_{\{\vec{r}_{l},\vec{r}_{m}\}}
\bar{G}^{(0)}(\vec{r}_{i},\vec{r}_{m})\cdot\bar{G}^{(0)}(\vec{r}_{m},\vec{r}_{l})\cdot
\bar{G}^{(0)}(\vec{r}_{l},\vec{r}_{j}),
\end{equation}
where restrictions apply over the indices of the sums to incorporate correlations.
The analog to the $\bar{T}(k)$ matrix of Eq.(\ref{laXenAes}) is just
\begin{equation}
\bar{T}(\vec{k},\vec{k}')=\sum_{\{\vec{r}_{i},\vec{r}_{j}\}}\bar{t}(\vec{r}_{i},\vec{r}_{j})
e^{i\:[\vec{k}\cdot\vec{r}_{i}-\vec{k}'\cdot\vec{r}_{j}]}.
\end{equation}
Because in our case the only externally excited dipole is that at $\vec{r}_{0}$, $\vec{p}_{i}=-\Bigl(\frac{k_{0}^{2}}{\epsilon_{0}}\Bigr)^{-1}\bar{t}(\vec{r}_{i},\vec{r}_{0})\cdot\vec{E}^{\omega}_{0}$. Inserting this formula into Eq.(\ref{pemit}) we obtain
\begin{equation}\label{Gdiscr}
\bar{\mathcal{G}}(\vec{r}_{0},\vec{r}_{0})=\frac{1}{-k_{0}^{2}\tilde{\alpha}}\sum_{\{\vec{r}_{i}\}}
\bar{G}^{(0)}(\vec{r}_{0},\vec{r}_{i})\cdot\bar{t}(\vec{r}_{i},\vec{r}_{0}),
\end{equation}
which is the analogous expression to that in Eq.(\ref{infunctionofT}) but for a fixed configuration of scatterers.
\section{The real-cavity scenario}\label{Sect4B}
The computation of the $\gamma$-factors in the real cavity scenario is generally quite more complicated as an effective medium cannot be strictly speaking defined and  translation invariant bulk propagator and suceptibility tensors do not exist. There are two main differences with respect to the virtual cavity scenario. The first one is that the emitter does not behave as an ordinary scatterer of the host medium regarding photon propagation. The second is that the presence of the cavity induces additional correlations among the scattering events experienced by photons near the cavity surface. That is, because scattering events are correlated to the emitter through the stochastic kernel of Eq.(\ref{sthocastic}), they indirectly correlate among themselves beside the inherent spatial correlation of scatterers in absence of cavity.\\
\indent There are nevertheless two situations in which approximate solutions can be found. The first one corresponds to the case in which the emitter hardly alters the host medium. For this to be the case, the polarizability of the emitter must be much weaker than that of the host scatterers and its cavity radius $R$ much less than the typical distance between scatterers, $R\ll\rho^{-1/3}$. In his situation the foreign emitter is said weakly polarizable and interstitial. The second situation which can be treated analytically is that in which the cavity is large, $R\gg\rho^{-1/3}$. The density function of host scatterers is altered in a large patch but remains unaltered locally within distances of the order of $\xi$ where correlations matter. In this case, the foreign emitter is said to be substantial.
\subsection{Interstitial weakly-polarizable impurity within a small cavity}\label{Sect4B1}
\begin{figure}[h]
\includegraphics[height=9.2cm,width=14.8cm,clip]{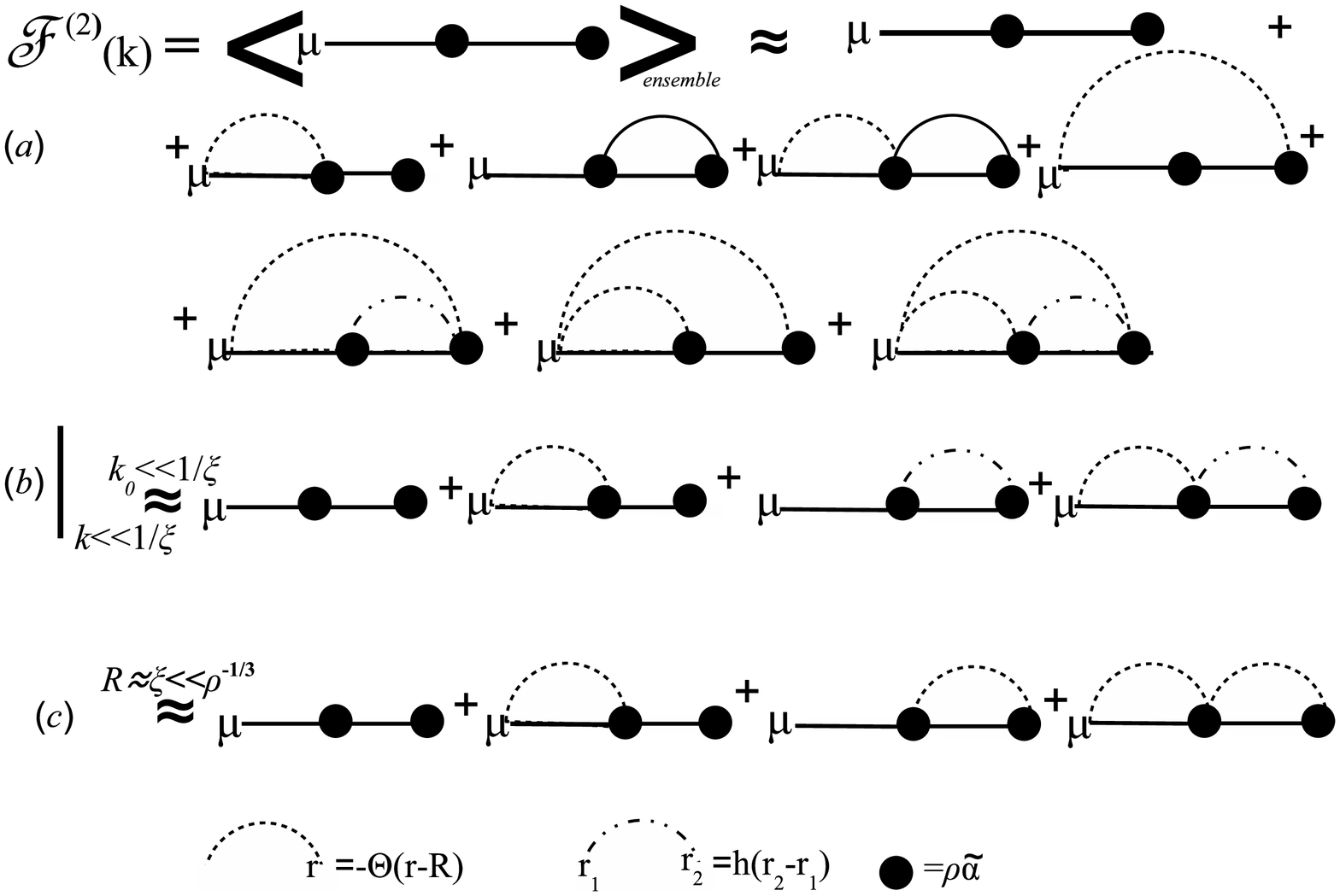}
\caption{($a$) Diagrammatic representation of Eq.(\ref{F2}).
 ($b$) Overlap approximation restricted to those modes $k\ll\xi^{-1}$ subject to $k_{0}\xi\ll1$. ($c$) Overlap approximation for shorter wavelengths valid only for $R\approx\xi$.}\label{fig14}
\end{figure}

\begin{figure}[h]
\includegraphics[height=6.0cm,width=14.0cm,clip]{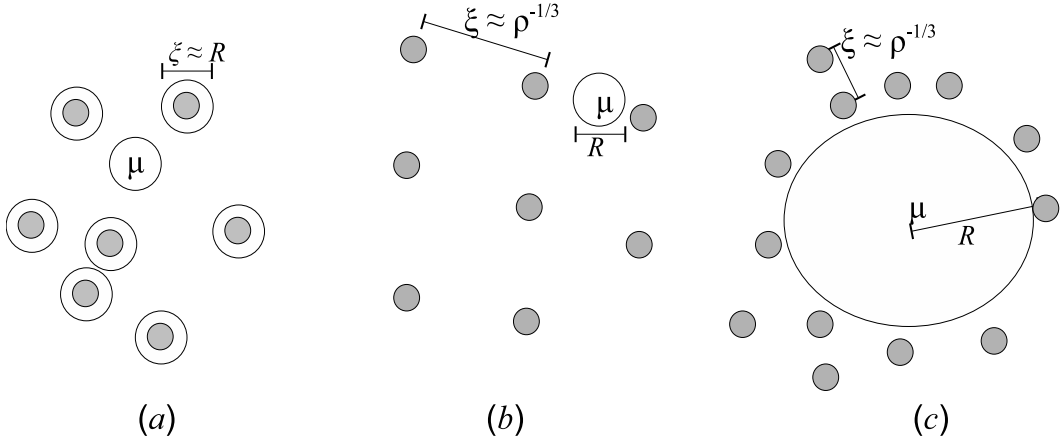}
\caption{Two dimensional sketch of the real cavity scenarios which can be treated analytically in some approximation. ($a$) Weakly correlated host medium (eg. gas) with an interstitial impurity $\mu$. ($b$) Strongly correlated host medium (eg. glass) with an interstitial impurity. $(c)$ Large cavity enclosing a substantial impurity.}\label{fig15}
\end{figure}

\indent The two-point correlation of the host scatterers to the foreign emitter reflects in this case on a slight modification of their average density function in the neighborhood of the impurity. The radial distribution density function with respect to the emitter site reads
\begin{equation}\label{densityri}
\bar{\rho}(r_{i})=\rho[1-\Theta(r_{i}-R)],\quad R\ll\rho^{-1/3},
\end{equation}
where $\vec{r}_{i}$ is the position vector of a generic scatterer with origin at the emitter location and $\rho$ is the 'would-be' uniform numerical density of host scatterers in absence of emitter. As an example, we depict in Fig.\ref{fig14}($a$) the diagrammatic representation of the scattering amplitude for the process in which a photon emitted by the foreign emitter is scattered by a host scatterer and absorbed by another one, $\mathcal{F}^{(2)}(k)$,
\begin{eqnarray}\label{F2}
\mathcal{F}^{(2)}(k)&=&\mu(\rho\tilde{\alpha})^{2}\int\textrm{d}^{3}r_{1}\textrm{d}^{3}r_{2}\:e^{i\vec{k}\cdot\vec{r}_{2}}
\Bigl[\textrm{Tr}\{\bar{G}^{(0)}(\vec{r}_{1})\cdot
\bar{G}^{(0)}(\vec{r}_{2}-\vec{r}_{1})\}\nonumber\\&\times&[1-\Theta(r_{1}-R)][1-\Theta(r_{2}-R)]
[1-h(|\vec{r}_{2}-\vec{r}_{1}|)\Bigr],
\end{eqnarray}
where $h(|\vec{r}_{2}-\vec{r}_{1}|)$ is the two-point correlation function between host scatterers with support within a correlation volume $\sim\xi^{3}$.
It is clear from Fig.\ref{fig14}($a$) that in general it is not possible to disentangle the correlation of the emitter to each scatterer event. However, for the case that $k_{0}\xi\ll1$, $k_{0}R\ll1$, one can apply the overlap approximation \cite{Feldoher} for those modes $k\ll\xi^{-1},R^{-1}$. This is so because those modes cannot distinguish between the exclusion volume of the host scatterers and the real cavity. That is what Fig.\ref{fig14}($b$) depicts. In principle, the overlap approximation in this scenario demands that the only relevant correlation in the medium be given by the Heviside function of exclusion volume. This is clearly the case of a weakly correlated host medium (eg. a diluted gas like that of Fig.\ref{fig15}($a$)) for which $h(|\vec{r}_{2}-\vec{r}_{1}|)\simeq-\Theta(|\vec{r}_{2}-\vec{r}_{1}|-\xi)$ with $\xi\ll\rho^{-1/3}$. For a highly correlated medium  with $\xi\sim\rho^{-1/3}$ (eg. a solid like that of Fig.\ref{fig15}($b$)) a term proportional to $\delta^{(1)}(r-\xi)$ amounting for first neighbors must be included in $h(|\vec{r}_{2}-\vec{r}_{1}|)$. Nevertheless, for $k_{0}\xi\ll1$ that term is irrelevant in comparison to the exclusion volume function and the overlap approximation for $k\ll\xi^{-1}$ is equally valid. Therefore, the condition $k_{0}\xi\ll1$ is generally sufficient for application of the overlap approximation. In addition, this situation corresponds to that in which an effective medium exists with $\chi_{eff}$  the constant effective susceptibility valid for those $k$-modes with $k\xi\ll1$. The contribution of transverse modes with $k\xi\ll1$ to LDOS, $\Gamma_{\perp}$ and $W_{\perp}$ is well approximated by the formulae of the VC scenario which we will compute in Section \ref{Sect5B2}. The corresponding $\gamma$-factor will be found to be
\begin{equation}
2\gamma_{\perp}|_{k,k_{0}\ll\xi^{-1}}\approx2\int\frac{\textrm{d}^{3}k}{(2\pi)^{3}}\frac{\chi_{eff}}{\rho\tilde{\alpha}}G_{\perp}^{eff}(k).
\end{equation}
\indent For those modes $k\gtrsim\xi^{-1}$, the equivalence with the virtual cavity formulae Eqs.(\ref{LDOSIperg},\ref{LDOSIparalg})  is only possible for the case that $R\approx\xi$ --see Fig.\ref{fig14}($c$). This is only applicable to a weakly-correlated host medium as otherwise $R$ would not satisfy the interstitial condition $R\ll\rho^{-1}$. That is the case of the Maxwell-Garnett dielectric which will be studied in Section \ref{Sect6A}. This has the nice implication that, if the impurity is an excited host scatterer of a diluted gas, the $\gamma^{VC}$ factors computed in Section \ref{Sect4A} are applicable in good approximation provided that the polarizability of the excited atom during its decay process is weak in comparison to that of the rest of scatterers.
\subsection{Substantial impurity within a large cavity}\label{Sect4B2}
\begin{figure}[h]
\includegraphics[height=6.2cm,width=12.cm,clip]{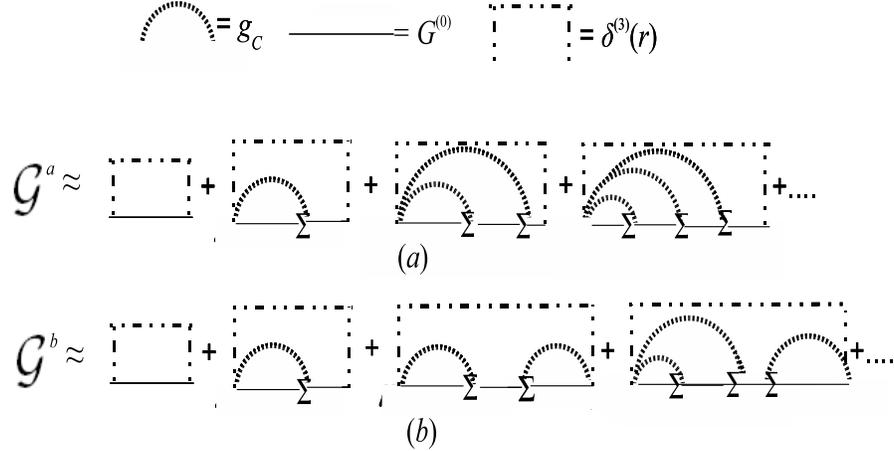}
\caption{Diagrammatic representation
of the self-polarization propagator $\mathcal{G}$
according to Eqs.(\ref{seriesa},\ref{gammaanper},\ref{gammaanpar}) --series ($a$)-- and Eqs.(\ref{seriesb},\ref{gammabnper},\ref{gammabnpar})) --series ($b$)-- in the large cavity scenario.}\label{FigIV01}
\end{figure}

\indent As mentioned above, the presence of an emitter which substitutes one or several host scatterers can modify sensitively the homogeneity of the host medium. In particular, this is the case of a host scatterer which is promoted to an excited atomic state in a strongly correlated medium. In comparison to the scenario of the previous Subsection, this is the same situation but with $\xi\simeq\rho^{-1/3}$. Because the polarizability of the excited scatterer is different to that of the rest, namely $\alpha_{exc.}$, it behaves as a foreign emitter which occupies an effective volume of the order of $\rho^{-1}$. Thus, the approximate average polarizability density at a point $\vec{r}$ away from the emitter is given by $\rho\tilde{\alpha}[1-\Theta(r-\rho^{-1/3})]+\rho\alpha_{exc.}\Theta(r-\rho^{-1/3})$. This makes the electrical susceptibility to be clearly not a homogeneous function. For instance, the two-scattering term would read
\begin{eqnarray}
\bar{\chi}^{(2)}(\vec{r}_{1},\vec{r}_{2})&\approx&\rho^{2}\Bigl[\tilde{\alpha}
[1-\Theta(r_{1}-\rho^{-1/3})]+\rho\alpha_{exc.}\Theta(r_{1}-\rho^{-1/3})\Bigr]\nonumber\\&\times&\Bigl[\tilde{\alpha}
[1-\Theta(r_{2}-\rho^{-1/3})]+\rho\alpha_{exc.}\Theta(r_{2}-\rho^{-1/3})\Bigr]\bar{G}^{(0)}(\vec{r}_{2}-\vec{r}_{1})
h(|\vec{r}_{2}-\vec{r}_{1}|).
\end{eqnarray}
That implies that the susceptibility function evaluated at $\vec{r}_{1},\vec{r}_{2}$ close to the emitter can be very different to that at $\vec{r}_{1},\vec{r}_{2}$ far from it.\\
\indent However, whatever the nature of the foreign emitter is, if the cavity radius is large in comparison to the mean distance between scatterers, $R\gg\rho^{-1/3}$, the medium will look statistically homogeneous at scales of the order of $\rho^{-1/3}$ as seen from the emitter. This is so because the electric field which propagates from the emitter is statistically uniform within the interval $[R,R+\rho^{-1/3}]$ for $R\gg\rho^{-1/3}$. Thus, the average value of the numerical density of host scatterers at a point $\vec{r}_{i}$ away from the emitter reads in good approximation, $\bar{\rho}(\vec{r}_{i})\simeq\rho[1-\Theta(r_{i}-R)]$. Because the typical correlation length between host scatterers is $\xi\lesssim\rho^{-1/3}\ll R$, the cavity exclusion volume factor $-\Theta(r-R)$ is common to all the scatterers which are correlated among themselves and enter the susceptibility function. This has the nice implication that solids with well-defined susceptibility function $\bar{\chi}(k)$ can be treated this way with no-need to know the individual polarizabilities of their constituents. Any fluctuation not yet considered for the case of point-host-dipoles (eg. excitons) is assumed integrated out in $\bar{\chi}(k)$, provided that the wavelength of those fluctuations is much shorter than $R$. In the following we develop a formalism which bases on the definition of quantities analogous to $\bar{\chi}(k)$ and $\bar{G}(k)$ but for a non-translation invariant medium.\\
\indent Let us formulate mathematically the above approximation. Let $g(r)$ be the  full two-point correlation function of the emitter with the host scatterers,
\begin{equation}
g(r)=1+h_{C}(r),
\end{equation}
where $h_{C}(r)$ is the irreducible piece. In general, because the cavity is macroscopical, it can distort the homogenous distribution of host scatterers in the surrounding medium. In such a case $h_{C}(r)$ must contain a maximum at $r\simeq\xi$ that makes account of the overdensity of first neighbors. For the sake of simplicity, we will restrict $h_{C}$ to a step function which stands for the cavity exclusion volume, $h_{C}(r)\approx-\Theta(r-R)$. The Fourier transform of $g(r)$ reads
\begin{equation}
g(k)=(2\pi)^{3}\delta^{(3)}(\vec{k}) + h_{C}(k).
\end{equation}
Because $h_{C}(r)$ has support in $r\leq R$, it is therefore expected that its Fourier transform gets support in  $k\lesssim 1/R$. On the other hand, in case that spatial dispersion in $\bar{\epsilon}(k)$ be relevant, such a dispersion must be of the order of $1/\xi$. Therefore, any convolution of $g(|\vec{k}-\vec{k}'|)$ with the self-energy operator $\bar{\Sigma}(k')$ and any generic function $f(k')$ can be approximated by
\begin{equation}\label{laprox}
\int\frac{\textrm{d}^{3}k}{(2\pi)^{3}}g(|\vec{k}'-\vec{k}|)\bar{\Sigma}(k)\cdot\bar{f}(k)\approx
\bar{\Sigma}(k^{'})\cdot\int\frac{\textrm{d}^{3}k}{(2\pi)^{3}}g(|\vec{k}'-\vec{k}|)\bar{f}(k).
\end{equation}
In Eq.(\ref{laprox}), the difference between $\vec{k}$ and  $\vec{k}'$ is negligible in comparison to $1/\xi$ in the range of momenta where $h_{C}(|\vec{k}-\vec{k}'|)$
 takes nearly constant value and $\bar{\Sigma}(k)$ is not zero. In turn, that implies that the correlation of the emitter with any 1PI (multiple)scattering process can be approximated by the correlation of the emitter to any of the scatterers involved in such a process. In particular, things get mathematically simpler if the correlation functions connect the impurity  either to the first or to the last scatterers within the 1PI diagrams of $\bar{\Sigma}(k)$. This allows to write the series for the RC propagator, $\bar{\mathcal{G}}^{RC}$,  as two apparently different expansions, $(a)$ and $(b)$, as depicted in Figs.\ref{FigIV01}$(a,b)$ respectively. The corresponding series read
\begin{eqnarray}
\mathcal{G}_{\perp}^{RC}&=&\sum_{n=0}^{\infty}\mathcal{G}_{\perp}^{a(n)},\qquad\qquad \mathcal{G}_{\parallel}^{RC}=\sum_{n=0}^{\infty}\mathcal{G}_{\parallel}^{a(n)},\label{seriesa}\\
\mathcal{G}_{\perp}^{RC}&=&\mathcal{G}^{(0)}_{\perp}+\sum_{n=1}^{\infty}\mathcal{G}_{\perp}^{b(n)},\quad \mathcal{G}_{\parallel}^{RC}=\mathcal{G}^{(0)}_{\parallel}+\sum_{n=1}^{\infty}\mathcal{G}_{\parallel}^{b(n)}\label{seriesb}.
\end{eqnarray}
In each series, the $n^{th}$ terms read respectively,
\begin{eqnarray}
\mathcal{G}^{a(n)}_{\perp}(k)&=&\kappa^{(n)}_{\perp}(k)\Sigma_{\perp}(k)G_{\perp}^{(0)}(k),
\label{gammaanper}\\
\mathcal{G}^{a(n)}_{\parallel}(k)&=&\kappa^{(n)}_{\parallel}(k)\Sigma_{\parallel}(k)G_{\parallel}^{(0)}(k),
\label{gammaanpar}
\end{eqnarray}
and
\begin{eqnarray}
\mathcal{G}^{b(n)}_{\perp}(k)=\kappa^{(n-1)}_{\perp}(k)\Sigma_{\perp}(k)G_{\perp}^{(0)}(k)
\Sigma_{\perp}(k)\kappa^{(1)}_{\perp}(k),
\label{gammabnper}\\
\mathcal{G}^{b(n)}_{\parallel}(k)=\kappa^{(n-1)}_{\parallel}(k)\Sigma_{\parallel}(k)G_{\parallel}^{(0)}(k)
\Sigma_{\parallel}(k)\kappa^{(1)}_{\parallel}(k).
\label{gammabnpar}
\end{eqnarray}
The recurrent formulae for the \emph{partial pseudo-susceptibilities} $\kappa^{(n)}_{\perp,\parallel}$ --the reason for this nomenclature will get clear later on-- read
\begin{eqnarray}
\kappa^{(n)}_{\perp}(k)&=&\frac{1}{2}\int\frac{\textrm{d}^{3}k'}{(2\pi)^{3}}g(|\vec{k}'-\vec{k}|)
\Bigl[(1+\cos^{2}{\theta})\nonumber\\&\times&\kappa^{(n-1)}_{\perp}(k')\Sigma_{\perp}(k')G^{(0)}_{\perp}(k')\nonumber\\
&+&\sin^{2}{\theta}\:\kappa^{(n-1)}_{\parallel}(k')\Sigma_{\parallel}(k')G^{(0)}_{\parallel}(k')\Bigr]
\quad\textrm{for }n\geq1,\label{chinper}\\
\kappa^{(n)}_{\parallel}(k)&=&\int\frac{\textrm{d}^{3}k'}{(2\pi)^{3}}g(|\vec{k}'-\vec{k}|)
\Bigl[\cos^{2}{\theta}\nonumber\\&\times&\kappa^{(n-1)}_{\parallel}(k')\Sigma_{\parallel}(k')G^{(0)}_{\parallel}(k')
\nonumber\\&+&\sin^{2}{\theta}\:\kappa^{(n-1)}_{\perp}(k')\Sigma_{\perp}(k')G^{(0)}_{\perp}(k')\Bigr]
\quad\textrm{for }n\geq1,\label{chinpar}\\
\textrm{and }\quad\kappa^{(0)}_{\perp,\parallel}(k)&\equiv&-\tilde{k}^{2}
\Sigma_{\perp,\parallel}(k)^{-1}=1/\chi_{\perp,\parallel}(k)\quad\textrm{for }n=0.\label{chin0}
\end{eqnarray}
In Eqs.(\ref{chinper},\ref{chinpar}), $\Sigma_{\perp,\parallel}(k')$  can be factored out of the integrals as $\Sigma_{\perp,\parallel}(k)$ in application of the approximation in Eq.(\ref{laprox}). Therefore, the building blocks of the series are $G^{(0)}_{\perp,\parallel}(k)$ and the irreducible pieces of $\kappa^{(1)}_{\perp,\parallel}(k)$. Hereafter we will refer to the 1PI parts of $\kappa^{(1)}_{\perp,\parallel}(k)$ as \emph{cavity factors},
\begin{eqnarray}
C_{\perp}(k)&\equiv&\frac{1}{2}\int\textrm{d}^{3}r\:e^{i\vec{k}\cdot\vec{r}}h_{C}(r)\textrm{Tr}
\{\bar{G}^{(0)}(r)[\mathbb{I}-\hat{k}\otimes\hat{k}]\}\nonumber\\&=&
\frac{1}{2}\int\frac{\textrm{d}^{3}k'}{(2\pi)^{3}}
h_{C}(|\vec{k}'-\vec{k}|)\Bigl[G_{\perp}^{(0)}(k')\nonumber\\&+&G_{\perp}^{(0)}(k')\cos^{2}{\theta}\:+\:G_{\parallel}^{(0)}(k')\sin^{2}{\theta}\Bigr],
\label{Xioperp}\\
C_{\parallel}(k)&\equiv&\int\textrm{d}^{3}r\:e^{i\vec{k}\cdot\vec{r}}h_{C}(r)\textrm{Tr}
\{\bar{G}^{(0)}(r)[\hat{k}\otimes\hat{k}]\}\nonumber\\&=&\int\frac{\textrm{d}^{3}k'}{(2\pi)^{3}}
h_{C}(|\vec{k}'-\vec{k}|)\nonumber\\&\times&\Bigl[G_{\parallel}^{(0)}(k')\cos^{2}{\theta}\:+\:G_{\perp}^{(0)}(k')\sin^{2}{\theta}\Bigr],
\label{Xioparall}
\end{eqnarray}
and so $\kappa^{(1)}_{\perp,\parallel}(k)=-k_{0}^{2}[ G^{(0)}_{\perp,\parallel}(k)+C_{\perp,\parallel}(k)]$.\\
\indent Also, Eqs.(\ref{gammabnper},\ref{gammabnpar}) resemble the formulae of the $\gamma$-factors for the virtual cavity scenario --Eqs.(\ref{LDOSIper},\ref{LDOSIparal}). That is, if we define the \emph{total pseudo-susceptibility} as
$\kappa_{\perp,\parallel}(k)\equiv\sum_{n=0}\kappa_{\perp,\parallel}^{(n)}(k)$
we can write
\begin{eqnarray}
\mathcal{G}^{RC}_{\perp}(k)&=&\kappa_{\perp}(k)\chi_{\perp}(k)\:G_{\perp}^{(0)}(k),\label{gammaperp}\\
\mathcal{G}^{RC}_{\parallel}(k)&=&\kappa_{\parallel}(k)\chi_{\parallel}(k)\:G_{\parallel}^{(0)}(k).\label{gammapara}
\end{eqnarray}
However, the above equations are not yet quite similar to those for the VC scenario. In particular, the propagator on the \emph{r.h.s.} of Eqs.(\ref{gammaperp},\ref{gammapara}) is that in free-space whereas it is the bulk propagator in the virtual cavity formulae. The reason being that $\kappa_{\perp,\parallel}(k)$ contain both 1PI and non-1PI processes. We can separate $\kappa_{\perp,\parallel}(k)$  into 1PI and non-1PI (N1PI) pieces, $\kappa_{\perp,\parallel}(k)=\kappa^{1PI}_{\perp,\parallel}(k)+\kappa^{N1PI}_{\perp,\parallel}(k)$
according to the following decomposition in partial pseudo-susceptibility functions,
\begin{eqnarray}
\kappa^{N1PI(n)}_{\perp,\parallel}(k)&=&\kappa^{(n-1)}_{\perp,\parallel}(k)\Sigma_{\perp,\parallel}(k)
G^{(0)}_{\perp}(k),\quad n\geq1,\nonumber\\
\kappa^{1PI(n)}_{\perp,\parallel}(k)&=&
\kappa^{(n)}_{\perp,\parallel}(k)-\kappa^{N1PI(n)}_{\perp,\parallel}(k),\quad n\geq1,\nonumber\\
\kappa^{1PI(0)}_{\perp,\parallel}(k)&=&\kappa^{(0)}_{\perp,\parallel}(k),\quad
\kappa^{N1PI(0)}_{\perp,\parallel}(k)=0.
\end{eqnarray}
Using this decomposition, we can write Eqs.(\ref{gammaperp},\ref{gammapara}) in the form,
\begin{eqnarray}
\mathcal{G}^{RC}_{\perp}(k)&=&\kappa^{1PI}_{\perp}(k)
\chi_{\perp}(k)\:G_{\perp}(k),\label{gammaperp2}\\
\mathcal{G}^{RC}_{\parallel}(k)&=&\kappa^{1PI}_{\parallel}(k)
\chi_{\parallel}(k)\:G_{\parallel}(k),\label{gammapara2}
\end{eqnarray}
where $G_{\perp,\parallel}(k)$ is the Dyson propagator given by Eqs.(\ref{DysonI},\ref{DysonII}) for a 'would-be' homogeneous medium in absence of the real cavity. As a matter of fact, in passing from Eqs.(\ref{gammaperp},\ref{gammapara}) to Eqs.(\ref{gammaperp2},\ref{gammapara2}) one can use the same arguments as those employed in the virtual cavity scenario to push the effective bulk propagator to the right as in the diagrams of Fig.\ref{FigI04}($b$). Thus, the above equations resemble those expressions for the VC scenario in Eqs.(\ref{LDOSIper},\ref{LDOSIparal}) with the replacement of $\frac{\chi_{\perp,\parallel}(k)}{\rho\tilde{\alpha}}$ by $\kappa^{1PI}_{\perp,\parallel}(k)
\chi_{\perp,\parallel}(k)$. Therefore, the relation between the propagators $\mathcal{G}^{RC}_{\perp,\parallel}(k)$ and $\mathcal{G}^{VC}_{\perp,\parallel}(k)$ for two identical random media which differ just by the presence/absence of a real cavity enclosing the emitter is given by
\begin{equation}
\frac{\mathcal{G}^{RC}_{\perp,\parallel}(k)}{\mathcal{G}^{VC}_{\perp,\parallel}(k)}=\rho\tilde{\alpha}\kappa^{1PI}_{\perp,\parallel}(k).
\end{equation}
\indent Finally, we write the Lippmann-Schwinger equations for the \emph{large} real cavity scenario,
\begin{eqnarray}
\mathcal{G}^{RC}_{\perp}(k)&=&G_{\perp}^{(0)}(k)\:+\:G_{\perp}^{(0)}(k)\:\Xi^{RC}_{\perp}(k)\:\mathcal{G}^{RC}_{\perp}(k),\label{LSIRC}\\
\mathcal{G}^{RC}_{\parallel}(k)&=&G_{\parallel}^{(0)}(k)\:+\:G_{\parallel}^{(0)}(k)\:\Xi^{RC}_{\parallel}(k)\:\mathcal{G}^{RC}_{\parallel}(k),
\label{LSIIRC}
\end{eqnarray}
where the stochastic kernel is
\begin{equation}\label{Xirc}
\Xi^{RC}_{\perp,\parallel}(k)=-\frac{1}{\kappa^{1PI}_{\perp,\parallel}(k)\:\chi_{\perp,\parallel}(k)\:G^{(0)}_{\perp,\parallel}(k)}
\Bigl[1\:-\:\kappa^{1PI}_{\perp,\parallel}(k)\:\chi_{\perp,\parallel}(k)\:+\:k_{0}^{2}\chi_{\perp,\parallel}(k)\:G^{(0)}_{\perp,\parallel}(k)\Bigr].
\end{equation}
\section{The nature of dipole emission and the\emph{ coherent vacuum}}\label{Sect5}
\indent In this section we first study  a simple model in which dipole emission can be studied microscopically in full detail. We compare the results with the usual Lorentz-Lorenz (LL) and Onsager-B\"{o}ttcher (OB) formulae  and other more recent works. Next, the same decomposition is performed over the general formulae of the previous sections. Our microscopical approach allows us to distinguish  between the transverse and the longitudinal, the coherent and the incoherent, the direct-coherent and the induced-coherent, and the dispersed-incoherent and the absorptive components of the dipole emission. Special attention is paid to the phenomenon of radiative/non-radiative energy transfer. Our study reveals why the nature of dipole emission was erroneously interpreted in previous works.\\
\indent In connection with QFT formalism, we define a \emph{coherent vacuum} associated to coherent emission. We find its relation to the sourceless vacuum in a random medium, which is itself associated to direct emission.
\subsection{Decomposition of dipole emission in the single scattering approximation}\label{Sect5A}
\begin{figure}[h]
\includegraphics[height=9.2cm,width=14.cm,clip]{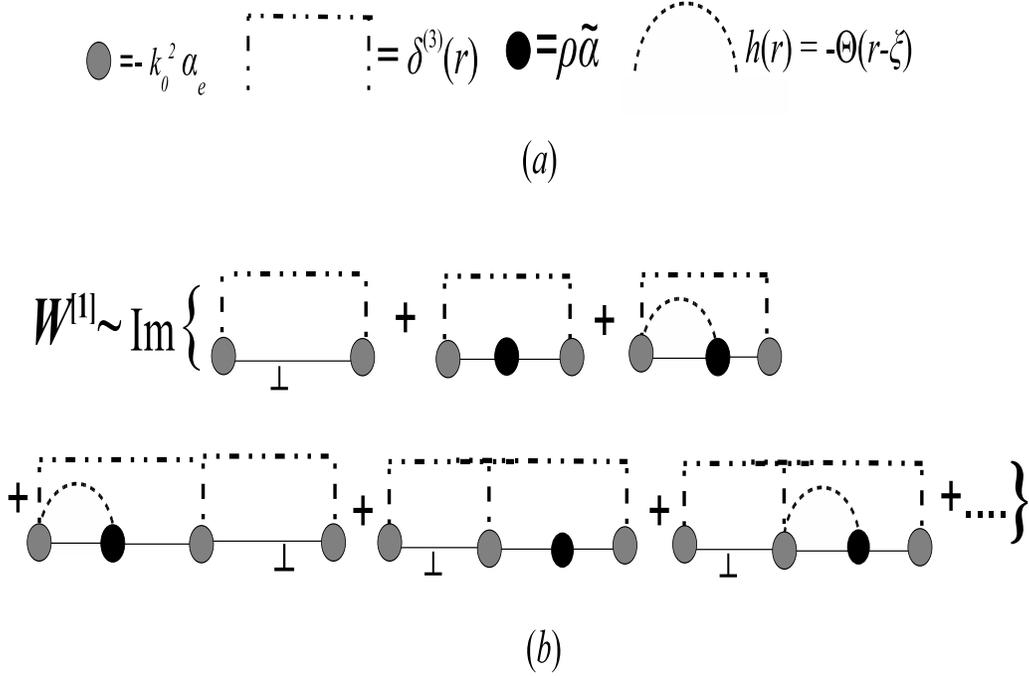}
\caption{($a$) Feynman's rules . ($b$) Diagrammatic representation
of $W^{[1]}$. The first three diagrams stand for those terms in Eq.(\ref{radapp}). The next three diagrams take account of the self-polarization terms of Eqs.(\ref{le2},\ref{obapp}).}\label{FIGnew7}
\end{figure}
\begin{figure}[h]
\includegraphics[height=7.4cm,width=11.2cm,clip]{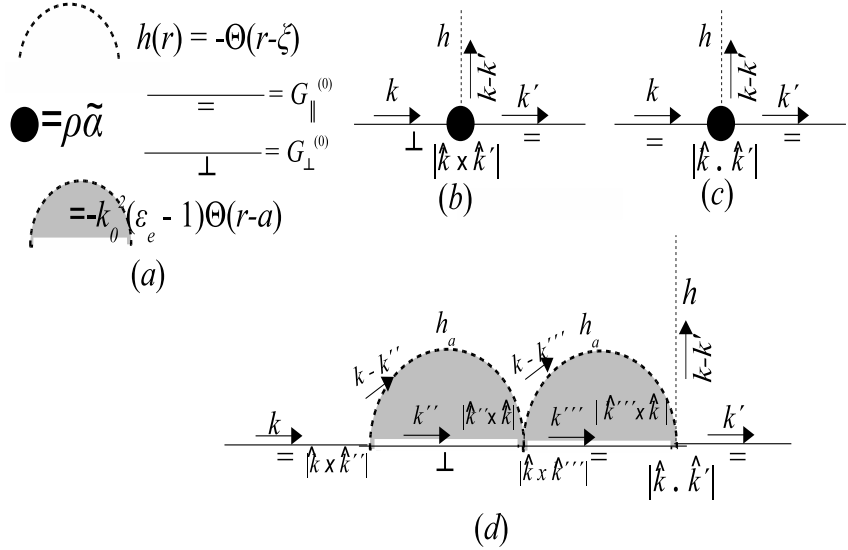}
\caption{($a$) Feynman's rules. ($b$) Common vertex derived from Eqs.(\ref{g1perpc},\ref{g1paralc}) which couples longitudinal to
transverse modes, with $|\hat{k}\times\hat{k}'|=\sin{\theta}$.
($c$) Longitudinal-longitudinal vertex from Eq.(\ref{g1paralb}), with
 $|\hat{k}\cdot\hat{k}'|=\cos{\theta}$. ($d$) Diagrammatic representation of the origin of the coupling of single scattering radiative corrections
  to longitudinal external modes.}\label{fig132}
\end{figure}

\indent Our simplified model consists of a stimulated emitter embedded in a diluted host medium made of spherical
inclusions of electrostatic bare polarizability $\alpha_{0}$. The scatterers present a minimum spherical exclusion volume of radius $\xi$ and density $\rho\ll\xi^{-3}$. The emitter is excited by an external field $\vec{E}^{\omega}_{0}$ whose frequency, $\omega=ck_{0}$ is such that $k_{0}\xi\ll1$. A similar model was studied in \cite{LagvanTig,BerMil,LuisRemiMole}. The single scattering approximation implies that virtual photons experience, at the most, a single scattering event with the scatterers in the host medium. This way, the only relevant
correlation is that given by the two-point function
$h(r)=-\Theta(r-R)$. If the cavity radius $R$ satisfies $R=\xi$ and the electrostatic polarizability of the emitter is the same as that of the host scatters, $\alpha_{0}$, this set-up corresponds to the VC  scenario. It corresponds to the RC scenario otherwise. Let us assume without loss of generality that $R=\xi$ and let us take $\alpha_{e}$ as the electrostatic polarizability of the emitter which, eventually, can be taken equal to $\alpha_{0}$. This implies that, in the single scattering approximation both scenarios are equivalent with just exchanging the cavity factors appearing in the RC scenario of Section \ref{Sect4B2} with the two-scattering term of the susceptibility in the VC scenario, $\bar{\chi}^{(2)}$. They relate through   $\chi^{(2)}_{\perp,\parallel}=-k_{0}^{2}(\rho\tilde{\alpha})^{2}C_{\perp,\parallel}|_{R=\xi}$, where $\tilde{\alpha}$ is given by Eq.(\ref{alpha1}) and $C_{\perp,\parallel}(k)$ are given in Eqs.(\ref{Xioperp},\ref{Xioparall}). The
susceptibility of this model reads $\chi^{[1]}_{\parallel,\perp}(k)=
\rho\tilde{\alpha}+\chi_{\parallel,\perp}^{(2)}(k)$. In the following, the superscript $[1]$ will denote single-scattering approximation.\\
\indent The quantity to compute is
\begin{equation}
W_{\omega}=\frac{-\omega^{3}\epsilon_{0}}{6c^{2}}\frac{|\alpha_{e}|^{2}}
{|1+\frac{1}{3}k_{0}^{2}\alpha_{e}[2\gamma^{(0)}_{\perp}+2\gamma_{\perp}+\gamma_{\parallel}]|^{2}}
\Im{\{2\gamma^{Tot}_{\perp}+\gamma^{Tot}_{\parallel}\}}
|E^{\omega}_{0}|^{2}\label{larr3}.
\end{equation}
Pictorially, $\gamma^{[1]}$ is given by the second and third
 diagrams of Fig.\ref{FIGnew7}($b$). It reads,
\begin{equation}\label{gammatot}
\gamma^{[1]}=\textrm{Tr}\Bigl\{\int\textrm{d}^{3}r_{1}
\:\bar{G}^{(0)}(\vec{r}_{1})(-k_{0}^{2}\rho\tilde{\alpha})
\bar{G}^{(0)}(\vec{r}_{1})\:[1-\Theta(r_{1}-\xi)]\Bigr\}.
\end{equation}
Let us disregard for now self-polarization cycles on the emitter. In doing so, the emitter cavity is said empty. In the limit $k_{0}\xi\ll1$, up to $\mathcal{O}(0)$ terms in $k_{0}\xi$ we get
\begin{equation}
W^{[1],emp}_{\omega}\simeq W^{0}_{\omega}\:\Bigl\{\:1\:+\:\frac{7}{6}\Re{\{\rho\tilde{\alpha}\}}+\Im{\{\rho\tilde{\alpha}\}}[\frac{1}{(k_{0}
\xi)^{3}}+\frac{1}{k_{0}\xi}]\:\Bigr\},\label{radapp}
\end{equation}
where $W^{0}_{\omega}=\frac{\omega^{4}\alpha_{e}^{2}\epsilon_{0}}{12\pi c^{3}}|E^{\omega}_{0}|^{2}$ is the power emitted in free space. The $\mathcal{O}(0)$ terms of Eq.(\ref{radapp}) equal the
usual Lorentz-Lorenz (LL) and Onsager--B\"{o}ttcher (OB) formulae
in absence of back-reaction effects \cite{Lorentz,Onsager,PRLdeVries} with $\epsilon^{[1]}=1+\rho\tilde{\alpha}$,
\begin{equation}
W^{[1]}_{LL}=W^{0}_{\omega}\Bigl(\frac{\Re{\{\epsilon^{[1]}\}}+2}{3}\Bigr)^{2}
\Re{\{\sqrt{\epsilon^{[1]}}\}},\qquad W^{[1],emp}_{OB}=W^{0}_{\omega}\Bigl(\frac{3\Re{\{\epsilon^{[1]}\}}}{2\Re{\{\epsilon^{[1]}\}}+1}\Bigr)^{2}
\Re{\{\sqrt{\epsilon^{[1]}}\}},\label{LLOB}
\end{equation}
\begin{equation}\label{laLLOB1}
W^{[1]}_{LL}=W^{[1],emp}_{OB}=W^{0}_{\omega}(1+\frac{7}{6}\Re{\{\rho\tilde{\alpha}\}}).
\end{equation}
Same results as those in Eqs.(\ref{gammatot}-\ref{laLLOB1}) were obtained by the authors in \cite{BerMil} using a Quantum Optics formalism. The factors $\Bigl(\frac{\Re{\{\epsilon\}}+2}{3}\Bigr)$ and
$\Bigl(\frac{3\Re{\{\epsilon\}}}{2\Re{\{\epsilon\}}+1}\Bigr)$ are referred to in the literature
as \emph{virtual cavity} and \emph{empty cavity} local field factors respectively. The factor $\sqrt{\Re{\{\epsilon\}}}$
is sometimes refereed to as \emph{bulk} factor. Both $W_{LL}$ and $W^{emp}_{OB}$ are commonly attributed in the literature to transverse emission.
Our next task is to clarify the actual nature of the terms in Eqs.(\ref{gammatot},\ref{radapp}) through a rigorous microscopical
analysis. In Fourier space, Eq.(\ref{gammatot}) can be decomposed into transverse and longitudinal parts,
\begin{eqnarray}
2\gamma^{[1]}_{\perp}&=&2\rho\tilde{\alpha}\int\frac{\textrm{d}^{3}k}{(2\pi)^{3}}\Bigl
[G_{\perp}^{(0)}(k)+C_{\perp}(k)\Bigl]
G_{\perp}^{(0)}(k)=2\rho\tilde{\alpha}\int\frac{\textrm{d}^{3}k}{(2\pi)^{3}}
\int\frac{\textrm{d}^{3}k'}{(2\pi)^{3}}\nonumber\\&\times&
\Bigl[G_{\perp}^{(0)}(k')\delta^{(3)}(\vec{k}'-\vec{k})G_{\perp}^{(0)}(k)\label{g1perpa}\\
&+&\frac{1}{2}G_{\perp}^{(0)}(k')h(|\vec{k'}-\vec{k}|)(1+\cos^{2}\theta)
G_{\perp}^{(0)}(k)\label{g1perpb}\\&+&\frac{1}{2}G_{\parallel}^{(0)}(k')h(|\vec{k'}-\vec{k}|)\sin^{2}\theta
G_{\perp}^{(0)}(k)\Bigr],\label{g1perpc}
\end{eqnarray}
\begin{eqnarray}
\gamma^{[1]}_{\parallel}&=&\rho\tilde{\alpha}\int\frac{\textrm{d}^{3}k}{(2\pi)^{3}}
\Bigl[G_{\parallel}^{(0)}(k)+C_{\parallel}(k)\Bigl]
G_{\perp}^{(0)}(k)=\rho\tilde{\alpha}\int\frac{\textrm{d}^{3}k}{(2\pi)^{3}}
\int\frac{\textrm{d}^{3}k'}{(2\pi)^{3}}\nonumber\\&\times&\Bigl[G_{\parallel}^{(0)}(k')
\delta^{(3)}(\vec{k}'-\vec{k})G_{\parallel}^{(0)}(k)\label{g1parala}\\&+&G_{\parallel}^{(0)}(k')h(|\vec{k'}-\vec{k}|)\cos^{2}\theta\:G_{\parallel}^{(0)}(k)\label{g1paralb}\\
&+&G_{\perp}^{(0)}(k')h(|\vec{k'}-\vec{k}|)\sin^{2}\theta\:G_{\parallel}^{(0)}(k)\Bigr]\label{g1paralc},
\end{eqnarray}
where $h(q)=\frac{4\pi}{q^3}[\sin{(q\xi)}-q\xi\cos{(q\xi)}]$
and we can write,
\begin{equation}
W_{\perp}^{[1]}=W^{0}_{\omega}[1-\frac{2\pi c}{\omega}\Im{\{2\gamma^{[1]}_{\perp}\}}],\qquad
W_{\parallel}^{[1]}=W^{0}_{\omega}[-\frac{2\pi c}{\omega}\Im{\{\gamma^{[1]}_{\parallel}\}}].
\end{equation}
In the integrals of Eqs.(\ref{g1perpa}-\ref{g1paralc}), the only imaginary pieces are those associated to the poles of $G_{\perp}^{(0)}(k)$ factors. Therefore,  $\Im{\{\gamma^{[1]}_{\perp,\parallel}\}}$ is composed of terms of the form $\Re\{\rho\tilde{\alpha}\}$ times residues computed at $k=k_{0}$ plus  terms of the form $\Im\{\rho\tilde{\alpha}\}$
times the real part of the integrals. Alternatively, we can write $2\gamma_{\perp}^{[1]}$
and $\gamma_{\parallel}^{[1]}$ in terms of spatial integrals,
\begin{eqnarray}
2\gamma^{[1]}_{\perp}&=&\rho\tilde{\alpha}\textrm{Tr}\Bigl\{\int\textrm{d}^{3}r
[\bar{G}_{rad.}^{(0)}(r)+\bar{G}_{stat.}^{(0)}(r)]\cdot\bar{G}_{rad.}^{(0)}(r)\:
[1-\Theta(r-\xi)]\Bigr\}=\rho\tilde{\alpha}\int\textrm{d}^{3}r\nonumber\\
&\times&\textrm{Tr}\Bigl\{\bar{G}_{rad.}^{(0)}(r)\cdot\bar{G}_{rad.}^{(0)}(r)\label{g1perrra}\\&-&
\bar{G}_{rad.}^{(0)}(r)\cdot\bar{G}_{rad.}^{(0)}(r)\Theta(r-\xi)\label{g1perrrb}\\&-&
\bar{G}_{stat.}^{(0)}(r)\cdot\bar{G}_{rad.}^{(0)}(r)\Theta(r-\xi)\Bigr\}\label{g1perrrc},
\end{eqnarray}
\begin{eqnarray}
\gamma^{[1]}_{\parallel}&=&\rho\tilde{\alpha}\textrm{Tr}\Bigl\{\int\textrm{d}^{3}r
[\bar{G}_{rad.}^{(0)}(r)+\bar{G}_{stat.}^{(0)}(r)]\cdot\bar{G}_{stat.}^{(0)}(r)\:
[1-\Theta(r-\xi)]\Bigr\}=\rho\tilde{\alpha}\int\textrm{d}^{3}r\nonumber\\
&\times&\textrm{Tr}\Bigl\{\bar{G}_{stat.}^{(0)}(r)\cdot\bar{G}_{stat.}^{(0)}(r)\label{g1parrra}\\&-&
\bar{G}_{stat.}^{(0)}(r)\cdot\bar{G}_{stat.}^{(0)}(r)\Theta(r-\xi)\label{g1parrrb}\\&-&
\bar{G}_{rad.}^{(0)}(r)\cdot\bar{G}_{stat.}^{(0)}(r)\Theta(r-\xi)\Bigr\}\label{g1parrrc}.
\end{eqnarray}
The correspondence between the integrals in Fourier space and those in spatial space is obvious. However,
such an obvious correspondence is only possible in the single scattering approximation. This is so because the
bulk propagator is identical to that in free space but for the substitution $k_{0}^{2}\rightarrow
\epsilon^{[1]}\:k_{0}^{2}$ so that it does not present spatial dispersion  and
 there exists a direct identification between the radiative and electrostatic bulk propagators in spatial space and transverse and longitudinal propagators in Fourier space respectively. Note also that, in absence of the correlation function, there is no coupling between radiative (transverse) and electrostatic (longitudinal) modes.\\
\indent Next, we search for the terms in Eqs.(\ref{g1perpa}-\ref{g1parrrc}) which amount to those of Eq.(\ref{laLLOB1}). The free space term $2\gamma^{(0)}_{\perp}$ together with that in Eq.(\ref{g1perpa}) (or Eq.(\ref{g1perrra})) yield $W^{0}_{\omega}(1+\frac{1}{2}\Re{\{\rho\tilde{\alpha}\}})$, which corresponds to the bulk factor $W^{0}_{\omega}\Re{\{\sqrt{\epsilon^{[1]}}\}}$. This is the emission directly radiated by the dipole into the medium --see Fig.\ref{fig13}($a$). Eq.(\ref{g1perpa}) (or Eq.(\ref{g1perrra})) corresponds also to the diagram of Fig.\ref{FIGnew2}($d_{2}$), which explains why of the two transverse propagators there.
Eq.(\ref{g1perpb}) (or Eq.(\ref{g1perrrb})) yields higher order terms. The common terms to $2\gamma^{[1]}_{\perp}$ and $\gamma^{[1]}_{\parallel}$ are those in  Eq.(\ref{g1perpc}) (or Eq.(\ref{g1perrrc})) and
Eq.(\ref{g1paralc}) (or Eq.(\ref{g1parrrc})) which amount to $W^{0}_{\omega}\frac{1}{3}\Re{\{\rho\tilde{\alpha}\}}$ each. They equal one local field factor each, $\Bigl(\frac{\Re{\{\epsilon^{[1]}\}}+2}{3}\Bigr)$ or
$\Bigl(\frac{3\Re{\{\epsilon^{[1]}\}}}{2\Re{\{\epsilon^{[1]}\}}+1}\Bigr)$. The first thing we learn from this analysis is that, contrarily to the common assumption, one of the local field factors is associated to longitudinal emission while the other one together with the bulk factor belong to transverse emission. The latter is associated to the poles of the bulk transverse propagator of the general expression in Eq.(\ref{LDOSIper}) and  form part of the coherent radiative emission --see next subsection. On the contrary, the former is associated to the imaginary part of $\chi_{\parallel}$ which contains a transverse bare propagator (see Eq.(\ref{Xioparall})). It is part of the incoherent emission which is dispersed. The existence of the common term in
Eqs.(\ref{g1perpc},\ref{g1paralc}) is a consequence of reciprocity. However,
while it gives rise to coherent emission when the coupling
reads $G_{\parallel}-G_{\perp}$, it
gives rise to dispersion as read in opposite direction as in Fig.\ref{fig132}($b$).\\
\indent Regarding the $\xi$-dependent terms of Eq.(\ref{radapp}), they both are proportional to $\Im{\{\rho\tilde{\alpha}\}}$. The leading order one, $\frac{\Im{\{\rho\tilde{\alpha}\}}}{(k_{0}\xi)^{3}}$, is given by Eq.(\ref{g1parala}) plus Eq.(\ref{g1paralb}) (or Eq.(\ref{g1parrra}) plus Eq.(\ref{g1parrrb})) and is associated to the imaginary parts of both the longitudinal bulk propagator --Eq.(\ref{g1parala})-- and the longitudinal susceptibility --Eq.(\ref{g1paralb}). The term $\frac{\Im{\{\rho\tilde{\alpha}\}}}{k_{0}\xi}$ has two identical contributions of value $\frac{1}{2}\frac{\Im{\{\rho\tilde{\alpha}\}}}{k_{0}\xi}$ coming from the common term to $2\gamma^{[1]}_{\perp}$ and $\gamma^{[1]}_{\parallel}$  in  Eq.(\ref{g1perpc}) and
Eq.(\ref{g1paralc}) respectively. They are associated to the imaginary parts of $\chi_{\perp}(k)$ and $\chi_{\parallel}(k)$ respectively. All these terms proportional to $\Im{\{\tilde{\alpha}\}}$
are related to absorbtion in the host scatterers.\\
\indent Finally we consider the back-reaction of the host medium on the
emitter with electrostatic polarizability $\alpha_{e}$.
To this aim and at first order in $\rho$ we have to compute the  self-polarization diagrams in the second row of Fig.\ref{FIGnew7}($b$). For the sake of simplicity, in the following we disregard absorbtion and radiative corrections in $\tilde{\alpha}$ and keep leading order terms in $\alpha_{0}$ and $k_{0}\xi$. That is, $\epsilon^{[1]}\simeq1+\rho\alpha_{0}$ is fully real and
\begin{equation}
\gamma^{[1]}_{\parallel}\simeq\frac{-k_{0}}{2\pi}\rho\alpha_{0}[\frac{i}{3}+\frac{1}{(k_{0}\xi)^{3}}],\:\:
2\gamma^{[1]}_{\perp}\simeq\frac{-k_{0}}{2\pi}[i+\rho\alpha_{0}\frac{5i}{6}].\label{le2}
\end{equation}
Plugging Eq.(\ref{le2}) into Eq.(\ref{larr3}), we obtain
\begin{equation}\label{obapp}
W^{[1],selfp}_{\omega}\simeq W_{OB}^{[1],emp}[1+\frac{4}{9}\frac{\alpha_{e}}{V_{\xi}}\rho\alpha_{0}],
\end{equation}
where $V_{\xi}$ is the volume of the cavity. One can verify the
agreement of the above expression with the OB formula which incorporate self-polarization terms at leading
order in $\alpha_{0},\alpha_{e}$ \cite{BulloughHynne2,PRLdeVries},
$W^{[1],selfp}_{OB}=W^{0}_{\omega}\Bigl(\frac{3\Re{\{\epsilon^{[1]}\}}}{2Re{\{\epsilon^{[1]}\}}
+1-\frac{2}{3}\frac{\alpha_{e}}{V_{\xi}}(Re{\{\epsilon^{[1]}\}}-1)}\Bigr)^{2}
\Re{\{\sqrt{\epsilon^{[1]}}\}}$.
\subsection{General decomposition of dipole emission}\label{Sect5B}
Next, we go beyond the single scattering approximation. Again, let us consider Eq.(\ref{larr3}) in absence of absorbtion in the emitter. In order to examine the nature of the emission, it is convenient to write the $\gamma$-factors in terms of the propagators of both the virtual and the real cavity scenarios using Eqs.(\ref{LDOSIper}-\ref{LDOSIparal}) and Eqs.(\ref{gammaperp2}-\ref{gammapara2}) respectively,
\begin{eqnarray}
W_{w}^{VC}&\propto&\int\frac{\textrm{d}^{3}k}{(2\pi)^{3}}
\Bigl[\Im{\{2\frac{\chi_{\perp}(k)}{\rho\tilde{\alpha}}\:
G_{\perp}(k)\}}+\Im{\{\frac{\chi_{\parallel}(k)}{\rho\tilde{\alpha}}\:G_{\parallel}(k)\}}\Bigr]\label{GVC},\\
W_{w}^{RC}&\propto&\int\frac{\textrm{d}^{3}k}{(2\pi)^{3}}\Bigl[\Im{\{2\kappa_{\perp}^{1PI}\chi_{\perp}(k)\:G_{\perp}(k)\}}+
\Im{\{\kappa_{\parallel}^{1PI}\chi_{\parallel}(k)\:G_{\parallel}(k)\}}\Bigr]\label{GRC}.
\end{eqnarray}
From now on, we will just work with $W_{w}^{VC}$ bearing in mind that the computations are in all equivalent to those  for $W_{w}^{RC}$ but for the replacement  $\chi_{\perp,\parallel}\rightarrow\rho\tilde{\alpha}\kappa^{1PI}_{\perp,\parallel}\chi_{\perp,\parallel}$. For the sake of simplicity we will drop the scripts $VC$ and $\omega$ and we will denote the proportionality constant omitted in the above equations by $W_{o}$.\\
\indent The first obvious decomposition is that between transverse, $2W_{\perp}$, and longitudinal emission, $W_{\parallel}$,
\begin{eqnarray}
2W_{\perp}&=&W_{o}\int\frac{\textrm{d}^{3}k}{(2\pi)^{3}}\:2\Im{\{\frac{\chi_{\perp}(k)}{\rho\tilde{\alpha}}G_{\perp}(k)\}},\label{eka}\\
W_{\parallel}&=&W_{o}\int\frac{\textrm{d}^{3}k}{(2\pi)^{3}}\:\Im{\{\frac{\chi_{\parallel}(k)}{\rho\tilde{\alpha}}G_{\parallel}(k)\}}.
\end{eqnarray}
\subsubsection{Coherent and incoherent emission; direct and induced coherent emission}\label{Sect5B1}
Coherent emission is that associated to the coherent propagating field. That is, the emission whose modes satisfy the same dispersion relations as the normal modes which propagate through the bulk \cite{Hopfield,Fano,Bullough},
\begin{eqnarray}
k_{0}^{2}\epsilon_{\perp}(k)-k^{2}|_{k=k^{prop.}_{\perp}}&=&0\qquad\textrm{for transverse modes}\label{dispper}\\
\textrm{and }\epsilon_{\parallel}(k)|_{k=k^{prop.}_{\parallel}}&=&0\qquad\textrm{for longitudinal modes.}\label{disppar}
\end{eqnarray}
These are the poles of $G_{\perp}(k)$ and $G_{\parallel}$ in the integrands of Eqs.(\ref{Gammapcoh},\ref{Gammalcoh}). Therefore, we identify from Eqs.(\ref{GVC},\ref{GRC}),
\begin{eqnarray}
2W^{Coh.}_{\perp}&=&W_{o}\int\frac{\textrm{d}^{3}k}{(2\pi)^{3}}\:2\Re{\{\frac{\chi_{\perp}(k)}{\rho\tilde{\alpha}}\}}\Im{\{G_{\perp}(k)\}},
\label{Gammapcoh}\\
2W^{Incoh.}_{\perp}&=&W_{o}\int\frac{\textrm{d}^{3}k}{(2\pi)^{3}}\:2\Im{\{\frac{\chi_{\perp}(k)}{\rho\tilde{\alpha}}\}}\Re{\{G_{\perp}(k)\}},
\label{Gammapincoh}\\
W^{Coh.}_{\parallel}&=&W_{o}\int\frac{\textrm{d}^{3}k}{(2\pi)^{3}}\:\Re{\{\frac{\chi_{\parallel}(k)}{\rho\tilde{\alpha}}\}}\Im{\{G_{\parallel}(k)\}},
\label{Gammalcoh}\\
W^{Incoh.}_{\parallel}&=&W_{o}\int\frac{\textrm{d}^{3}k}{(2\pi)^{3}}\:\Im{\{\frac{\chi_{\parallel}(k)}{\rho\tilde{\alpha}}\}}\Re{\{G_{\parallel}(k)\}}.
\label{Gammalincoh}
\end{eqnarray}
Because longitudinal normal modes need of material support to propagate and are subject to very specific arrangements of scatterers \cite{Citrus}, $W^{Coh.}_{\parallel}$ can be ignored for practical purposes.\\
\indent Let us examine firstly $W^{Coh.}$ in the framework of Classical Optics. It is given by
\begin{equation}\label{Wcoher}
W^{Coh.}=\frac{-\omega^{3}}{6c^{2}\epsilon_{0}}|\vec{p}_{0}|^{2}\int \textrm{d}^{3}r'\textrm{Tr}_{-}\Bigl\{\Re{\{\bar{\chi}(\vec{r}-\vec{r}')/\rho\tilde{\alpha}\}}\cdot
\Im{\{\bar{G}(\vec{r}',\vec{r};\omega)\}}\Bigr\},
\end{equation}
where both $\bar{\chi}$ and $\bar{G}$ are written in the spatial space representation for convenience and $\vec{p}_{0}=\epsilon_{0}\tilde{\alpha}\vec{E}^{\omega}_{0}(\vec{r})$ is the dipole moment induced by the external field $\vec{E}^{\omega}_{0}(\vec{r})$ on the emitter. Considering the fields classically, the fluctuation-dissipation relation reads
\begin{equation}\label{Disiclass}
\Im{\{\bar{G}_{\omega}(\vec{r}',\vec{r})\}}=-\frac{\pi\epsilon_{0}}{\hbar k_{0}^{2}}
\langle\vec{E}^{\omega}_{D}(\vec{r}')\:\vec{E}^{\omega*}_{D}(\vec{r})\rangle,
\end{equation}
where $\vec{E}^{\omega}_{D}(\vec{r})$ is the $\omega$-mode of the coherent-Dyson field and the script $D$ stands both for Dyson and for direct emission. Using Eq.(\ref{Disiclass}) and writing $\Re{\{\bar{\chi}(\vec{r}-\vec{r}')/\rho\tilde{\alpha}\}}=[\Re{\{\bar{\chi}(\vec{r}-\vec{r}')/\rho\tilde{\alpha}\}}
-\delta^{(3)}(\vec{r}-\vec{r}')\mathbb{I}]\:+\:\delta^{(3)}(\vec{r}-\vec{r}')\mathbb{I}$ in Eq.(\ref{Wcoher}), we separate explicitly the field emitted directly by the source dipole from that which is  emitted by the induced  surrounding dipoles,
\begin{eqnarray}
W^{Coh.}&=&\frac{\pi\omega}{6\hbar}|\vec{p}_{0}|^{2}\langle\vec{E}^{\omega}_{D}(\vec{r})\cdot\vec{E}^{\omega*}_{D}(\vec{r})\rangle\label{E1coh}\\
&+&\frac{\pi\omega}{6\hbar}|\vec{p}_{0}|^{2}\textrm{Tr}_{-}\Bigl\{\int\textrm{d}^{3}r'\Bigl\langle[\Re{\{\bar{\chi}(\vec{r}-\vec{r}')/\rho\tilde{\alpha}\}}
-\delta^{(3)}(\vec{r}-\vec{r}')\mathbb{I}]\cdot\vec{E}^{\omega}_{D}(\vec{r}')\:\vec{E}^{\omega*}_{D}(\vec{r})\Bigr\rangle\Bigr\}.\label{E1E2coh}
\end{eqnarray}
The term in Eq.(\ref{E1coh}) is the coherent power carried by the field directly emitted by the source dipole into the bulk as if it were an external source. In Eq.(\ref{E1E2coh}) we can identify the field emitted at $\vec{r}'$ by the induced dipoles sited around the source which propagates towards the source located at $\vec{r}$,
\begin{equation}
E_{I}^{\omega}(\vec{r})=\int\textrm{d}^{3}r'[\Re{\{\bar{\chi}(\vec{r}-\vec{r}')/\rho\tilde{\alpha}\}}
-\delta^{(3)}(\vec{r}-\vec{r}')\mathbb{I}]\cdot\vec{E}^{\omega}_{D}(\vec{r}'),
\end{equation}
where the subscript $I$ stands for either \emph{induced} or \emph{indirect}.
Therefore, we can write
\begin{equation}\label{Wcohy}
W^{Coh.}=W_{D}^{Coh.}+W_{I}^{Coh.}=\frac{\pi\omega}{6\hbar}|\vec{p}_{0}|^{2}\Bigl[\langle|\vec{E}^{\omega}_{D}(\vec{r})|^{2}\rangle\:+\:
\Re{\{\langle\vec{E}^{\omega}_{I}(\vec{r})\cdot\vec{E}^{\omega*}_{D}(\vec{r})\rangle\}}\Bigr].
\end{equation}
As expected, the first term is the coherent power emitted directly by the dipole source while the second term is the coherent power associated to the interference of the field emitted by the source and that emitted by the induced dipoles.\\
\indent As an example, let us consider $W_{I}^{[1],Coh.}$ in the single-scattering model of Section \ref{Sect5A}. It is given by Eq.(\ref{g1perrrc}) --modulo the appropriate prefactors-- and its diagram is that of Fig.\ref{fig13}($b$). Making the identification  $[\Re{\{\bar{\chi}^{[1]}(\vec{r}-\vec{r}')/\rho\tilde{\alpha}\}}-\delta^{(3)}(\vec{r}-\vec{r}')\mathbb{I}]
\simeq k_{0}^{2}\rho\alpha_{0}\Theta(|\vec{r}-\vec{r}'|-\xi) \Re{\{\bar{G}_{stat.}^{(0)}(\vec{r}-\vec{r}')\}}$, and applying the fluctuation-dissipation relation, it reads
\begin{equation}\label{w1co1}
W_{I}^{[1],Coh.}=W_{\omega}^{0}\frac{2\pi^{2}}{k_{0}\hbar}\textrm{Tr}\Bigl\{\int\textrm{d}^{3}r'
 \bar{G}_{stat.}^{(0)}(\vec{r}-\vec{r}')\cdot[\Theta(|\vec{r}-\vec{r}'|-\xi)\epsilon_{0}\rho\alpha_{0}
\langle\vec{E}^{\omega}_{D}(\vec{r}')]\vec{E}^{\omega*}_{D}(\vec{r})\rangle\Bigr\},
\end{equation}
where the Dyson field is fully transverse. The quantity within square brackets is
the polarization density induced by the coherent field on the surrounding host scatterers,
$\vec{p}^{\omega}(\vec{r}')=\Theta(|\vec{r}-\vec{r}'|-\xi)\epsilon_{0}\rho\alpha_{0}
\vec{E}^{\omega}_{D}(\vec{r}')$. The electrostatic components of the fields emitted by the induced dipoles within the sphere of radius $\xi$ add up at the emitter location, $\vec{r}$, in the form of and induced effective field,
\begin{equation}
\vec{E}^{'\omega}_{I}(\vec{r})=\frac{\k_{0}^{2}}{\epsilon_{0}}\int\textrm{d}^{3}r'\bar{G}_{stat.}^{(0)}
(\vec{r}-\vec{r}')\cdot\vec{p}^{\omega}(\vec{r}').
\end{equation}
It is the transverse component of the above field that interferes with $\vec{E}^{D}(\vec{r})$,
\begin{equation}\label{w1co2}
W_{I}^{[1],Coh.}=W_{\omega}^{0}\frac{2\pi^{2}\epsilon_{0}}{k^{3}_{0}\hbar}
\langle\vec{E}^{'\omega}_{I}(\vec{r})\cdot\vec{E}^{\omega*}_{D}(\vec{r})\rangle.
\end{equation}
Because $\vec{E}^{'\omega}_{I}(\vec{r})$ has its source in $\vec{p}^{\omega}(\vec{r}')$ and this is excited by $\vec{E}^{\omega}_{D}(\vec{r}')$, the interference is coherent.\\
\indent In experiments, it is possible in principle to measure both the total and the coherent intensity. One can think for instance of a medium made of dipole antennas in which one of them is excited by some external fixed field of frequency $\omega=ck_{0}$. Part of the power supplied is absorbed directly by the source antenna located at $\vec{r}$, which is given by the term of Eq.(\ref{la3}). The rest is radiated. Again, part of the radiation will be extinguished, but $W^{Coh.}(\vec{r})$ as given in Eq.(\ref{Wcoher}) will propagate coherently through the medium and can be collected, eventually, by some distant receiver. If the medium can be treated as an effective medium, the coherent intensity collected by a receiver located at $\vec{r'}$ will be given by
\begin{equation}\label{WCohprop}
W^{Coh.}(|\vec{r'}-\vec{r}|)=W^{Coh.}(\vec{r})\exp{[-2k_{0}|\vec{r'}-\vec{r}|\bar{\kappa}]},
\end{equation}
where $\bar{\kappa}\equiv\Im{\{\sqrt{\chi_{eff}\}}}$ is the extinction coefficient which contains both dispersion and absorbtion by the host scatterers.\\
\indent It is also instructive to compute  $W^{Coh.}$ for the stimulated emission of one of the atoms of a monoatomic dielectric close to the resonance. Let us assume the polarizability of the atoms adjusts to the Lorentzian function of Eq.(\ref{alphaLorentz}). Using Eq.(\ref{la2}), we can write the coherent power emitted by the stimulated atom sited at $\vec{r}$ as
\begin{equation}\label{gammacohat}
W^{Coh.}_{atom}=-\frac{\omega^{3}\epsilon_{0}}{6c^{2}\rho}|\tilde{\alpha}(\tilde{k})|^{2}|\vec{E}_{0}^{\omega}|^{2}\textrm{Tr}_{-}
\Bigl\{\int\textrm{d}^{3}r'\:\Re{\{\frac{\bar{\chi}(\vec{r},\vec{r}')}{\tilde{\alpha}}\}}
\Im{\{\bar{G}(\vec{r}',\vec{r};\omega)\}}\Bigr\}\Bigr|_{\tilde{k}=\omega/c}
\end{equation}
Close to the resonance, $\tilde{\alpha}(k_{res})\approx i\:c k_{res}\alpha_{0}/\Gamma$
is nearly pure imaginary and the total power supplied reads
\begin{equation}
W_{Tot.}^{\omega_{res}}=\frac{\omega_{res}\epsilon_{0}}{2}|\vec{E}_{0}^{\omega_{res}}|^{2}
\Im{\{\tilde{\alpha}(k_{res})\}}\approx\frac{\omega_{res}\epsilon_{0}}{2}|\vec{E}_{0}^{\omega_{res}}|^{2}
|\tilde{\alpha}|.
\end{equation}
We can write $\Im{\{\bar{G}(\vec{r}',\vec{r};\omega)\}}$
in terms of $\hat{\vec{E}}^{\omega}_{D}$ using the fluctuation-dissipation relation of Eq.(\ref{Disiclass})
and, making use of the constitutive relation for coherent fields, write the polarization vector as
\begin{equation}\label{MacroPolarizability}
\vec{P}^{\omega}(\vec{r})=\epsilon_{0}\int\textrm{d}^{3}r'\:\bar{\chi}(\vec{r},\vec{r}')\cdot\vec{E}^{\omega}_{D}(\vec{r}').
\end{equation}
Finally, by inserting the above results into Eq.(\ref{gammacohat}) we obtain the more familiar expression
\begin{equation}
W^{Coh.}_{\omega_{res}}=-\frac{\pi}{3\hbar}W_{Tot.}^{\omega_{res}}
\Im{\{\langle\frac{1}{\rho}\vec{P}^{*\omega_{res}}(\vec{r})\cdot\vec{E}_{D}^{\omega_{res}}(\vec{r})\rangle\}},
\end{equation}
where $\frac{1}{\rho}\vec{P}^{*\omega_{res}}(\vec{r})$ is the average molecular dipole moment.
Therefore, we see that in the VC scenario there is a well defined ratio between the coherent
 emission of a stimulated dipole close to resonance and the total power supplied. That ratio being given as a function of coherent fields.
\subsubsection{The coherent vacuum and the additional polarization of the sourceless vacuum}\label{Sect6B2}
In the following we interpret the coherent emission in QFT terms.
From Eqs.(\ref{Disiclass},\ref{Wcohy}), we conclude that the states accessible to direct coherent emission are the normal modes which satisfy Eqs.(\ref{dispper},\ref{disppar}). Therefore, in an effective manner, the vacuum though which these modes propagate is the sourceless vacuum, $|\Omega\rangle^{s.l.}$. This effective equivalence is analogous to that found in Section \ref{Sect2} between $|\Omega\rangle^{s.p.}_{1D}$ and $|0\rangle$ for a unique dipole in free space. $|\Omega\rangle^{s.l.}$ is made of the superposition of EM vacua, $\{|\phi_{m}\rangle^{s.l.}\}$, in which the normal modes of Maxwell's equations with permitivities $\{\epsilon_{m}(\vec{r})\}$ propagate. Following the formalism of Section \ref{Sect2C}, the $m^{th}$ pair of vacuum state $|\phi_{m}\rangle^{s.l.}$ and dielectric permitivity $\epsilon_{m}(\vec{r})$ is associated in a one-to-one correspondence to the $m^{th}$ configuration of host scatterers such that $|\Omega\rangle^{s.l.}=\sum_{m}\sqrt{M_{mm}}|\phi_{m}\rangle^{s.l.}$,
\begin{eqnarray}\label{Dysoninterpret}
\sum_{m,n}\:^{s.l.}\langle\phi_{m}|\hat{M}\hat{\vec{E}}^{\omega}(\vec{r})\hat{\vec{E}}^{\omega\dag}
(\vec{r})|\phi_{n}\rangle^{s.l.}
&=&^{s.l.}\langle\Omega|\hat{\vec{E}}^{\omega}(\vec{r})\hat{\vec{E}}^{\omega\dag}(\vec{r})|\Omega\rangle^{s.l.}
\nonumber\\&=&
-\hbar\frac{\omega^{2}}{\epsilon_{0}\pi c^{2}}\Im{\{\bar{G}(\vec{r},\vec{r}';\omega)\}}.
\end{eqnarray}
\indent On the other hand, it is plain from Eqs.(\ref{Gammapcoh},\ref{Gammalcoh}) that the modes which contribute to $W^{Coh.}$ are the same as those which propagate in $|\Omega\rangle^{s.l.}$. However, the amplitude of the spectrum of fluctuations which enter $W^{Coh.}$ differs w.r.t. that of the fluctuations in $|\Omega\rangle^{s.l.}$ by a multiplicative factor $\Re{\{\bar{\chi}/\rho\tilde{\alpha}\}}$. In other words, the EM vacuum in which coherent emission propagates, $|\Omega\rangle^{Coh}$, is additionally polarized w.r.t. $|\Omega\rangle^{s.l.}$. The  renormalization function is given by $Z^{\omega}_{\perp,\parallel}=\Re{\Bigl\{\frac{\chi^{\omega}_{\perp,\parallel}}{\rho\tilde{\alpha}}
\Bigr\}}$, which takes account of the polarization due to the closest scatters surrounding the emitter. We will see later on that $Z^{\omega}_{\perp,\parallel}$ is the so-called local field factor in the theory of the effective medium. In the framework of QFT,
$Z^{\omega}_{\perp,\parallel}$ can be interpreted as the field-strength renormalization factors. In a scalar QFT it is the analytical structure of $G(k)$ together with $Z$ in the complex plane
that determine the amplitude of the vacuum fluctuations and the mass spectrum of the particles propagating in space-time \cite{Peskin}. Correspondingly, it is
the analytical structure of $G_{\perp,\parallel}(k)$ together with $\Re{\{\frac{\chi_{\perp,\parallel}(k)}{\rho\tilde{\alpha}}\}}$ that determine the amplitude of the EM vacuum  fluctuations and the $k_{\perp,\parallel}^{prop.}$-spectrum of the coherent photons propagating
in a random medium.
Following up the QFT interpretation, $Z_{\perp,\parallel}^{\omega}(k)$ yields the probability density for creating a photon of frequency $\omega$ from vacuum at the emitter site, $Z^{\omega}(k)=|^{Coh}\langle\Omega|E_{\perp,\parallel}(\vec{r})|\gamma^{\omega}_{k}\rangle|^{2}$. Out of those photons, the coherent ones are selected and propagated through the bulk by $G^{\omega}_{\perp}$. Also, the electric field  gets renormalized  and the field so renormalized is nothing but the Dyson field, $\vec{E}_{D}^{\omega}|_{\perp,\parallel}=Z_{\perp,\parallel}^{-1/2}\vec{E}^{\omega}_{\perp,\parallel}$. This leads to the equivalence relation,
\begin{eqnarray}\label{ED}
^{s.l.}\langle\Omega|\hat{\vec{E}}^{\omega}(\vec{r})\hat{\vec{E}}^{\omega\dag}(\vec{r})
|\Omega\rangle^{s.l.}&=&
^{Coh}\langle\Omega|\hat{\vec{E}}_{D}^{\omega}(\vec{r})\hat{\vec{E}}_{D}^{\omega\dag}(\vec{r})|\Omega\rangle^{Coh}
\nonumber\\&=&
-\hbar\frac{\omega^{2}}{\epsilon_{0}\pi c^{2}}\Im{\{\bar{G}(\vec{r},\vec{r}';\omega)\}}.
\end{eqnarray}
Thus, in passing from $\vec{E}^{\omega}_{D}$ to $\vec{E}^{\omega}$, any correlation function between vacuum states picks up a factor $[Z^{\omega}]^{1/2}$ per field operator.\\
\indent All this suggests that an effective electromagnetic theory in random media can be formulated in terms of the renormalized quantities --see eg. Ch.7-10 of \cite{Peskin}. The host scatterers play the role of bare vertices of photon interactions as depicted in the diagrams of Fig.\ref{fig132}. The correlations, depicted by dashed lines in our diagrams, play an analogous role to that of the gauge fields which mediate interactions between charged fields in QFTs. If correlations are assumed not to have dynamics in accordance to the weak-coupling-Markovian approximation, they must be intended as a back-ground field.
\subsubsection{Radiation in an effective medium}\label{Sect5B2}
\indent In this section we deal with the effective medium theory, which is implicitly used in the majority of works on dipole emission. In this context, and only in this context, an analog contribution to the radiative emission in free space can be defined.\\
\indent The effective medium theory is defined such that, for some range of frequencies and for $k\xi\ll1$, an effective homogeneous complex dielectric constant can be defined as $\epsilon_{eff}\equiv\:\textrm{Lim.}\:\{\epsilon_{\perp,\parallel}(k)\}$ as $k\xi\rightarrow0$, so that the  Dyson propagator $G_{\perp}(k)$ can be approximated by the effective propagator $G_{\perp}^{eff}(k)$ for some range of frequencies,
\begin{equation}\label{effective}
G_{\perp}^{eff}(k)\equiv\frac{1}{\epsilon_{eff}k_{0}^{2}-k^{2}},
\qquad G_{\parallel}^{eff}(k)\equiv\frac{1}{\epsilon_{eff}k_{0}^{2}}.
\end{equation}
The field with propagator $\bar{G}_{eff}$ will be referred to as Dyson-effective field or \emph{macroscopic} field, $\vec{E}^{eff}_{D}$. Correspondingly, the field with propagators
\begin{equation}\label{effectivemath}
\mathcal{G}_{\perp}^{eff}(k)\equiv\frac{\chi_{eff}}{\rho\tilde{\alpha}}\frac{1}{\epsilon_{eff}k_{0}^{2}-k^{2}},
\qquad\mathcal{G}_{\parallel}^{eff}(k)\equiv\frac{\chi_{eff}}{\rho\tilde{\alpha}}\frac{1}{\epsilon_{eff}k_{0}^{2}},
\end{equation}
will be referred to as local-effective field $\vec{E}_{loc}^{eff}$. Further on, in order for $G_{\perp}^{eff}(k)$ to be the Green function of a propagating field of wave number $k_{0}$ it is required $k/k_{0}\gtrsim1$ at the same time that $k\xi\ll1$, which implies $k_{0}\xi\ll1$ as well.\\
\indent To what computation
of power emission is concerned, the use of macroscopic fields implies the neglect in the integrals of
Eqs.(\ref{Gammapcoh}-\ref{Gammalincoh}) of the poles of order $\gtrsim1/\xi$. The neglected  modes correspond to the near field dispersion and absorbtion carried out by the scatterers surrounding the source --see  Fig.\ref{fig13} and also \cite{Feldoher}. In turn, this implies the neglect of all longitudinal emission and the restriction of the integration domain in Eq.(\ref{eka}) for transverse emission,
\begin{eqnarray}
2W_{\perp}^{eff}&\equiv&2W_{\perp}|^{k_{0}\xi\ll1}_{k\ll\xi^{-1}}\simeq 2W_{o}
\int\frac{\textrm{d}^{3}k}{(2\pi)^{3}}\:\Im{\Bigl\{\frac{\chi_{eff}}{\rho\tilde{\alpha}}
\frac{1}{\epsilon_{eff}k_{0}^{2}-k^{2}}\Bigr\}}\label{rad1}\\
&=&2\tilde{W}_{o}\Im{\Bigl\{\frac{\chi_{eff}}{\rho\tilde{\alpha}}
\sqrt{\epsilon_{eff}}\Bigr\}},\qquad\tilde{W}_{o}\equiv-\frac{k_{0}}{4\pi}W_{o}.\label{rad2}
\end{eqnarray}
The coherent component of the above formula, $2W_{\perp}^{eff}|^{Coh.}$ can be written in an analogous manner to the radiative emission in free space,
\begin{eqnarray}
2W_{\perp}^{eff}|^{Coh.}=2\tilde{W}_{o}\Re{\Bigl\{\frac{\chi_{eff}}{\rho\tilde{\alpha}}\Bigr\}}
\Re{\{\sqrt{\epsilon_{eff}}\}}.\label{radc2}
\end{eqnarray}
$2W_{\perp}^{eff}|^{Coh.}$ can be given in closed form for a Maxwell-Garnett (MG) fluid. In such a fluid the only relevant correlation between host particles is the negative correlation due to the exclusion volume $\sim(4\pi/3)\xi^{3}$ which prevents scatterers from overlap. Provided $k_{0}\xi\ll1$,
$\epsilon_{eff}$ obeys the MG formula \cite{Maxwell}, $\rho\tilde{\alpha}=\frac{\epsilon_{eff}-1}{(\epsilon_{eff}+2)/3}$. It was firstly proved in \cite{Feldoher} and then in \cite{vanTigg} following a diagrammatic approach that the MG formula is correct at all orders in $\rho\tilde{\alpha}$. Note however that the usage of the renormalized polarizability $\tilde{\alpha}$ incorporates in our case the effects of recurrent scattering. Using the MG formula, Eq.(\ref{rad2}) reads
\begin{equation}\label{GammaRADMG}
2W_{\perp}^{MG}|^{Coh.}=2\tilde{W}_{o}\Re{\Bigl\{\frac{\epsilon_{eff}+2}{3}\Bigr\}}\Re{\{\sqrt{\epsilon_{eff}}\}}.
\end{equation}
\indent It is in order to make a couple of comments regarding the use of effective fields in the computation of LDOS,
 $W$ and $\Gamma$. First, in Eq.(\ref{rad1}) the symbol $\simeq$ stands for the approximation
$\int_{0<k<A^{-1}\xi^{-1}}\frac{\textrm{d}^{3}k}{(2\pi)^{3}}\simeq\int_{k>0}\frac{\textrm{d}^{3}k}{(2\pi)^{3}}$
under the assumption $(k_{0}\xi)^{-1}>A\gg1$. That implies that the integral of Eq.(\ref{rad1}),
which is the inverse Fourier transform of $\mathcal{G}^{eff}_{\perp}(k)$ evaluated at $r=0$, does
not actually carry information about the fluctuations of the microscopical local self-polarization field evaluated around the emitter location, $r\leq \xi$. Instead, the fluctuations there are those of the field $\vec{E}_{loc}^{eff}$,
which is the local field spatially-averaged in volumes of the order of $\xi^{3}$ over which also
the spatial average of $\epsilon(r)$ is performed in order to get $\epsilon_{eff}$. Therefore, because the contribution to $W_{\parallel}$ comes from high frequency modes in the range $k\gtrsim1/\xi$, it is not possible to take the effective dielectric constant limit $k\xi\ll1$ for its evaluation and it is in general unavoidable to incorporate the spatial dispersion in $\chi_{\parallel}(k)$. Differently to normal propagating modes, the electric field of the source dipole is not approximately uniform within a cavity volume of the order of $\sim\xi^{3}$ and the emitted photons do not see an effective (uniform) dielectric constant there. In the real large-cavity scenario however, it might still happen that the field at distances $1/R$ is approximately the same as that at $1/(R+\xi)$, $\xi$ being the typical dispersion length scale of $\bar{\chi}$ in the host medium. It is only in that case and for $k_{0}\xi\ll1$ that it is possible to consider $\bar{\chi}$ uniform outside the real cavity.\\
\indent Second, let us assume for a moment that the effective field approximation is good enough.
It is known long ago that the effective field which appears in Fermi's Golden rule in the form of
the equal-point correlation functional, $\langle \vec{E}(r)\vec{E}(r)\rangle$, is not $\vec{E}_{D}^{eff}(\vec{r})$
but $\vec{E}_{loc}^{eff}(\vec{r})$. Further on, assuming the medium is passive, emission restricts to
 $2W_{\perp}^{eff}|^{Coh.}$. In application of the fluctuation-dissipation theorem,
\begin{eqnarray}
2W_{\perp}^{eff}|^{Coh.}&=&
W_{o}\int\frac{\textrm{d}^{3}k}{(2\pi)^{3}}\Re{\{\frac{\chi_{eff}}{\rho\tilde{\alpha}}\}}\Im{\{G_{\perp}^{eff}(k)\}}\nonumber
\\&=&\frac{-\pi\epsilon_{0}}{\hbar k_{0}^{2}}W_{o}\Re{\{\frac{\chi_{eff}}{\rho\tilde{\alpha}}\}}\langle \vec{E}^{eff}_{\perp D}(r)\vec{E}_{\perp D}^{eff}(r)\rangle.
\end{eqnarray}
Notice that it contains just a factor $\Re{\{\frac{\chi_{eff}}{\rho\tilde{\alpha}}\}}$, which  equals a Lorentz-Lorentz local field factor for the case that MG formula applies, $\mathcal{L}_{LL}=\Re{\{\frac{\epsilon_{eff}+2}{3}\}}$. It is a common mistake to think that $\mathcal{L}_{LL}$ should appear as squared in front of $\langle \vec{E}_{D}^{eff}(r)\vec{E}_{D}^{eff}(r)\rangle$ in  Fermi's Golden rule. The reason being that while the propagator of $\vec{E}_{loc}^{eff}(r)$ is that of $\vec{E}_{D}^{eff}(r)$ multiplied by $\mathcal{L}_{LL}$, $\mathcal{G}_{eff}\sim\mathcal{L}_{LL}G_{eff}$, the fluctuation-dissipation theorem applied over $\mathcal{G}_{eff}$ yields a contribution quadratic in $\vec{E}_{D}$ but linear in $\mathcal{L}_{LL}$. This contradicts the paradigm which states that \emph{the effect of a dielectric host is to multiply each occurrence of the dipole moment of a a two-level atom by a local-field factor} \cite{Crenshaw} --see Section \ref{Sect7} for further discussion.
\subsubsection{On the radiative and non-radiative energy transfer}\label{Sect5B3}
\begin{figure}[h]
\includegraphics[height=11.8cm,width=14.8cm,clip]{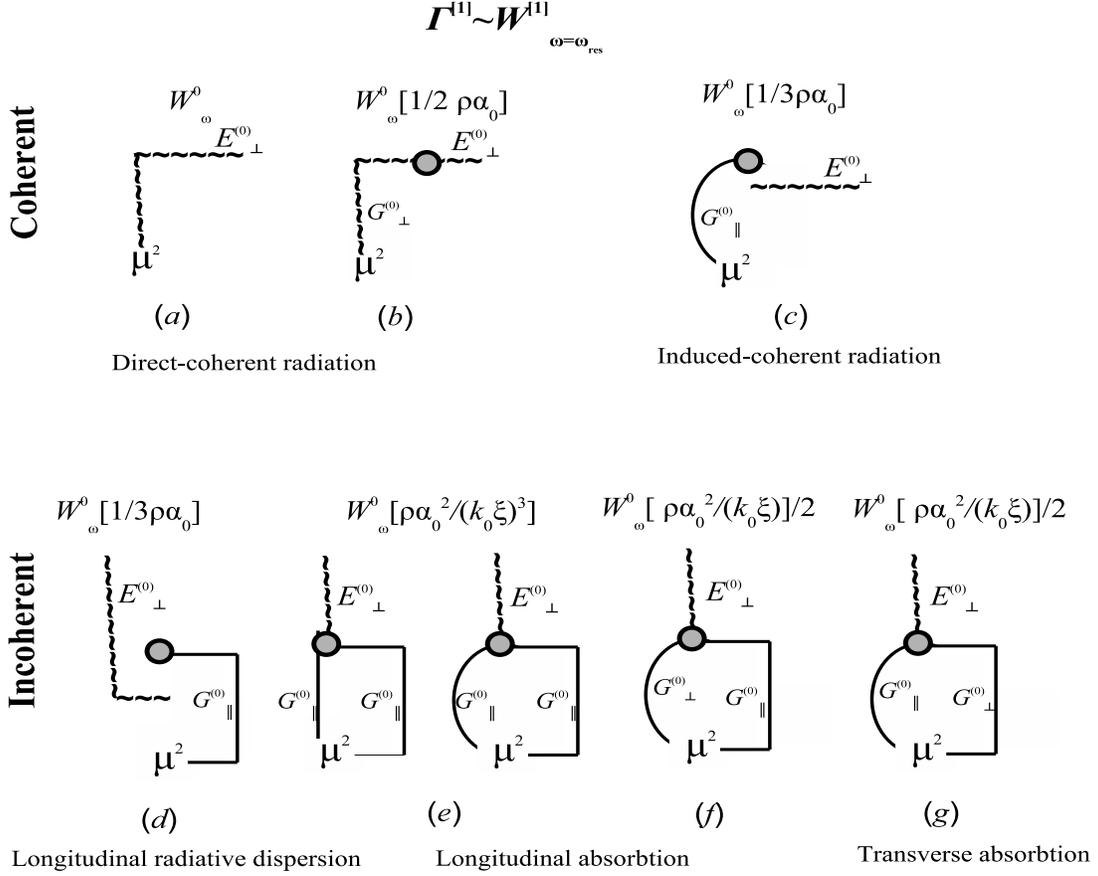}
\caption{($a$) Diagrammatic representation of the stimulated emission/spontaneous emission of an emitter in the single-scattering approximation --Eq.(\ref{radapp}). Explanation is given within the text.}\label{fig13}
\end{figure}

Taking advantage of the above study, we can interpret the radiative and non-radiative nature of the processes involved in the dipole emission of the single scattering model. Let us consider the diagrams in Fig.\ref{fig13}. They represent either the stimulated emission or the spontaneous emission of an emitter in the single-scattering approximation --Eq.(\ref{radapp}). In the case of stimulated emission, $\mu^{2}$ stands for the polarizability assigned to the emitter. The propagators on the right hand side of the emitter are part of the bulk propagator $G$. Actual emission is denoted by wavy lines while  induction is mediated by propagators depicted by solid lines. Curved propagators on the left hand side of the emitter indicate being constrained by spatial correlations within $\chi$. Emission with origin at the emitter is direct. It is induced emission otherwise.\\
\indent Since the work of Andrews \cite{Andrews2,Andrews}, it is well appreciated that the radiative and non-radiative energy transfer between a donor and an acceptor correspond to a unique quantum mechanical process. When no induction exists, the emission is direct and there is no transfer of energy. It corresponds to the diagrams ($a$), for in-free-space emission; and ($b$), which stands for single scattering in the bulk --Eq.(\ref{g1perrra}).\\
\indent Induced emission processes, either coherent or incoherent, are characterized by two features. First, there exists some correlation between the donor and the acceptor. Second, it is one of the transverse virtual photons of frequency $\omega_{res}$ which mediated the interaction donor-acceptor, that is made actual when   satisfying the dispersion relation of Eq.(\ref{dispper}). The pair donor-acceptor behaves as an effective dipole. That in diagram ($c$) is coherent and so is its (transverse) radiation, as correlations apply to the electrostatic induction --Eq.(\ref{g1perrrc}). On the contrary, that in diagram ($d$) is incoherent, as correlations apply over the transverse induction whose photon is the one becoming actual --Eq.(\ref{g1parrrc}). As explained above, the processes in $(c)$, ($d$) carry one local field factor each in the single scattering approximation. They are the processes which can be considered to amount to radiative transfer.\\
\indent The non-radiative transfer processes are those in diagrams ($e,f,g$). They are characterized by two features. First, the actual photons have their origin either in radiative corrections over the acceptor bare polarizability or on the imaginary part of the bare polarizability. Hence, the corresponding terms are all proportional to $\Im{\{\tilde{\alpha}\}}$. Only radiative corrections are depicted in Fig.\ref{fig13}. Second, the transference of energy takes place through either purely --diagrams in ($e$)-- or partially --diagrams ($f,g$)-- longitudinal loops which remain closed during the re-emission process. In general the spectrum of energy levels of the acceptor presents a number of channels into which the quantum of energy absorbed, $\hbar\omega_{res}$, can be re-emitted. Therefore, re-emission takes place in a variety of frequencies and the associated decay rate is said non-radiative. Note however that this implies a slight inconsistency in the case that donor and acceptor be identical two-level atoms. That is so because in such a case the acceptor can only re-emit in the same frequency as that of the direct radiation. Nevertheless, because  two-level dipoles are unrealistic idealizations, we will consider those processes as non-radiative without loss of generality. The diagrams in ($e$) stand for the terms in Eqs.(\ref{g1parrra},\ref{g1parrrb}). They would diverge if considered separately. They represent the phenomenon commonly referred to as F\"{o}ster  Resonant Energy Transfer (FRET) \cite{Foster}. The diagrams ($f$) and ($g$) correspond to the absorptive longitudinal and transverse  terms of Eq.(\ref{g1parrrc}) and Eq.(\ref{g1perrrc}) respectively.
\section{Analytical calculations in Maxwell-Garnett dielectrics}\label{Sect6}
An MG dielectric is one made of point dipoles with well defined single particle polarizabilities. The correlation length $\xi$ among dipoles satisfies $k_{0}\xi\ll1$ for the frequencies of interest. Under these conditions, the only relevant correlation function between its constituents is given by an exclusion volume correlation function. In such a medium, an effective dielectric constant can be defined and adjusts to the MG formula,
\begin{equation}
\rho\tilde{\alpha}=\frac{\epsilon_{eff}-1}{(\epsilon_{eff}+2)/3}.\label{MGeq}
\end{equation}
Hereafter we will denote $\epsilon_{eff}$ with $\epsilon_{MG}$ instead.
In this scenario it is possible to obtain analytical formulae for the virtual cavity $\gamma$-factors. We apply those formulae to two situations of experimental interest. In the first one, we compute the decay rate of an excited atom which can be either an interstitial optical center or one of the molecules of the dielectric itself. In the second situation, we compute the effective dielectric constant for an MG dielectric close to the resonance frequency.
\subsection{Spontaneous decay rate of a weakly-polarizable excited atom}\label{Sect6A}
Let us consider the explicit computation of $\Gamma$ for an excited atom. This can be either a foreign ion   placed within one of the molecules of the dielectric or a molecule of the dielectric itself. In the former case, the manner the emission rate of an optic center gets modified when placed within the ligand is a difficult issue as, in general, one should consider the internal atomic structure of the host molecule \cite{Chew,Ruskies,Duan2}. To simplify matters, we assume here that the bare dipole transition amplitude of the ion and its resonance frequency get renormalized to known values $\mu$, $\omega_{res}$ as a result of the interactions with other atoms within the ligand. This scenario is that of Section \ref{Sect3D}. The electrostatic polarizability of the doped molecule, $\alpha_{0}$, is carried out by some other polarizable atoms, being $\alpha_{0}$ equivalent to the bare polarizability of the rest of the host scatterers. The total dipole moment associated to the particle which hosts the ion is the combination of a fixed and an induced dipole, $\vec{p}_{host}=\vec{\mu}+\frac{\omega^{2}}{c^{2}}\tilde{\alpha}\bar{\mathcal{G}}_{\omega}(\vec{r},\vec{r})\vec{\mu}$,
$\tilde{\alpha}$ being the renormalized polarizbility of the host
particle.\\
\indent The case of an excited host particle can be treated in the same fashion provided that the additional  condition $\xi\ll\rho^{-1/3}$ is satisfied and the effective polarizability of the excited molecule is negligible --see Section \ref{Sect4B1}. That corresponds to the setup of Section \ref{Sect3D} with $\alpha_{0}=0$ and assuming previous knowledge of $\mu$ and $\omega_{res}$.\\
\indent In this scenario, the only correlation function which matters is the two-point function which amounts to the exclusion volume of each atom, $h(r)=-\Theta(r-\xi)$. For higher correlation orders and following \cite{Feldoher} we will make use of the overlap approximation. Ultimately, because Eq.(\ref{MGeq}) establishes a relationship between  $\chi_{MG}$ and $\tilde{\alpha}$, and the $\gamma$-factors can be expressed as power series of $\rho\tilde{\alpha}$, a relationship can be given between $\Gamma$, LDOS and the effective parameters which determine the transport of coherent light. Those are the index of refraction, $\bar{n}$, and the extinction mean free path, $l_{ext}$, which relate to $\chi_{MG}$ through
\begin{equation}
\bar{n}=\Re{\{\sqrt{1+\chi_{MG}}\}},\qquad l_{ext}^{-1}=2\frac{\omega}{c}\Im{\{\sqrt{1+\chi_{MG}}\}}.
\end{equation}
\indent First, we compute $\mathcal{G}_{\perp}$. In the overlap approximation,
\begin{equation}
\chi_{\perp}(k)=\frac{\rho\tilde{\alpha}}{1-\chi_{\perp}^{(2)}(k)/(\rho\tilde{\alpha})},
\end{equation}
where at resonance,
\begin{eqnarray}
\chi^{(2)}_{\perp}(k)&=&\frac{-(\tilde{\alpha}\rho)^{2}k_{res}^{2}}{2}\int\textrm{d}^{3}r\:e^{i\vec{k}\cdot\vec{r}}h(r)\textrm{Tr}
\{\bar{G}^{(0)}(r)[\mathbb{I}-\hat{k}\otimes\hat{k}]\}\nonumber\\&=&
-\frac{(\tilde{\alpha}\rho)^{2}k_{res}^{2}}{2}\int\frac{\textrm{d}^{3}k'}{(2\pi)^{3}}
h(|\vec{k}'-\vec{k}|)\Bigl[G_{\perp}^{(0)}(k')\nonumber\\&+&G_{\perp}^{(0)}(k')\cos^{2}{\theta}\:+\:G_{\parallel}^{(0)}(k')\sin^{2}{\theta}\Bigr].
\label{Xiperp}
\end{eqnarray}
For an MG dielectric, $h(Q)=-4\pi\xi^{3}\:j_{1}(Q)/Q=4\pi\xi^{3}\frac{Q\cos{Q}-\sin{Q}}{Q^{3}}$ with $Q\equiv|\vec{k}'-\vec{k}|\xi$ and $j_{1}$ the spherical Bessel function of first order. Applying next Eq.(\ref{LDOSIper}) at the resonance frequency, we find
\begin{equation}\label{laG}
\mathcal{G}_{\perp}(k)=G_{\perp}^{(0)}(k)
\Bigl[1+\rho\tilde{\alpha}[k_{res}^{2}G_{\perp}^{(0)}(k)-\chi^{(2)}_{\perp}(k)/(\rho\tilde{\alpha})^{2}]\Bigr]^{-1}.
\end{equation}
From the above expression it is not straightforward to differentiate between the coherent and the incoherent components of $2\Gamma_{\perp}$ as given in Eqs.(\ref{Gammapcoh},\ref{Gammapincoh}) for the power emission. However, it is possible to go around this problem by first computing Eq.(\ref{laG}) order by order in $\zeta\equiv k_{res}\xi$ and then find out the origin of those terms in view of Eqs.(\ref{Gammapcoh},\ref{Gammapincoh}). First, we write $2\gamma^{Tot.}_{\perp}$ as
\begin{equation}\label{laGama}
2\gamma^{Tot.}_{\perp}=\frac{k_{res}}{\pi^{2}\zeta}\Bigl\{\int_{0}^{\infty}\textrm{d}Q\frac{Q^{2}}{\zeta^{2}-Q^{2}}
\sum_{n=0}^{\infty}(-\rho\tilde{\alpha})^{n}[\frac{\zeta^{2}}{\zeta^{2}-Q^{2}}-(\rho\tilde{\alpha})^{-2}\chi^{(2)}_{\perp}(\zeta,Q)]^{n}\Bigr\},
\end{equation}
where $Q\equiv k\xi$. Careful examination of the integrand of Eq.(\ref{laGama}) yields the following decomposition which includes all the terms up to $\mathcal{O}(0)$ in $\zeta$,
\begin{eqnarray}
2\gamma^{Tot.}_{\perp}&\simeq&\frac{k_{res}}{\pi^{2}\zeta}\Bigl\{
\sum_{n=0}^{\infty}(-\rho\tilde{\alpha})^{n}\int_{0}^{\infty}\textrm{d}Q\frac{Q^{2}}{\zeta^{2}-Q^{2}}
[\frac{\zeta^{2}}{\zeta^{2}-Q^{2}}-1/3]^{n}\label{laGamaCoh1}\\
&+&\sum_{n=1}^{\infty}(-\rho\tilde{\alpha})^{n}\int_{0}^{\infty}\textrm{d}Q[j_{1}(Q)/Q]^{n}\Bigr\}\label{laGamainCoh1}\\
&=&2\gamma_{\perp}^{Tot.A}+2\gamma_{\perp}^{Tot.B}.
\end{eqnarray}
The integrand of Eq.(\ref{laGamaCoh1}) contains the poles of $G_{\perp}$ which satisfy the dispersion relations of coherent modes. Eq.(\ref{laGamainCoh1}) carries those very high frequency modes, $k\gg k_{res}$, which contain the near field contribution. By taking the limit $\zeta\rightarrow0$, it is immediate to identify in Eq.(\ref{laGamaCoh1}) the factor $-1/3\:(\rho\tilde{\alpha})^{2}$ with the zero mode of
$-\chi^{(2)}_{\perp}(k)$. In the overlap approximation, that leads to the MG formula for the effective susceptibility. Hence, it is $\chi^{(2)}_{\perp}(k=0)=1/3\:(\rho\tilde{\alpha})^{2}$ that enters in the place of $\chi^{(2)}_{\perp}(k)$ in Eq.(\ref{laG}) for $\mathcal{G}_{\perp}^{eff}$. We take advantage of this to arrange the leading order terms of Eq.(\ref{laGamaCoh1}) (which are those of $\mathcal{O}(0)$ in $\zeta$). We obtain,
\begin{eqnarray}
2\Gamma_{\perp}^{A}&=&2\tilde{\Gamma}_{o}\Im{\{-2\gamma_{\perp}^{Tot.A}\}}\frac{2\pi}{k_{res}}\approx2\tilde{\Gamma}_{o}\Re{\Bigl\{\frac{\epsilon_{MG}+2}{3}\sqrt{\epsilon_{MG}}\Bigr\}},\label{laGamaA}\\
\tilde{\Gamma}_{o}&=&\frac{k_{res}^{3}}{6\epsilon_{0}\pi\hbar}|\mu|^{2}\Big{|}1-\frac{i}{6\pi}k_{0}^{3}\alpha_{0}+
\frac{1}{3}k_{0}^{2}\alpha_{0}[2\gamma_{\perp}+\gamma_{\parallel}]\Big{|}^{-2},\label{grenorm}
\end{eqnarray}
where $\tilde{\Gamma}_{o}$ contains the $\gamma$-dependent factor which renormalizes $|\mu|^{2}$ due to the polarization of the host particle.
It is also immediate to verify that the coherent contribution of $2\Gamma_{\perp}^{A}$ is precisely $2\Gamma_{\perp}^{MG}|^{Coh.}$ as given in Eq.(\ref{GammaRADMG}) (modulo prefactors),
\begin{equation}\label{laGamaACoh}
2\Gamma_{\perp}^{A,Coh.}=2\Gamma_{\perp}^{MG}|^{Coh.}=2\tilde{\Gamma}_{o}\Re{\Bigl\{\frac{\epsilon_{MG}+2}{3}\Bigr\}}\Re{\{\sqrt{\epsilon_{MG}}\}},
\end{equation}
while
\begin{equation}\label{laGamaAinCoh}
2\Gamma_{\perp}^{A,Incoh.}=-2\tilde{\Gamma}_{o}\Im{\Bigl\{\frac{\epsilon_{MG}+2}{3}\Bigr\}}\Im{\{\sqrt{\epsilon_{MG}}\}}.
\end{equation}
Therefore, at leading order in $\zeta=k_{res}\xi$, we conclude that the coherent emission --and only! the coherent emission-- can be computed in the framework of the effective dielectric constant approximation.\\
\indent On the contrary, incoherent emission does not only contains the term $2\Gamma_{\perp}^{A,Incoh.}$, but also the terms in  $2\Gamma_{\perp}^{B}$ which are proportional to $\Im{\{2\gamma_{\perp}^{Tot.B}\}}$ given in  Eq.(\ref{laGamainCoh1}). $2\gamma_{\perp}^{Tot.B}$ takes account of those non-radiative high frequency modes which scale as $1/\zeta$. Eq.(\ref{laGamainCoh1}) contains terms of the form
\begin{eqnarray}
I^{(1)}_{p,q}&\equiv&\int_{0}^{\infty}\textrm{d}Q\:\frac{\sin{(pQ)}}{Q^{2q+1}}\quad\textrm{with }p<2q+1,\:2n\leq p\leq3n,\label{inte1}\\
I^{(2)}_{p,q}&\equiv&\int_{0}^{\infty}\textrm{d}Q\:\frac{\cos{(pQ)}}{Q^{2q}}\quad\textrm{with }p<2q,\:2n\leq p\leq3n\label{inte2}.
\end{eqnarray}
With increasing value of $n\gg1$ it is easy to verify that the above integrals converge quickly to zero. Hence,
\begin{eqnarray}
I^{(1)}_{p,q}&=&\frac{\pi}{2}\frac{(-1)^{q}p^{2q}}{(2q)!}\label{intel1}\\
I^{(2)}_{p,q}&=&\frac{\pi}{2}\frac{(-1)^{q}p^{2q-1}}{(2q-1)!}\label{intel2}.
\end{eqnarray}
Therefore, for any finite value of $\rho\tilde{\alpha}$, the corresponding series in equation  Eq.(\ref{laGamainCoh1}) is rapidly convergent. We give below the first five terms of that series,
\begin{equation}\label{laGamaB}
2\gamma_{\perp}^{B}=\frac{-k_{res}}{2\pi}\:\frac{1}{\zeta}\Bigl[\frac{1}{2}\rho\tilde{\alpha}-
\frac{2}{15}(\rho\tilde{\alpha})^{2}+\frac{47}{1280}(\rho\tilde{\alpha})^{3}-\frac{334}{31185}(\rho\tilde{\alpha})^{4}+
\frac{6891623}{2145927168}(\rho\tilde{\alpha})^{5}+ ...\Bigr].
\end{equation}
\indent Let us consider next the computation of $\gamma_{\parallel}$. The Green function for longitudinal modes reads
\begin{equation}\label{laGpar}
\mathcal{G}_{\parallel}(k)=\frac{1}{k_{res}^{2}}\frac{\chi_{\parallel}(k)}{\rho\tilde{\alpha}}[1+\chi_{\parallel}(k)]^{-1},
\end{equation}
where in the overlap approximation,
\begin{equation}
\chi_{\parallel}(k)=\frac{\rho\tilde{\alpha}}{1-\chi_{\parallel}^{(2)}(k)/(\rho\tilde{\alpha})},
\end{equation}
with
\begin{eqnarray}
\chi^{(2)}_{\parallel}(k)&=&-(\alpha\rho)^{2}k_{0}^{2}\int\textrm{d}^{3}r\:e^{i\vec{k}\cdot\vec{r}}h(r)\textrm{Tr}
\{\bar{G}^{(0)}(r)[\hat{k}\otimes\hat{k}]\}\nonumber\\&=&-(\alpha\rho)^{2}k_{0}^{2}\int\frac{\textrm{d}^{3}k'}{(2\pi)^{3}}
h(|\vec{k}'-\vec{k}|)\nonumber\\&\times&\Bigl[G_{\parallel}^{(0)}(k')\cos^{2}{\theta}\:+\:G_{\perp}^{(0)}(k')\sin^{2}{\theta}\Bigr].
\label{Xiparall}
\end{eqnarray}
The above equation can be computed in closed form,
\begin{equation}\label{fzeta}
\chi^{(2)}_{\parallel}(Q)=(\rho\tilde{\alpha})^{2}[1+\frac{j_{1}(Q)}{Q}f(\zeta)],\qquad f(\zeta)=2i\:e^{i\zeta}(i+\zeta).
\end{equation}
Likewise, it is possible to give a closed expression for $\mathcal{G}_{\parallel}(Q)$ using the MG relationship for $\epsilon_{MG}$ in function of $\rho\tilde{\alpha}$,
\begin{equation}
\mathcal{G}_{\parallel}(Q)=\frac{1}{k_{res}^{2}}[1+6\frac{\epsilon_{MG}-1}{\epsilon_{MG}+2}\frac{j_{1}(Q)}{Q}]^{-1}.
\end{equation}
The use of this expression in the computation of $\Gamma_{\parallel}$ implies the computation of an infinite series of roots of $1+6\frac{\epsilon_{MG}-1}{\epsilon_{MG}+2}\frac{j_{1}(Q)}{Q}$ \cite{Feldoher}. Instead of that, we rather expand Eq.(\ref{laGpar}) in powers of $\rho\tilde{\alpha}$ as this allows to keep control over the convergence of the series,
\begin{equation}
\mathcal{G}_{\parallel}(Q)=\frac{1}{k_{res}^{2}}\sum_{n=0}^{\infty}(\rho\tilde{\alpha})^{n}(-1)^{n}
f^{n}(\zeta)[j_{1}(Q)/Q]^{n}.
\end{equation}
Finally, we arrive at
\begin{equation}\label{laqtoca}
\gamma_{\parallel}=\frac{k_{res}}{2\pi^{2}}\frac{1}{\zeta^{3}}\sum_{n=1}^{\infty}(-1)^{n+1}
(\rho\tilde{\alpha})^{n}f^{n}(\zeta)\int_{0}^{\infty}\textrm{d}Q\:Q^{2}[j_{1}(Q)/Q]^{n}.
\end{equation}
Note the similarity between the above equation and that for $2\gamma_{\perp}^{Tot.B}$ in Eq.(\ref{laGamainCoh1}).
This is at the root of the coincidence of the terms of order $\rho\tilde{\alpha}/\zeta$ in the single-scattering  model studied previously. The convergence the series in Eq.(\ref{laqtoca}) can be verified using similar arguments to those employed for Eq.(\ref{laGamainCoh1}). $\gamma_{\parallel}$ contains however additional terms of order $1/\zeta^{3}$ and order zero in $\zeta$ as a result of exponenzing the function $f(\zeta)$ in Eq.(\ref{fzeta}). We give below the first five terms of the orders $\mathcal{O}(\zeta^{-3})$, $\mathcal{O}(\zeta^{-1})$ and $\mathcal{O}(\zeta^{0})$,
\begin{eqnarray}
\gamma_{\parallel}&\simeq&\frac{-k_{res}}{2\pi}\frac{1}{\zeta^{3}}[\rho\tilde{\alpha}+\frac{2}{3}(\rho\tilde{\alpha})^{2}
+\frac{5}{24}(\rho\tilde{\alpha})^{3}+\frac{272}{2835}(\rho\tilde{\alpha})^{4}+\frac{40949}{870912}(\rho\tilde{\alpha})^{5}
+...]\label{laO1}\\
&-&\frac{k_{res}}{2\pi}\frac{1}{\zeta}[\frac{1}{2}\rho\tilde{\alpha}+\frac{2}{3}(\rho\tilde{\alpha})^{2}
+\frac{5}{16}(\rho\tilde{\alpha})^{3}+\frac{544}{2835}(\rho\tilde{\alpha})^{4}+\frac{204745}{1741824}(\rho\tilde{\alpha})^{5}
+...]\label{laO2}\\
&-&i\frac{k_{res}}{2\pi}[\frac{1}{3}\rho\tilde{\alpha}+\frac{4}{9}(\rho\tilde{\alpha})^{2}
+\frac{5}{24}(\rho\tilde{\alpha})^{3}+\frac{1088}{8505}(\rho\tilde{\alpha})^{4}+\frac{204745}{2612736}(\rho\tilde{\alpha})^{5}
+...].\label{laO0}
\end{eqnarray}
For an MG dielectric, it is possible to distinguish between dispersion and absorbtion in the same manner we did for the simple single-scattering model in Section \ref{Sect5B3}. In particular, the terms in Eq.(\ref{laO1}) are the analog of a FRET process for the case that the transfer of energy takes place between the emitter and an MG dielectric. Below, we write the complete formulae for the different contributions to the decay rate (up to renormalization prefactors) in function of the effective dielectric constant,
\begin{eqnarray}
\Gamma_{MG}^{Coh.}&=&2\tilde{\Gamma}_{o}\Re{\Bigl\{\frac{\epsilon_{MG}+2}{3}\Bigr\}}\Re{\{\sqrt{\epsilon_{MG}}\}},\label{GMG1}\\
\Gamma_{MG}^{\parallel,Disper.}&=&
2\tilde{\Gamma}_{o}\Re{\{\frac{1}{3}\chi_{MG}
+\frac{1}{3}\chi^{2}_{MG}-0.051\:\chi^{3}_{MG}+0.055\:\chi^{4}_{MG}
-0.015\:\chi^{5}_{MG}+...\}}\label{laLONG}\\
\Gamma_{MG}^{Absorb.}&=&-2\tilde{\Gamma}_{o}\Im{\Bigl\{\frac{\epsilon_{MG}+2}{3}\Bigr\}}\Im{\{\sqrt{\epsilon_{MG}}\}}\\
&+&\frac{2\tilde{\Gamma}_{o}}{\zeta}\Im{\{\frac{1}{2}\chi_{MG}
+\frac{1}{2}\chi^{2}_{MG}-0.076\:\chi^{3}_{MG}+0.083\:\chi^{4}_{MG}
-0.022\:\chi^{5}_{MG}+...\}}\label{tuntun}\\
&+&\frac{2\tilde{\Gamma}_{o}}{\zeta^{3}}\Im{\{\chi_{MG}
+\frac{1}{3}\chi^{2}_{MG}-\frac{1}{8}\chi^{3}_{MG}+0.073\:\chi^{4}_{MG}
-0.028\:\chi^{5}_{MG}+...\}}
\end{eqnarray}
Note however that when multiple-scattering processes are involved, it might well be that part of the decay attributed to absorbtion is actually due to dispersion by clusters of correlated host dipoles. A microscopical analysis of the origin of $\Im{\{\chi_{MG}\}}$ would be needed to discriminate between both  extinction processes.\\
\indent At this point, it is interesting to compare the order $\mathcal{O}(\zeta^{0})$ term of  $\Gamma^{MG}_{rad.}$ to that usually employed in the literature which includes the square of the Lorentz-Lorenz local field factor,
$\Gamma_{LL}=2\tilde{\Gamma}_{o}\Bigl(\frac{\epsilon_{MG}+2}{3}\Bigr)^{2}\sqrt{\epsilon_{MG}}$. In our computation, $\Gamma^{MG}_{rad.}=\Gamma_{MG}^{Coh.}+\Gamma_{MG}^{\parallel,Disper.}$.
These terms are the relevant ones for emission in absence of absorbtion, $\Im{\{\epsilon_{MG}\}}\ll\Re{\{\epsilon_{MG}\}}$.
Expanding in powers of $\chi_{MG}=\epsilon_{MG}-1$, we obtain up to order five,
\begin{eqnarray}
\Gamma^{MG}_{rad.}&=&2\tilde{\Gamma}_{o}\Re{\Bigl\{1+\frac{7}{6}\chi_{MG}
+\frac{3}{8}\chi^{2}_{MG}-0.030\:\chi^{3}_{MG}+0.037\:\chi^{4}_{MG}
-0.0007\:\chi^{5}_{MG}+...\Bigr\}}\label{Wminever}\\
\Gamma_{LL}&=&2\tilde{\Gamma}_{o}\Re{\Bigl\{1+\frac{7}{6}\chi_{MG}+\frac{23}{72}\chi^{2}_{MG}
+0.035\:\chi^{3}_{MG}-0.01\:\chi^{4}_{MG}
+0.008\:\chi^{5}_{MG}+...\Bigr\}}.\label{WLLver}
\end{eqnarray}
Therefore, in absence of extinction, the agreement is almost perfect up to order $\chi^{2}_{MG}$. As we will argue in the next section, this agreement is just accidental. Therefore, in order to differentiate between both formulae in non-absorptive media, it is necessary that higher order terms be relevant. That is, the index of refraction has to satisfy $\bar{n}\gtrsim\sqrt{2}$. Nevertheless, we must point out  that it has been assumed implicitly in Eqs.(\ref{Wminever},\ref{WLLver}), by using a unique value for $\tilde{\Gamma}_{o}$, that the renormalization factor which accompanies $\mu$ is  the same  in both equations. For the decay of a weakly-polarizable host molecule this is not a problem. However, for the decay of a foreign particle within a host molecule this might not be so. Even for a non-dissipative medium the terms of orders $\mathcal{O}(\zeta^{-3})$ and $\mathcal{O}(\zeta^{-1})$ are relevant (see Eq.(\ref{latercera1})) and they do not appear in the Lorentz-Lorenz approach. Therefore, any try to differentiate between both approaches  must necessarily assume some \emph{ad hoc} common value for the renormalization factor. We give below the $\gamma$-factors which enter the renormalization factor in Eq.(\ref{grenorm}) bearing in mind that such a renormalization factor must be set to 1 for the case of a weakly-polarizable excited host molecule,
\begin{eqnarray}
2\gamma_{\perp}+\gamma_{\parallel}&=&\gamma_{\zeta^{0}}+\gamma_{\zeta^{-1}}
+\gamma_{\zeta^{-3}}\label{g0}\\
&=&-i\frac{k_{res}}{2\pi}[\frac{7}{6}\chi_{MG}
+\frac{3}{8}\chi^{2}_{MG}-0.030\:\chi^{3}_{MG}+0.037\:\chi^{4}_{MG}
-0.0007\:\chi^{5}_{MG}+...]\label{g1}\\
&-&\frac{k_{res}}{2\pi}\frac{1}{\zeta}[\chi_{MG}
+\frac{1}{5}\chi^{2}_{MG}+0.105\:\chi^{3}_{MG}-0.027\:\chi^{4}_{MG}
+0.006\:\chi^{5}_{MG}+...]\label{g2}\\
&-&\frac{k_{res}}{2\pi}\frac{1}{\zeta^{3}}[\chi_{MG}
+\frac{1}{3}\chi^{2}_{MG}-\frac{1}{8}\chi^{3}_{MG}+0.07\:\chi^{4}_{MG}
-0.03\:\chi^{5}_{MG}+...]\label{g3}.
\end{eqnarray}
\indent The above formulae are thought to fit the experimental results analyzed quantitatively in \cite{Duan1}. There, Ce$^{+3}$ ions of low polarizability replace low-polarizability cations of several host media. The life-time of the transition $5d\rightarrow4f$ is measured
in several hosts and the data are fitted in \cite{Duan2} to the usual VC and RC formulae.
The authors find 'relative' agreement with the VC model. However, in their fits they leave the value of $\mu$ to be fitted as well, introducing this way an additional degree of freedom which improves artificially the fits.
Therefore, their results are far from conclusive.
\subsection{The dielectric constant of a Maxwell-Garnett dielectric}\label{Sect6B}
The problem in this case is one of self-consistency. On the one hand, we must compute the renormalized single-particle polarizability of the molecules, $\tilde{\alpha}$, due to their embedding in the dielectric. On the other hand, the renormalization of $\tilde{\alpha}$ gives rise to a renormalization of the dielectric constant itself with respect to the MG formula involving in-free-space polarizabilities. Following the renormalization scheme developed in Section \ref{Sect3C}, beside the value of $\alpha_{0}$ and the in-free-space resonance wave number, $k_{0}$, knowledge of the exclusion volume (i.e. the correlation length $\xi$) and of non-radiative effects (i.e. the collisional  shift $\Delta k^{2}_{coll.}$ and collisional line broadening $\Gamma_{coll.}$) are needed. While the latter effect are thought to be relevant at high temperature, radiative effects are expected to dominate at low temperature and for frequencies close to the resonance.\\
\indent A possible  setup which adjusts to this scenario corresponds to that in the experiment carried out in \cite{PRLMakietal}. There, selective-reflection techniques are employed to measure the frequency shift in a high temperature potassium gas. At the frequencies of interest, the bare electrostatic polarizability reads $\alpha_{0}=f\:4\pi r_{e}k_{0}^{-2}$ where $r_{e}=\frac{e^{2}}{4\pi\epsilon_{0}m_{e}c^{2}}=2.82\:10^{-6}nm$ is the electron radius, the resonance wavelength is $\lambda_{0}=2\pi k_{0}^{-1}=770.1nm$ and the strength factor is $f=0.339$
for the transition $4\:^{2}\textrm{S}_{1/2}\rightleftarrows4\:^{2}\textrm{P}_{1/2}$ \cite{Hilborn}. The range of atomic densities in the experiment is $10^{23}m^{-3}\lesssim\rho\lesssim2\cdot10^{20}m^{-3}$. We can keep the renormalization scheme of Section \ref{Sect3C} by just replacing $\alpha_{0}$ with a renormalized static polarizability which includes both the regularization of $2\Re{\{\gamma^{(0)}_{\perp}\}}$ and collision factors,
\begin{equation}\label{alfapr}
\alpha_{stat.}^{coll.}(\tilde{k})=\alpha_{0}k_{0}^{2}[k_{0}^{2}+\Delta k^{2}_{coll.}-i\:\Gamma_{coll.}c\tilde{k} -\tilde{k}^{2}]^{-1}.
\end{equation}
This leads to the Lorentzian renormalized single-particle polarizability,
\begin{equation}\label{Lory}
\tilde{\alpha}(\tilde{k})=\alpha_{0}k_{0}^{2}
\Bigl[k_{0}^{2}+\Delta k^{2}_{coll.}-i\:\Gamma_{coll.}c\tilde{k} -\tilde{k}^{2}+\frac{1}{3}\alpha_{0}k_{0}^{2}\tilde{k}^{2}(i\Im{\{2\gamma_{\perp}^{(0)}\}}+
2\gamma_{\perp}+\gamma_{\parallel})_{\tilde{k}}\Bigr]^{-1}.
\end{equation}
Next, in application of MG formula, the effective dielectric constant reads
\begin{eqnarray}\label{laeps}
\tilde{\epsilon}_{MG}(\tilde{k})&=&1+\frac{\rho\tilde{\alpha}(\tilde{k})}
{1-\frac{1}{3}\rho\tilde{\alpha}(\tilde{k})}=1+\chi_{MG}(\tilde{k})=1\\
&+&\rho\alpha_{0}k_{0}^{2}
\Bigl[k_{0}^{2}+\Delta k^{2}_{coll.}-i\:\Gamma_{coll.}\tilde{k}/c -\tilde{k}^{2}+\frac{1}{3}\alpha_{0}k_{0}^{2}\tilde{k}^{2}(i\Im{\{2\gamma_{\perp}^{(0)}\}}+
2\gamma_{\perp}+\gamma_{\parallel})_{\tilde{k}}-\frac{1}{3}\alpha_{0}k_{0}^{2}\rho\Bigr]^{-1}.\nonumber
\end{eqnarray}
The $\gamma$-factors are those of Eqs.(\ref{g0}-\ref{g3}). Because they depend on $\chi_{MG}$, the problem becomes one of self-consistency. Self-consistency can be seen also as a consequence of including recurrent scattering in the derivation of the MG formula. In physical grounds, it reflects the complementarity of the double role played by the actual dipoles. That is, on the one hand they polarize the sourceless EM vacuum characterized by $\epsilon_{MG}$. On the other hand, they renormalize their own polarizability, $\tilde{\alpha}$.\\
\indent Once the self-consistency problem has been solved in terms of $\chi_{MG}$, the $\gamma$-factors can be computed and the renormalized values of $\Gamma$, $\omega_{res}$ and $\alpha_{0}$ follow the equations of Section \ref{Sect3C}. Note that the resonance frequency of $\tilde{\epsilon}_{MG}$ differs w.r.t. to that of $\tilde{\alpha}$ in the so-called Lorentz-shift, $\Delta k^{2}_{L}=-\frac{1}{3}\alpha_{0}k^{2}_{0}\rho$.\\
\indent The authors of \cite{PRLMakietal} however did not include all the possible radiative effects in their analysis. They restricted themselves to the Lorentz-shift, $\Delta k^{2}_{L}$. Because $\Delta k^{2}_{L}$ does not depend on $\chi_{MG}$, no self-consistency was demanded. In the experimental setup of \cite{PRLMakietal} no further radiative effects are relevant despite the fact that the medium is highly opaque. The reason being that collisional effects dominate. A detailed analysis of the problem --see \cite{tobepublished}-- reveals that this is the case. The main reason is that those $\zeta$ dependent terms in Eqs.(\ref{g2},\ref{g3}), which would be expected to become large for $|\chi_{MG}|\sim1$, $k_{0}\xi\ll1$, turn out to be irrelevant in comparison to the collisional contributions. For a cold gas of potassium, a  estimate for $\xi$ would be the van-der-Waals radius of $K$ atoms, which is roughly $0.5nm$.  However, at high temperature $\xi$ is rather determined by the collision cross section between atoms, which turns out to be much greater. It would still be possible to increase the relative weight of the radiative terms with respect to the collisional ones by lowering the atomic density. However, for a precise calculation we would still need a good estimation for $\xi$.\\
\indent To the opposite extreme belongs the scenario in which the medium is an atomic glass. In such a case, there is the advantage that $\xi$ can be accurately determined. In fact, $\xi\sim\rho^{-1/3}$. However, as pointed out in Section \ref{Sect2C}, atomic orbitals overlap in such a way that the electronic band configurations can be very different to those of the individual atoms isolated. On top of that, there might be contributions of free electrons, which introduce an additional source of uncertainty in computing that  component of the susceptibility due to plasma oscillations \cite{Potter}.\\
\indent In view of the difficulties that high temperature atomic gases and solids present in evaluating the radiative contributions to the renormalization of the dielectric constant, we propose  alternative setups. They must be such that undesired dynamical and collisional effects be reduced to a minimum. In addition, the electronic structure of the dipole-constituents must be hardly altered by their cohesive interactions. A cold atomic gas offers a scenario which adjusts to these requirements \cite{Cold}. The other possibility is the preparation of colloidal liquids \cite{PRLRojasFranck}. There, particles of well defined polarizability are in suspension on a background solvent. The particles are ionized in such a way that the spatial correlation among them is determined by the Coulombian long-ranged mutual repulsion and crystallization is suppressed. The degree of order and the value of $\xi$ can be tuned to satisfy the conditions of an MG dielectric.
\section{Spontaneous decay of a low-polarizability substantial impurity}\label{Sect7P}
\begin{figure}[h]
\includegraphics[height=5.5cm,width=10.2cm,clip]{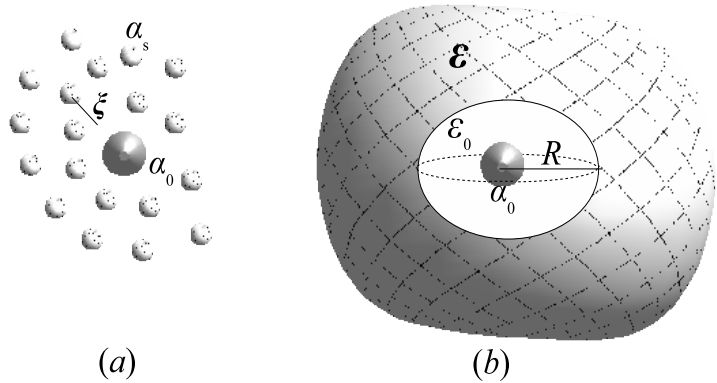}
\caption{($a$) Cermet topology scenario. The emitter with polarizability $\alpha_{0}$ is
 surrounded by a disconnected net of scatterers with correlation length $\xi$. ($b$) Simply-connected-non-contractible topology in which the emitter is placed at the center of a cavity of radius $R$ surrounded by a continuous medium of dielectric constant $\epsilon$.}\label{fig32}
\end{figure}
In this section we compute the decay rate, $\Gamma$, of a fluorescent weakly-polarizable emitter with transition dipole moment $\mu$ and resonance frequency $\omega_{res}=ck_{0}$. The emitter locates at the center of a large cavity of radius $R$ much larger than the correlation length $\xi$ among host scatterers. The polarizability of the emitter is low in comparison to that of the host scatterers. The long-wavelength limit applies on the host medium, $k_{0}\xi\ll1$, such that it behaves locally as an effective medium of uniform dielectric constant $\epsilon=1+\chi$. The setup corresponds to that of Section \ref{Sect4B2}. The topology associated to the embedding of the emitter is that of a simply-connected-non-contractible manifold --see Fig.\ref{fig32}($b$). Its experimental counter part corresponds to that of $Eu^{3+}$ ions in random media \cite{SchuEu}.
Our master formula is
\begin{equation}
\Gamma=\frac{-2\omega_{res}^{2}\epsilon_{0}}{3c^{2}\hbar}|\mu|^{2}
\Im{\{2\gamma^{RC}_{\perp}+\gamma^{RC}_{\parallel}\}},\label{latercera12}
\end{equation}
where the $\gamma$-factors are those of the RC scenario given by Eqs.(\ref{seriesa}-\ref{gammapara2}). Following those equations, we will compute the $\gamma$-factors as power series of $\chi$. Our computation is exact up to order two in $\chi$. In order to compare with the conventional empty-cavity model formulation, we will restrict ourselves to the case $k_{0}R\ll1$, although exact formulae at all the orders in $k_{0}R$ can be obtained beyond this approximation.\\
\indent The $\gamma$-factors are given by
\begin{eqnarray}
2\gamma^{RC}_{\perp}(\tilde{k})&=&-i\frac{\tilde{k}}{2\pi}-2\tilde{k}^{2}\int\frac{\textrm{d}^{3}k}{(2\pi)^3}
[C_{\perp}+G_{\perp}^{(0)}]\chi_{\perp}G_{\perp}^{(0)}(k)\nonumber\\&+&
2\tilde{k}^{4}\int\frac{\textrm{d}^{3}k}{(2\pi)^3}[C_{\perp}+G^{(0)}_{\perp}]^{2}
G_{\perp}\chi_{\perp}^{2}(k),\label{OB2a}\\
\gamma^{RC}_{\parallel}(\tilde{k})&=&-\tilde{k}^{2}\int\frac{\textrm{d}^{3}k}{(2\pi)^3}
[C_{\parallel}+G_{\parallel}^{(0)}]\chi_{\parallel}G_{\parallel}^{(0)}(k)\nonumber\\&+&
\tilde{k}^{4}\int\frac{\textrm{d}^{3}k}{(2\pi)^3}[C_{\parallel}+G^{(0)}_{\parallel}]^{2}
G_{\parallel}\chi_{\parallel}^{2}(k),\label{OB2b}
\end{eqnarray}
where the cavity factors are those of Eqs.(\ref{Xioperp},\ref{Xioparall}).
\begin{figure}[h]
\includegraphics[height=12.2cm,width=15.4cm,clip]{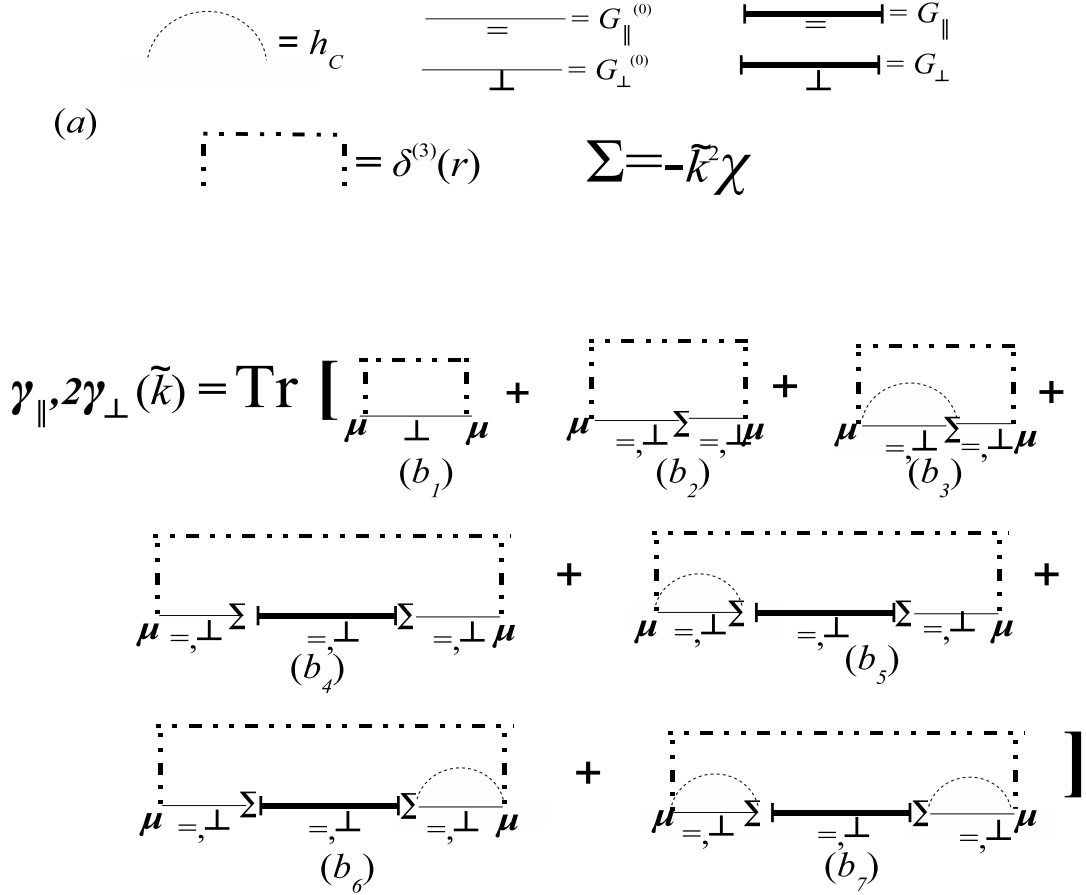}
\caption{($a$) Feynman's rules . ($b$) Diagrammatic representation
of the transverse and longitudinal $\gamma$-factors,
$2\gamma^{RC}_{\perp}$ and $\gamma^{RC}_{\parallel}$, as given by
Eqs.(\ref{OB2a},\ref{OB2b}).}\label{fig31}
\end{figure}
Considering that the polarizability of the emitter is negligible and that the interaction of the emitter with the surrounding particles does not modify its energy levels, neither changes on $\mu$ nor in the resonance frequency hold. That way, we can perform the substitution of $\tilde{k}$ with $k_{0}$ in the above formulae. Up to zero order in $k_{0}R$ we obtain,
\begin{eqnarray}\label{gamis}
2\gamma^{RC}_{\perp}+\gamma^{RC}_{\parallel}\simeq&-&\frac{i}{2\pi}k_{0}\Bigl[\Bigl(\frac{\epsilon+2}{3}\Bigr)
^{2}\sqrt{\epsilon}-\frac{4}{9}\frac{(\epsilon-1)^{2}}{\epsilon}\Bigr]\nonumber\\
&-&\frac{k_{0}}{2\pi}\Bigl[\Bigl(\frac{1}{(k_{0}R)^{3}}+\frac{1}{k_{0}R}\Bigr)
[(\epsilon-1)\frac{\epsilon+2}{3\epsilon}]\Bigr].
\end{eqnarray}
Normalizing by the in-free-space decay rate, $\Gamma_{0}=\frac{\omega_{res}^{3}}{3\pi\epsilon_{0} c^{3}\hbar}|\mu|^{2}$, $\Gamma$ reads
\begin{eqnarray}
\Gamma&=&\Gamma_{0}\Bigl[\Re{\{\Bigl(\frac{\epsilon+2}{3}\Bigr)^{2}\sqrt{\epsilon}\}}-\frac{4}{9}\Re{\{\frac{(\epsilon-1)^{2}}{\epsilon}\}}
\label{firstyp}\\
&-&\Bigl(\frac{1}{(k_{0}R)^{3}}+\frac{1}{k_{0}R}\Bigr)\Im{\{\frac{2(\epsilon-1)^{2}}{3\epsilon}-(\epsilon-1)\}}\Bigr].\label{thirdyp}
\end{eqnarray}
The terms in Eq.(\ref{thirdyp}) are
associated to absorbtion in the host medium \cite{MeII}.
We recognize in the first term of the \emph{r.h.s} of Eq.(\ref{firstyp}) the usual
bulk term corrected by Lorentz-Lorenz (LL) local field factors,
$\Gamma^{LL}=\Gamma_{0}\Re{\{\Bigl(\frac{\epsilon+2}{3}\Bigr)^{2}\sqrt{\epsilon}\}}$.
The second term there includes corrections of order
$\gtrsim\chi^{2}$. It is also remarkable that if the empty-cavity Onsager--B\"{o}ttcher (OB) local field factors are used instead, $\Gamma^{emp}_{OB}=\Gamma_{0}\Re{\{\Bigl(\frac{3\epsilon}{2\epsilon+1}\Bigr)^{2}\sqrt{\epsilon}\}}$
does agree with the two radiative terms of Eq.(\ref{firstyp}) up to order $\chi^{2}$,
\begin{equation}
\Gamma^{emp}_{OB}\simeq\Gamma_{0}(1+\frac{7}{6}\chi-\frac{1}{8}\chi^{2}+...).
\end{equation}
Our computation is in fact exact up to $\mathcal{O}(2)$. However, the nature of the emission associated to the terms of Eq.(\ref{firstyp}) is different to the one attributed in the conventional empty-cavity model. In the latter, $\Gamma^{emp}_{OB}$ is considered as fully transverse. We show next that transverse-coherent emission is just part of it, being the rest incoherent longitudinal emission associated to dispersion. We
can read the contribution of coherent emission from the diagrams in Fig.\ref{fig31},
\begin{eqnarray}
\Gamma^{Coh.}=\Gamma_{0}\frac{-2\pi}{k_{0}}\Im{\{2\gamma_{\perp}^{P}\}}=
\Gamma_{0}\frac{-2\pi}{k_{0}}\Bigl[2\int\Im{\{G_{\perp}(k)\}}\frac{\textrm{d}^{3}k}{(2\pi)^3}\nonumber\\
+2\tilde{k}^{2}\Re{\{\chi
C_{\perp}(k_{0})\}}\int\Im{\{G^{(0)}_{\perp}(k)\}}\frac{\textrm{d}^{3}k}{(2\pi)^3}\nonumber\\
+2k_{0}^{4}\Re{\{[\chi
C_{\perp}(\sqrt{\epsilon}k_{0})]^{2}\}}\int\Im{\{G_{\perp}(k)\}}\frac{\textrm{d}^{3}k}{(2\pi)^3}\nonumber\\
-4k_{0}^{4}\Re{\{\chi
C_{\perp}(\sqrt{\epsilon}k_{0})\}}\int\Im{\{G_{\perp}(k)\}}\frac{\textrm{d}^{3}k}{(2\pi)^3}\Bigr],\label{firstp}
\end{eqnarray}
where the first term on the \emph{r.h.s} of Eq.(\ref{firstp}) corresponds to direct-coherent emission --transverse components of diagrams ($b_{1}$),
($b_{2}$), ($b_{4}$) in Fig.\ref{fig31}. In the above expressions, $C_{\perp}(\sqrt{\epsilon}k_{0})$ stands for  $C_{\perp}(k)|_{k=\sqrt{\epsilon}k_{0}}$. $\Gamma^{Coh.}$ can be written in a more compact form as
\begin{eqnarray}
\Gamma^{Coh.}&=&\Gamma_{0}\frac{-2\pi}{k_{0}}\Bigl[2\Re{\{[1-k_{0}^{2}\chi C_{\perp}(\sqrt{\epsilon}k_{0})]^{2}\}}\int\Im{\{G_{\perp}(k)\}}\frac{\textrm{d}^{3}k}{(2\pi)^3}\\
&+&2\Re{\{k_{0}^{2}\chi C_{\perp}(\sqrt{\epsilon}k_{0})\}}\int\Im{\{G^{(0)}_{\perp}(k)\}}\frac{\textrm{d}^{3}k}{(2\pi)^3}\Bigr],
\end{eqnarray}
which immediately leads to
\begin{equation}
\Gamma_{RC}^{Coh.}=\Gamma_{0}\Bigl[\Re{\{\sqrt{\epsilon}\}}\Re{\Bigl\{\Bigl(\frac{\epsilon+2}{3}\Bigr)^{2}\Bigr\}}
-\frac{1}{3}\Re{\{\epsilon-1\}}\Bigr],
\end{equation}
at leading order in $k_{0}R$. For the dispersive and absorptive terms we get,
\begin{eqnarray}
\Gamma_{RC}^{\parallel,Disper.}&=&\Gamma_{0}\Re{\Bigl\{\frac{\epsilon-(\epsilon-2)^{2}}{9\epsilon}\Bigr\}},\\
\Gamma_{RC}^{Absorb.}&=&-\Gamma_{0}\Bigl[\Im{\{\sqrt{\epsilon}\}}\Im{\Bigl\{\Bigl(\frac{\epsilon+2}{3}\Bigr)^{2}\Bigr\}}\\
&+&\Bigl(\frac{1}{(k_{0}R)^{3}}+\frac{1}{k_{0}R}\Bigr)\Im{\{\frac{2(\epsilon-1)^{2}}{3\epsilon}-(\epsilon-1)\}}\Bigr]
\end{eqnarray}
Once more, the same reserves after Eq.(\ref{tuntun}) about the distinction between absorbtion and dispersion apply.
\section{Comparison with previous approaches}\label{Sect7}
We have proved that our computation of LDOS$^{emission}_{\omega}$ of Eq.(\ref{gammaLDOS}) with the $\gamma$-factors given by Eqs.(\ref{LDOSIperg},\ref{LDOSIparalg}) is exact for a microscopically linear and statistically homogeneous dielectric in which the electrical susceptibility tensor can be expanded as in  Eq.(\ref{laXenAes}). Here we review on previous works, compare them with ours and emphasize where the errors in those approaches are. Because the literature about complex media concentrates in setups where an effective medium can be defined, we will restrict ourselves to such a particular scenario. Our analytical study of an MG dielectric of Section \ref{Sect6} belongs to that kind.
\subsection{The erroneous inclusion of local field factors}
The majority of approaches on non-dissipative media agree in giving an effective density of states of the form
\begin{equation}
LDOS^{emission}_{\omega}=\frac{\epsilon_{0}c}{2\hbar\omega}\langle
\vec{E}^{\omega}_{Loc.}(\vec{r})\cdot\vec{E}^{\omega\dagger}_{Loc.}(\vec{r})\rangle
\approx|_{k\ll\xi^{-1}}\mathcal{L}^{2}_{f}\:\bar{n}\:LDOS^{0}_{\omega},\label{lagual}
\end{equation}
where $\mathcal{L}_{f}$ is a local field factor. The approximation symbol denotes the restriction to long-wavelength modes with $k\ll\xi^{-1}$ which is implicit in the use of the effective medium theory. $\langle\vec{E}^{\omega}_{Loc.}(\vec{r})\rangle$ is interpreted as the electric field felt by the emitter located at $\vec{r}$ and $\langle
\vec{E}^{\omega}_{Loc.}(\vec{r})\vec{E}^{\omega\dagger}_{Loc.}(\vec{r})\rangle$ is the spectrum of its vacuum fluctuations. As pointed out in Section \ref{Sect5B2}, because modes with $k>\xi^{-1}$ are not included in Eq.(\ref{lagual}), such a field is actually not microscopic but assumed uniform within the cavity of radius $\xi$. Hereafter we use the nomenclature of Section \ref{Sect5B2} where $\vec{E}^{eff}_{loc.}(\vec{r})$ stands for $\vec{E}^{\omega}_{Loc.}(\vec{r})$ in the effective medium theory. Likewise, $\vec{E}^{eff}_{D}(\vec{r})$ denotes the macroscopic field in the bulk. The usual procedure in the literature is to compute $\vec{E}^{eff}_{loc.}(\vec{r})$ classically by imposing boundary conditions at the cavity surface. Those boundary conditions depend on whether the medium inside the cavity is equivalent or not to the medium in the bulk. In both cases, effective media are assumed to exist on both sides and the local and macroscopic fields are related via the induced polarization field \cite{Jackson} --see also \cite{BulloughHynne2} for a microscopic rigorous approach. In either case one obtains $\vec{E}^{eff}_{loc.}=\mathcal{L}_{f}\vec{E}^{eff}_{D}$, in the small cavity limit. In the case the emitter is a dipole equivalent to all the rest, the cavity is virtual and  $\mathcal{L}_{f}=\mathcal{L}_{LL}=\frac{\epsilon_{MG}+2}{3}$ is the Lorentz-Lorenz local field factor \cite{Lorentz}. For the case that the cavity contains an emitter of much weaker polarizability than that of the host scatterers, the cavity is assumed to be empty and
$\mathcal{L}_{f}=\mathcal{L}_{emp}=\frac{3\epsilon}{2\epsilon+1}$ \cite{Onsager}. Same approach is followed in more complicated dielectric configurations \cite{Lavallard}.\\
\indent It is obvious that the above procedure cannot be the correct one as it ignores the modes $k\gtrsim\xi^{-1}$ and hence the $\zeta$-dependent terms in LDOS$^{emission}_{\omega}$. Nevertheless, let us assume for the moment that the above procedure is legitimum. The first erroneous assumption which appears implicitly in Eq.(\ref{lagual}) is that \emph{...the local field factor appears squared, as $\Gamma$} (or LDOS) \emph{can be expressed in terms of and expectation value of the product of the two electric field operators} ($\vec{E}_{Loc.}$) \cite{Schuurmans1} -- also in \cite{Schuurmans2,LaudonPRL,Toptygin,Crenshaw}. The point being that the fact that the classical values of $\vec{E}^{eff}_{loc}$ and $\vec{E}^{eff}_{D}$ are related by a constant of proportion $\mathcal{L}_{f}$, does not imply by any means that there is a constant of proportion $\mathcal{L}^{2}_{f}$ between their quadratic fluctuations in vacuum,
\begin{equation}\label{falla}
\vec{E}^{eff}_{loc}(\vec{r})=\mathcal{L}_{f}\vec{E}^{eff}_{D}(\vec{r})\nRightarrow
\langle\vec{E}^{eff}_{loc}(\vec{r})\cdot\vec{E}^{eff\dagger}_{loc}(\vec{r})\rangle=
\mathcal{L}^{2}_{f}\:\langle\vec{E}^{eff}_{D}(\vec{r})\cdot\vec{E}^{eff\dagger}_{D}(\vec{r})\rangle.
\end{equation}
The proof of Eq.(\ref{falla}) was given in Section \ref{Sect5B2}. We found there that the propagator of $\vec{E}^{eff}_{loc}$ reads
\begin{equation}\label{lageff}
\mathcal{G}^{eff}_{\parallel\perp}(k)=\textrm{Lim.}\{\mathcal{G_{\parallel,\perp}}\}|_{k\xi\rightarrow0}=
\frac{\chi^{eff}_{\parallel\perp}}{\rho\tilde{\alpha}}G^{eff}_{\parallel,\perp}(k),
\end{equation}
with $\frac{\chi^{eff}_{\perp}}{\rho\tilde{\alpha}}=\mathcal{L}_{LL}=\frac{\epsilon_{MG}+2}{3}$, while $G^{eff}_{\perp}$ defined in Eq.(\ref{effective}) is the propagator of $\vec{E}^{eff}_{D}$. Neglecting extinction, $\Im{\{\epsilon_{MG}\}}\ll\Re{\{\epsilon_{MG}\}}$, and in application of the fluctuation-dissipation theorem,
\begin{eqnarray}
\langle\vec{E}^{eff}_{loc}(\vec{r})\cdot\vec{E}^{eff\dagger}_{loc}(\vec{r})\rangle&=&
-\frac{\hbar\omega^{2}}{\epsilon_{0}\pi c^{2}}\:\mathcal{L}_{LL}\:\int\frac{\textrm{d}^{3}k}{(2\pi)^{3}} \Im{\{G^{eff}_{\perp}(k)\}}\nonumber\\&=&\mathcal{L}_{LL}\:
\langle\vec{E}^{eff}_{D}(\vec{r})\cdot\vec{E}^{eff\dagger}_{D}(\vec{r})\rangle\propto
\mathcal{L}_{LL}\:\bar{n},
\end{eqnarray}
which is proportional to the MG coherent emission computed in Eq.(\ref{GammaRADMG}) instead. An alternative manner to understand where the confusion resides consists of interpreting the macroscopic vector fields as operators acting on the coherent vacuum defined in Section \ref{Sect5B1}. In that framework, two factors $Z^{1/2}=\mathcal{L}^{1/2}_{LL}$ appear in the two-field vacuum expectation value in passing from
$\langle\vec{E}^{eff}_{D}(\vec{r})\cdot\vec{E}^{eff\dagger}_{D}(\vec{r})\rangle$ to $\langle\vec{E}^{eff}_{loc}(\vec{r})\cdot\vec{E}^{eff\dagger}_{loc}(\vec{r})\rangle$.
That is,
\begin{eqnarray}
\langle\vec{E}^{eff}_{loc}(\vec{r})\cdot\vec{E}^{eff\dagger}_{loc}(\vec{r})\rangle&=&
^{Coh}\langle\Omega|\hat{\vec{E}}^{eff}(\vec{r})\cdot\hat{\vec{E}}^{eff\dagger}(\vec{r})|\Omega\rangle^{Coh}=
\frac{\chi_{eff}}{\rho\tilde{\alpha}}\:^{s.l.}\langle|\Omega|\hat{\vec{E}}^{eff}(\vec{r})\cdot\hat{\vec{E}}^{eff\dagger}(\vec{r})
|\Omega\rangle^{s.l.}\nonumber\\&=&
\mathcal{L}_{LL}\:\langle\hat{\vec{E}}^{eff}_{D}(\vec{r})\cdot\hat{\vec{E}}^{eff\dagger}_{D}(\vec{r})\rangle.
\end{eqnarray}
The non-necessary implication expressed in Eq.(\ref{falla}) was first appreciated by the authors of \cite{KnollBarnett}. The key argument given there being the lack of noise polarization. The authors of that work introduced an operator to account for such an effect and then required consistency. In our approach, we compute exactly the Green's function of the self-polarization field. Application of the fluctuation-dissipation theorem takes care of the polarization noise in an exact manner.\\
\indent In some old works it is claimed that it is only the bulk density of states, $\sim\bar{n}$, the one responsible for the spontaneous decay which enters Fermi's Golden rule. That is based on the macroscopic quantization carried out for the first time in \cite{Nonhenhaimen}. Such interpretation has been modified since then as further studies on the role and origin of the local field factors have been performed. However, still recent works \cite{RMPdeVries,RemiMole,Tip} appeal to the erroneous interpretation that the density of states accessible to the photons emitted according to Fermi's Golden rule is just given by the bulk term. The latter corresponds to density of bulk normal modes instead which we have denoted by LDOS$_{\omega}^{sourceless}$. The relation between the coherent (and only coherent!) emission spectrum and LDOS$_{\omega}^{sourceless}$ was made clear in Section \ref{Sect6B2}.
\subsection{The actual nature of radiation and the need to go beyond the effective medium approximation}\label{Sect9B}
It is a common error in the literature to identify the total radiative emission with transverse emission. As a matter of fact, the expression in Eq.(\ref{lagual}) is intended as proportional to transverse spontaneous emission, eg. \cite{LaudonJPB,KnollBarnett,Juzel}. This error is related to the use of macroscopic fields and so to the neglect of short-wavelengths.\\
\indent In the first place there is a problem in imposing continuity conditions at the cavity surface. This is particularly problematic in the VC scenario. On the one hand, one assumes that an effective medium exists on both sides of the cavity, which implies some length scale over which the susceptibility is averaged in space. On the other hand, by imposing some 'matching' conditions at the cavity surface one is assuming that the width of the surface is negligible in comparison to the length scale over which the spatial average has been performed on each side. However, that width cannot be less than the typical correlation length between the scatterers, $\xi$. That is precisely the radius of the inner cavity in which the spatial average is performed. Therefore, matching conditions are expected to fail in general but for those coherent modes which propagate in the 'effective' medium. Those are, the ones in $W_{\perp}^{Coh.}$ and $\Gamma_{\perp}^{Coh.}$ on Eqs.(\ref{laGamaACoh},\ref{GammaRADMG}). Because the rest of the transverse emission contains $\zeta$-dependent terms, it is clear that Eq.(\ref{lagual}) cannot be fully transverse. However, we have seen in Section \ref{Sect6A} that Eq.(\ref{lagual}) yields almost the right result for the $\zeta$-independent total decay rate up to order two in $\chi$ --Eqs.(\ref{Wminever},\ref{WLLver}). This accidental coincidence is due the addition of the longitudinal dispersive terms of Eq.(\ref{laLONG}). We showed in the single-scattering model that, at leading order, the $\zeta$-independent longitudinal term equals the contribution of a local field factor. The resultant radiation was proved to be incoherent --see Section \ref{Sect5}. Nevertheless, it is still possible to distinguish between ours and the usual formula without going to higher orders in $\chi$. If instead of using an integration sphere to collect the total radiation in far field the coherent field were measured, the difference could be verified either in experiments or in numerical simulations.\\
\indent One possible way to go around the problem with the boundary conditions would be to impose them
not over the classical fields but over the Green's functions. In the VC model this is
not a solution as one would face again the problem with the length scales. In the RC model things
can be more promising provided $R\gg\xi$. However, one has still to assume some form for the Green's functions
on both sides as done in \cite{Welsh}. Unfortunately, in doing so all the intermediate correlations between
the emitter and the medium are neglected. Only the first scattering after the emission of a virtual photon
and the last scattering before its absorbtion get correlated to the emitter this way.\\
\indent It is also an error to attribute to the whole absorptive $\zeta$-dependent
decay rate a longitudinal nature. This was already noticed in \cite{Welsh}. In the overlap approximation, some
of the absorptive terms of Eq.(\ref{tuntun}) correspond to the transverse $\gamma$-factor of Eq.(\ref{laGamaB}).
\subsection{Why the approach of \cite{PRLdeVries} is not microscopic enough}
\begin{figure}[h]
\includegraphics[height=7.8cm,width=9.8cm,clip]{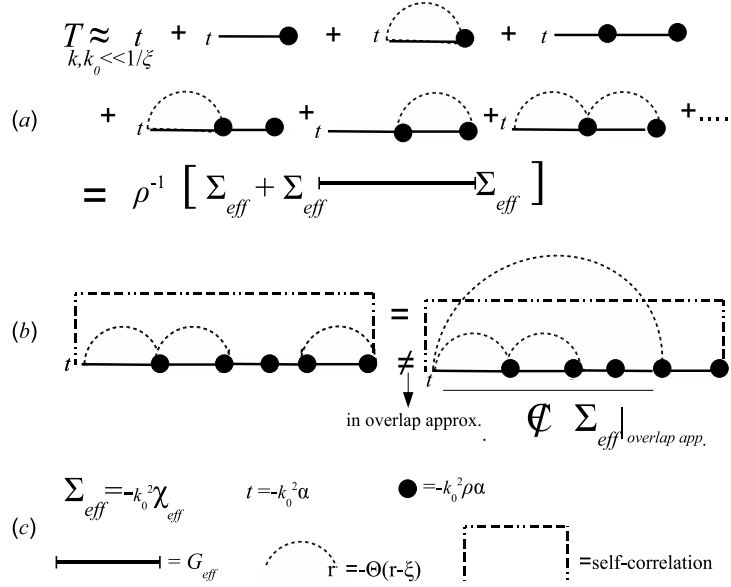}
\caption{($a$) Diagrammatic representation of the $T$-matrix as computed in \cite{PRLdeVries}, Eq.[(6)], in the long-wavelength limit of Eq.[(7)], $k\xi\ll1$. That limit is effectively equivalent to taking the overlap approximation. ($b$) Diagrammatic  representation of two 'apparently' different diagrams in $\bar{\mathcal{G}}(\vec{r},\vec{r})$. They  come to be equivalent when self-correlations are implemented. In the overlap approximation of ($a$), because the 1PI piece underlined  on the \emph{r.h.s.} is not included in the effective self-energy function under the overlap approximation, such an equivalence disappears in the approach of \cite{PRLdeVries}. ($c$) Feynman rules.}\label{fig17}
\end{figure}

\indent We comment on the approach of \cite{PRLdeVries} specifically because it is referred by some authors as a microscopical proof of the usual formulae of the RC and VC models based on Eq.(\ref{lagual}). In the following and for the sake of brevity we will not write all the equations in \cite{PRLdeVries} but only those essential for our arguments. To avoid confusion, we will quote them within double brackets.\\
\indent The same as ours, the approach there bases on the computation of the  Green function of a system which consists of an impurity placed within a host medium. They calculate the renormalized polarizability of the impurity following a similar procedure to that we used in Section \ref{Sect3C}. The host medium considered in \cite{PRLdeVries} is a cubic lattice with lattice spacing $\xi\approx\rho^{-1/3}$. The medium is therefore strongly correlated. However, because for the frequencies of interest $k_{0}\xi\ll1$, an effective medium can be defined with $\chi_{eff}\simeq\chi_{MG}$ for the reasons argued in Section \ref{Sect4B1}. This is used by the authors to extrapolate their results to random media. The propagator of the local field is denoted by $\mathcal{G}_{m}$ in Eq.[(13)].  The authors first compute a transference matrix  and a propagator denoted by $\mathcal{G}$, in the long-wavelength limit $k\xi\ll1$. The resultant propagator $\mathcal{G}(\vec{R},\vec{R}')$ of Eq.[(11)] links any two distant lattice sites and contains only long-wavelength modes.  Afterwards, the authors proceed to compute the $t$-matrices  of impurities which are embedded in the host medium. To do so, they calculate $\mathcal{G}_{m}(\vec{r}_{m},\vec{r}_{m})$, where $\vec{r}_{m}$ denotes the position vector of the impurity. In that calculation, the propagator of Eq.[(11)] is attached to the impurities in two different manners, depending on whether the impurity occupies a lattice site --in which case it is said substantial, with subscript $s$-- or not --in which case it is said interstitial, with subscript $i$. That is done in Eqs.[(16,17)] through the computation of the $\mathcal{T}_{m}$-matrices.\\
\indent In the first place, the approach is not  microscopical enough as the propagator in Eq.[(11)] does not contain the necessary resolution to probe distances shorter than $\xi$, and the distance from either an interstitial or substantial impurity to the first neighbors in the lattice is $\leq\xi$. Therefore, only transverse long-wavelength modes can be accurately accounted for.\\
\indent Nevertheless, for the case of an interstitial impurity their approach may be a good approximation under some conditions, as the impurity is connected through free space propagators to the $T$-matrix of the lattice and in the resultant function $\mathcal{G}_{i}(\vec{r}_{i},\vec{r}_{i})$, $\vec{r}_{i}$ is not a lattice vector. However, the correlation of the impurity to the lattice that way reduces to the correlation to the nearest lattice scatterers (the ones at the extremes of the $T$-matrix). An exact treatment would require the implementation of additional correlations as we did in Section \ref{Sect4B2}.\\
\indent The situation is worse for the case of a substantial impurity. This can be seen  by comparison with the exact result we obtained in the virtual cavity scenario. That is, for the case that the polarizability of the substantial "impurity" is equivalent to that of the rest of particles of the lattice, the resultant setup would correspond to the strict virtual cavity scenario. Therefore, in the long-wave length limit, one would expect to obtain a result proportional to Eq.(\ref{GammaRADMG}) for the renormalized long-wavelength decay rate --which is identified in \cite{PRLdeVries} with the radiative decay rate after Eq.[(15)].  Instead, they obtain an additional local field factor which supports the erroneous introduction of one local field factor per electric field operator. The reason for the mistake is that the authors make an erroneous usage of the $T$-matrix and the propagator $\mathcal{G}(\vec{R},\vec{R}')$ of Eq.[(11)]. The $T$-matrix in Eq.[(6)] just includes terms which link any two dipoles in the lattice. However, if one is to compute  $\mathcal{G}(\vec{R},\vec{R})$ out of it, self-correlations must be explicitly considered. That is, negative correlations are correctly implemented in Eq.[(6)] by excluding  the vector $\vec{R}=\vec{0}$
in all the summations, which leads to the equivalence with the overlap approximation in a random medium.
However, when distant scatterers are also correlated --eg. through self-correlations-- such an additional  correlation must be
incorporated as conditions over the lattice vectors of the sums. The translation invariance assumed in the sums gets 'virtually' broken when introducing self-correlation. That is the information which is missing in the approach of \cite{PRLdeVries}. As a consequence, the evaluation of $\mathcal{G}(\vec{R},\vec{R}')$ at $\vec{R}'=\vec{R}$ does not incorporate self-correlation in the right manner and its use in Eqs.[(14,16)] is incorrect.\\
\indent In the following, we analyze the formulae of \cite{PRLdeVries}, compare them with ours and spot the errors. We have illustrated the failure in Fig.\ref{fig17}.
 Eq.[(6)] with the approximation of Eq.[(7)] is equivalent to
\begin{equation}
\bar{T}(k\xi\rightarrow0)=-k_{0}^{2}\alpha\Bigl[1+k_{0}^{2}\rho\alpha
[\bar{G}^{(0)}(k)+\bar{C}(k\xi\rightarrow0)]\Bigr]^{-1},
\end{equation}
where $\bar{C}(k\xi\rightarrow0)$ is the cavity tensor with components given in Eqs.(\ref{Xioperp},\ref{Xioparall}). In the long-wavelength limit, $\bar{C}(k\xi\rightarrow0)=-\frac{\rho^{-1}}{3k_{0}}\:\mathbb{I}$.  $\alpha=\alpha_{0}[1+i\frac{k_{0}^{3}}{6\pi}\alpha_{0}]$ is the in-free-space radiatively corrected single particle polarizability --the authors of \cite{PRLdeVries} denote it by $\tilde{\alpha}$ instead. Only the two-point correlation function, $h_{C}(r)=-\Theta(r-\xi)$ with $\xi\approx\rho^{-1/3}$, enters the cavity factors. That implies that, effectively, the same negative correlation function which enters the computation of $\chi_{MG}$ in a random medium does so for the computation of the $T$-matrix in a cubic lattice in the limit  $k\xi\ll1$.  Expanding next the denominator of $\bar{T}$ as a perturbative series in $k_{0}^{2}\rho\alpha
[\bar{G}^{(0)}(k)+\bar{C}(k\xi\rightarrow0)]$, it is easy to verify that the overlap approximation is implicit in this expansion such that the susceptibility there is that of MG in the limit $k\xi\rightarrow0$. Thus,  $\bar{T}_{eff}(k)\equiv\bar{T}(k)|_{k\xi\ll1}$ reads,
\begin{equation}\label{Teffo}
\bar{T}_{eff}(k)=\rho^{-1}[\chi_{MG}+\chi_{MG}\bar{G}^{eff}(k)\chi_{MG}],
\end{equation}
where $\bar{G}_{eff}(k)$ is given in Eq.(\ref{effective}) and is denoted by $\mathcal{G}_{diel}$ in \cite{PRLdeVries}. Hence, for any two distant lattice points $\vec{R}\neq\vec{R}'$, Eq.[(11)] can be otained in the long-wavelength limit by removing the bare vertices $\alpha_{0}$, $\rho\alpha_{0}$ attached to the extremes of Eq.(\ref{Teffo}),
\begin{equation}
\bar{\mathcal{G}}^{eff}_{\vec{R}\neq\vec{R}'}(\vec{R},\vec{R}')=(\alpha/\alpha_{0})^{2}\mathcal{L}^{2}_{LL}\mathcal{G}_{diel}(\vec{R},\vec{R}').
\end{equation}
If one is to compute $\bar{\mathcal{G}^{eff}}(\vec{R},\vec{R})$ using the general expression Eq.[(4)],
the $\bar{\mathcal{T}}$-matrix there is not that in  Eq.[(6)], which corresponds to the one which enters the Dyson equation Eq.(\ref{Dysonalter}) for a cubic lattice. The point being that that in Eq.[(6)] was computed without considering the possibility that the first and the last dipoles in the diagrams were the same. As a matter of fact,
application of the overlap approximation from the very beginning prevents from correlating two distant scattering events --see Fig.\ref{fig17}-- and hence the diagrammatic 'trick' we used in Section \ref{Sect4A} to deduce the exact formulae turns inapplicable. Hence, if one aims to write the equation for $\bar{\mathcal{G}}(\vec{R},\vec{R})$ in the form of Eq.[(4)], one should use a modified $\bar{\mathcal{T}}$-matrix whose form derives from our Eq.(\ref{infunctionofT}),
\begin{equation}\label{laTmod}
\mathcal{T}_{\perp,\parallel}(k)=-[G^{(0)}_{\perp,\parallel}]^{-1}[1+\frac{1}{k_{0}^{2}\rho\tilde{\alpha}}T_{\perp,\parallel}].
\end{equation}
\indent Because a lattice is a strong correlated system, the exact formulation of the problem in terms of the $\bar{t}$-matrix is more convenient as the correlation length can be extended eventually further than $\xi$. The formula to compute  the propagator $\bar{\mathcal{G}}(\vec{R},\vec{R})$  is that in Eq.(\ref{Gdiscr}),
\begin{equation}\label{discr1}
\bar{\mathcal{G}}(\vec{R},\vec{R})=\frac{-1}{k_{0}^{2}\tilde{\alpha}}\sum_{\{\vec{r}_{i}\}}\bar{G}^{(0)}(\vec{r}_{i})
\cdot\bar{t}(\vec{r}_{i}),
\end{equation}
Note also that, for the sake of consistency, the fully renormalized $\tilde{\alpha}$ polarizabilities enter this formula, and neither the in-free-space radiatively corrected one, $\alpha$, nor the static one, $\alpha_{0}$.
The components of $\bar{T}_{\vec{k}}$ of Eq.[(6)] in \cite{PRLdeVries} relate to $\bar{t}(\vec{r}_{i})$ in our nomenclature through
$\bar{T}_{\vec{k}}=\sum_{\{\vec{r}_{i}\}}\bar{t}(\vec{r}_{i})e^{i\vec{k}\cdot\vec{r}_{i}}$,
\begin{equation}
\bar{T}_{\vec{k}}=-k_{0}^{2}\tilde{\alpha}\Bigl[\mathbb{I}+k_{0}^{2}\tilde{\alpha}\sum_{\{\vec{r}_{i}\}}
\bar{G}^{(0)}(\vec{r}_{i})e^{i\vec{k}\cdot\vec{r}_{i}}\Bigr]^{-1},\quad \{\vec{r}_{i}\}=\xi(a\hat{i}+b\hat{j}+c\hat{k}),\quad a,b,c\in\mathbb{Z},\quad\vec{r}_{i}\neq\vec{0}.
\end{equation}
Writing the $\bar{t}$-matrix in the alternative form, $\bar{T}(\vec{r})=\sum_{\{\vec{r}_{i}\}}\bar{t}(\vec{r}_{i})\delta^{(3)}(\vec{r}-\vec{r}_{i})$, Eq.(\ref{discr1}) reads
\begin{equation}\label{discr2}
\mathcal{G}(\vec{R},\vec{R})=\frac{-1}{k_{0}^{2}\tilde{\alpha}}\int\textrm{d}^{3}r\:\bar{G}^{(0)}(\vec{r})\cdot
\bar{T}(\vec{r})=\frac{-1}{k_{0}^{2}\tilde{\alpha}}\int\frac{\textrm{d}^{3}k}{(2\pi)^{3}}\:\bar{G}^{(0)}(k)\cdot
\bar{T}_{\vec{k}}.
\end{equation}
For the long-wavelength modes,
the same result as in an MG dielectric holds at leading order in $k_{0}\xi$,
\begin{equation}
\bar{\mathcal{G}}(\vec{R},\vec{R})|_{k\xi\ll1}
=\frac{-ik_{0}}{6\pi}\mathbb{I}\:\frac{\epsilon_{MG}+2}{3}\sqrt{\epsilon_{MG}}.
\end{equation}
For higher frequency modes, it is unavoidable to compute the first terms of $\bar{T}_{\vec{k}}$.
\section{The zero-temperature electromagnetic pressure}\label{Sect9}
\begin{figure}[h]
\includegraphics[height=10.8cm,width=11.8cm,clip]{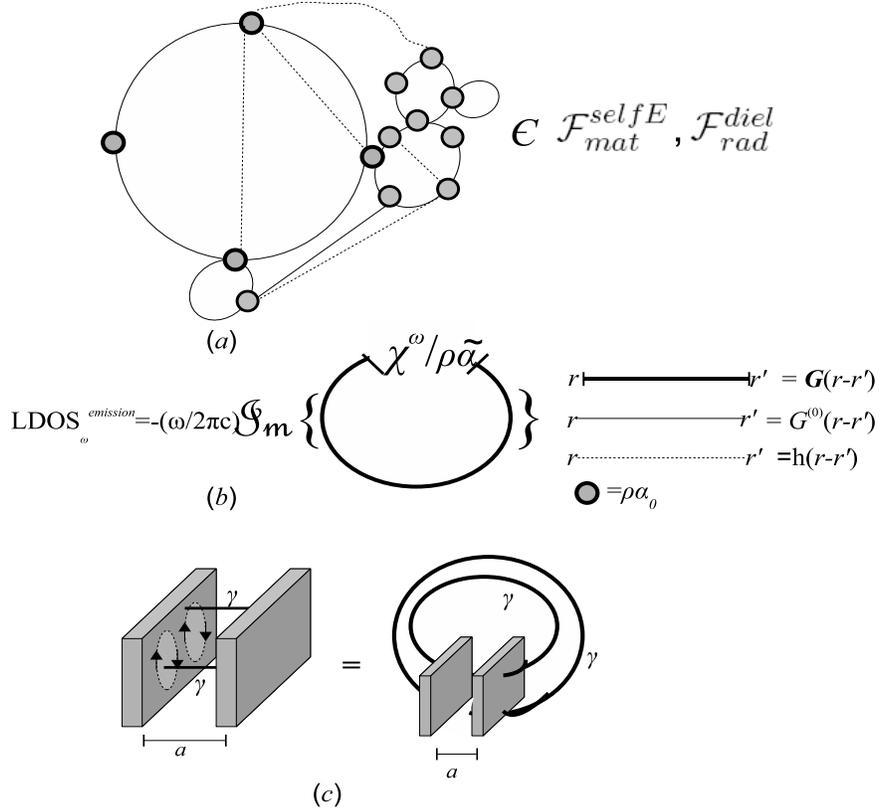}
\caption{$(a)$ Typical 1PI diagram contributing to both $\mathcal{F}^{diel}_{rad}$ and $\mathcal{F}^{selfE}_{mat}$. It presents a characteristic fractal structure. $(b)$ Diagrammatic representation of LDOS$^{emission}_{\omega}$. Translation invariance around the loop is broken at the position of the emitter. ($c$) Sketch of the Casimir self-energy which renormalizes the mass of two perfect conducting plates. On the \emph{l.h.s.}, current-loops are depicted. On the \emph{r.h.s.} the representation mimics the formalism we used for the radiative corrections which renormalize the bare polarizability of small dipoles.}\label{fig18}
\end{figure}

\indent The role of the zero-point-energy (ZPE) in a variety of physical phenomena has received renewal attention in the last years. In particular, a strong motivation is to estimate its contribution to the cosmological constant, $\Lambda$, in connection with the possibility of being measured in the laboratory. The need for the existence of the cosmological constant is itself a controversial matter which is out of the scope of this paper --see eg. \cite{Weinberg,Peeble,Straumann1}. Standard cosmology predicts a small value for the current $\Lambda$, which needs of an apparently unnatural fine-tuning. Leaving aside some exotic contributions to dark energy, at least two main contributions to $\Lambda$ must be considered. The first one comes from the Standard Model of particles in the form of condensates of quarks and gluons and the Higgs itself. The second one comes from the ZPE of quantum fields. Eventually, these energies give rise to negative pressure which sources the repulsive gravity force that leads the expansion of the universe. The main difficulty in dealing with  ZPE computations is that the integrals involve ultraviolate divergences. One could think of the need of being regulated by some physical cut-off. However, this is not yet fully justified \cite{PeebleRatra,Brodsky}. Nevertheless, it is customary --as quoted in \cite{Jaffetpaper}-- to present genuine quantum electromagnetic effects like the Casimir effect, the van der Waals forces, the Lamb-shift, etc. as evidences of the existence of the ZPE fluctuations of the EM field. Further on, some  papers have suggested the possibility of measuring the ZPE of the electromagnetic field in Josephson junctions \cite{Beck}. As pointed out by Jetzer and Straumann \cite{JetStrau1,JetStrau2}, the only physical observable which can be measured in absence of coupling to gravity is the variation on the vacuum energy as a response to external couplings. Any  physical observable measured this way remains invariable to any prescription on the normalization or regularization of the vacuum energy. Hence, the scheme we followed to regularize the shifted-electrostatic polarizability, $\alpha^{0}_{stat}$, was merely phenomenological. Infinite quantities were swept under the carpet and remained unaffected by the rest of the renormalization procedure. That regularization scheme only served to us to accommodate in the same footing further finite contributions. It is analogous to the regularization of the divergences in the electron self-energy which appear in QED in computing the experimental mass of the electron \cite{Peskin,Greiner}. Figs.\ref{FIGnew2}($a,b,c_{2},d_{2}$) depicts how, after formal integration of those EM modes which give rise to the bound atomic state $A$, one can interpret that the 'remaining' vacuum fluctuations which amount to the electromagnetic ZPE interact with such a state giving rise to radiative corrections. However, this is just a result of the order in which fluctuations are integrated out. It is plain that there is no transfer of energy from the ZPE into the dipoles self-energy.\\
\indent Jaffe pointed out in \cite{Jaffetpaper} that the Casimir force can be computed out of the derivation of the density of electromagnetic states with respect to some parameter on which the background of matter fields depend. In the case of the force between two metallic plates, the background fields are the electric currents confined on the plates separated by some distance. The diagrams involved contain external legs representing the fields on the plates --those are the current loops depicted on the left plate of Fig.\ref{fig18}($c$). This way, the diagrams with photons attached to the plates cannot be interpreted as vacuum-diagrams. An alternative interpretation is given by Milton \emph{et al.} in \cite{Milton}. The authors identify the interaction energy between the plates with the self-energy which renormalizes their (divergent) bare masses. In Fig.\ref{fig18}($c$) the photon propagators carry the radiative corrections which give rise to the plates self-energy. In the interpretation of \cite{Jaffetpaper}, it is the variation of the density of those photon states which yield the Casimir force. In addition, in the context of Lifschitz formalism \cite{Lifschitz} Milton identifies \cite{Milton2,Miltonbook} a \emph{bulk energy} which depends on the dielectric medium between the plates and is proportional to the volume.\\
\indent We proceed next to describe where energies and forces reside in our formalism. It was emphasized throughout Section \ref{Sect2} that actual dipoles play a double role as polarizing the EM vacuum and renormalizing their own polarizability. In energetic terms, this gives rise to a change on the energy carried out by the fluctuations of sourceless EM modes and to a shift on the resonance frequency of the dipoles. The former is the analog of the bulk energy of \cite{Milton2} while the latter is the analog of the self-energy of the metallic plates of \cite{Milton}. We will denote the energy density on sourceless EM modes by $\mathcal{F}_{rad}$ and that stored in the dipoles by $\mathcal{F}_{mat}$. The corresponding bare energy densities are the divergent ZPE of Eq.(\ref{disi3}) and $\mathcal{F}_{0}=\rho\hbar\omega_{A}$ for dipoles in the atomic state $A$. While the latter, associated to actual matter, can be detected in EM phenomena, the former eventually needs of coupling to gravity to be detected \cite{MiltonGrav}.\\
\indent Once the dipoles are considered bound states, statistical translation invariance holds at length scales greater than $a$. It is in this sense that we can talk of a proper vacuum, $|\Omega\rangle^{s.l.}$, in contrast to the scenario of the parallel metallic plates. As it was illustrated in Section \ref{Sect2B}, the diagrams contributing to the energy of the polarized EM vacuum are topologically identical to those contributing to the dipole self-energy. The only difference being a change on the reference frame. Fig.\ref{fig18}($a$) shows a typical diagram of both $\mathcal{F}_{rad}$ and $\mathcal{F}_{mat}$. It contributes to $\mathcal{F}_{rad}$ when all the points on the diagram are equivalent. It contributes to $\mathcal{F}_{mat}$ when the origin and end of the diagram is chosen to be a point scatterer as in Fig.\ref{fig18}($b$). The main difference between $\mathcal{F}_{rad}$ and
$\mathcal{F}_{mat}$ is that while the former can be computed at zero-order in perturbation theory together with $\mathcal{F}_{0}$, the remaining of $\mathcal{F}_{mat}$ is a second-order contribution. That is,
\begin{eqnarray}
\mathcal{F}_{rad}+\mathcal{F}_{0}&=&\Bigl\langle\Omega,\sum_{j=1}^{N}\psi^{j}_{A}(\vec{r})\:\Big{|}\mathcal{V}^{-1}\int\textrm{d}^{3}r\:
\frac{\epsilon_{0}}{2}|\hat{\vec{E}}(\vec{r})|^{2}+\mathcal{V}^{-1}\hat{H}_{0}(\vec{r})\:\Big{|}\sum_{j=1}^{N}\psi^{j}_{A}(\vec{r}),\Omega\Bigr\rangle\label{E1}\\
&=&\frac{\epsilon_{0}}{2}\:^{s.l.}\langle\Omega|
|\hat{\vec{E}}(\vec{r})|^{2}|\Omega\rangle^{s.l.}+\mathcal{F}_{0}=-\frac{\hbar}{2\pi c^{2}}
\int\omega^{2}\textrm{Tr}\Bigl\{\Im{\{G_{\perp}(\vec{r},\vec{r})\}}\Bigr\}\textrm{d}\omega+\mathcal{F}_{0}\label{E2}\\
&=&c^{-1}\int\hbar\omega\:\textrm{LDOS}_{\omega}^{sourceless}\textrm{d}\omega\:+\:\mathcal{F}_{0},\label{E3}
\end{eqnarray}
where $\hat{H}_{0}$ is given in Eq.(\ref{Hmat}) and $\mathcal{V}$ is the volume occupied by the dielectric and $N$ is the number of dipoles. The atomic wave functions of the dipoles in state $A$ have been explicitly separated so that their contributions to $\mathcal{F}_{0}$ are additive. The radiative term on Eq.(\ref{E2}) assumes knowledge of $\chi^{\omega}_{\perp}$ for any frequency. However, we have already argued that our approach has a natural frequency cut-off  at $c/a$. Also, in general, the dipoles contain a number of atomic resonances so that $\mathcal{F}_{0}$ should contain a sum over those resonances instead. For the sake of simplicity and in order to give a closed formula, we will restrict ourselves to the computation of $\mathcal{F}_{rad}$ up to frequencies where an effective medium exists, $\omega<c/\xi$, yielding,
\begin{equation}\label{fdiel}
\mathcal{F}_{rad}|_{\omega<c/\xi}\simeq\mathcal{E}^{0}+
\frac{\hbar}{4\pi^{2}c^{3}}\int\textrm{d}\omega\:\omega^{3}[\bar{n}(\omega)-1],
\end{equation}
where $\bar{n}(\omega)=\Re{\{\sqrt{\epsilon^{\omega}_{eff}}\}}$.  LDOS in Eq.(\ref{E3}) is referred to in some papers as \emph{radiative density of states} \cite{RMPdeVries,RemiMole}. However, as stressed in the previous section, it must not be confused with that entering Fermi's Golden rule. The second term in Eq.(\ref{fdiel}) is only non-zero within the volume occupied by the dielectric. Hereafter we will denote it by $\mathcal{F}_{rad}^{diel}\equiv\mathcal{F}_{rad}-\mathcal{E}^{0}$.\\
\indent For the computation of the remaining $\mathcal{F}_{mat}$ we have to go to second-order in perturbation theory to calculate the real part of the self-energy of each dipole. Formally, it reads
\begin{eqnarray}
\mathcal{F}_{mat}-\mathcal{F}_{0}&=&\mathcal{V}^{-1}\int\textrm{d}^{3}r\sum_{\{I_{j}\},\{\omega_{j}\}}\Re{}\Bigl\{\Bigl\langle \Omega,\sum_{j=1}^{N}\psi^{j}_{A}(\vec{r})\:\Big{|}\hat{H}_{int}(\vec{r},t)\Big{|}\:
\sum_{j=1}^{N}\psi^{j}_{I_{j}}(\vec{r}),\gamma_{\omega_{j}}\Bigr\rangle\\
&\times&\Bigl\langle\sum_{j=1}^{N}\psi^{j}_{I_{j}}(\vec{r}),\gamma_{\omega_{j}}\:\Bigl{|}\hat{H}_{int}(\vec{r},t)\Bigr{|}\:
\sum_{j=1}^{N}\psi^{j}_{A}(\vec{r}),\Omega\Bigr\rangle
\prod_{l=1}^{N}[\hbar\omega_{A}-\hbar\omega_{I_{l}}-\hbar\omega_{l}]^{-1}\Bigr\}
\label{ee1}\\
&=&\rho\sum_{I,\omega}\Re{}\Bigl\{\:^{s.p.}\langle\Omega,\psi_{A}(\vec{r})|\hat{H}_{int}(\vec{r},t)|
\psi_{I}(\vec{r}),\gamma_{\omega}\rangle
\langle \psi_{I}(\vec{r}),\gamma_{\omega}|\hat{H}_{int}(\vec{r},t)|
\psi_{A}(\vec{r}),\Omega\rangle^{s.p.}\nonumber\\
&\times&[\hbar\omega_{A}-\hbar\omega_{I}-\hbar\omega]^{-1}\Bigr\},\label{ee2}
\end{eqnarray}
where $\hat{H}_{int}$ is given by Eq.(\ref{dipole}). We will simplify matters by considering just two-level dipoles so that $I_{j}$ in the previous equations can take only the values $A$ or $B$, with $\omega_{0}=\omega_{A}-\omega_{B}$ the resonance frequency and $\mu$ the transition dipole amplitude. Next, we make use of the renormalization scheme developed in Section \ref{Sect3C} and identify
\begin{equation}
\mathcal{F}_{mat}-\mathcal{F}_{0}=\rho\hbar c\:\Delta k_{res}=\frac{2\rho}{3\epsilon_{0}}|\mu|^{2}\tilde{k}^{2}
\Re{\{2\gamma_{\perp}^{\tilde{k}}+\gamma_{\parallel}^{\tilde{k}}\}}|_{\tilde{k}=k_{res}}\label{ee3}.
\end{equation}
$\Delta k_{res}$ and $k_{res}$ deduce from  Eq.(\ref{kres}). Our renormalization procedure takes care of possible divergences in the computation of Eq.(\ref{ee2}). Hereafter, we will denote the dielectric self-energy by $\mathcal{F}^{selfE}_{mat}\equiv\mathcal{F}_{mat}-\mathcal{F}_{0}$. At leading order in $\rho\alpha_{0}$, the van-der-Waals energy $\mathcal{F}^{vdW}$ between two random dipoles in an MG gas is given by Eq.(\ref{ee3}) replacing
$\Re{\{2\gamma_{\perp}^{\tilde{k}}+\gamma_{\parallel}^{\tilde{k}}\}}|_{\tilde{k}=k_{res}}$ with the leading order term of $\Re{\{\gamma^{[1]}_{\parallel}(\tilde{k})\}}$ given by Eqs.(\ref{g1parrra},\ref{g1parrrb}). That energy corresponds to the electrostatic interaction between two fluctuating random dipoles,
\begin{eqnarray}\label{FvdW}
\mathcal{F}^{vdW}&=&\frac{2\rho}{3\epsilon_{0}}|\mu|^{2}\tilde{k}^{2}\Re{\{\gamma^{[1]}_{\parallel}(\tilde{k})\}}\nonumber\\
&=&\rho\frac{-1}{3\pi\epsilon_{0}}|\mu|^{2}\alpha_{0}\:\rho/\xi^{3}=-\rho\frac{\hbar\omega_{0}}{6\pi}\alpha_{0}^{2}\rho/\xi^{3}.
\end{eqnarray}
The mutual induction is given by the $2^{nd}$ and $3^{rd}$ diagrams of Fig.\ref{FIGnew7}($b$) with the propagators carrying only longitudinal photons.\\
\indent The dielectric and the corresponding polarized vacuum are characterized in our case by
the density of dipoles and a typical  correlation length $\xi$ between them. The former plays an analogous role to that of the separation between the metallic plates in the Casimir setup. In a van-der-Waals gas, $\xi$ is the van-der-Waals radius. Therefore, internal forces between dipoles originate as a response to variations in both/either $\rho$ and/or $\xi$. In high correlated systems they both are related. For a gas, $\xi$ is fixed and  the pressure reads,
\begin{eqnarray}
\mathcal{P}_{tot}&=&-\frac{\partial}{\partial\mathcal{V}}\{\mathcal{V}\:(\mathcal{F}^{diel}_{rad}+\mathcal{F}^{selfE}_{mat})\}\nonumber\\
&=&[\rho\frac{\partial\mathcal{F}^{diel}_{rad}}{\partial\rho}-\mathcal{F}^{diel}_{rad}]+
[\rho\frac{\partial\mathcal{F}^{selfE}_{mat}}{\partial\rho}-\mathcal{F}^{selfE}_{mat}]
\equiv\mathcal{P}_{rad}+\mathcal{P}_{mat}\label{P}.
\end{eqnarray}
Because the medium is statistically isotropic and homogeneous, the pressure is a scalar which tends to expand or contract isotropically the distance between particles. Note that only $\mathcal{P}_{mat}$ is considered when introducing van-der-Waals-Casimir forces in the free energy of a complex medium.
No reference is made to the electromagnetic pressure, $\mathcal{P}_{rad}$.
$\mathcal{P}_{rad}$ is interpreted as the pressure exerted  over the dipoles by the vacuum fluctuations of the normal EM modes. The fact that $\Gamma$, $\Delta k_{res}$ and $\mathcal{P}_{mat}$ are all proportional to $\gamma$-factors and their derivatives implies that none of them can be attributed to transference of energy from EM vacuum fluctuations. They are rather associated to changes in the self-energy of material degrees of freedom.\\
\indent It is now straightforward to compute the zero temperature pressure of a van-der-Waals gas. That is usually written as \cite{Huang} $P^{vdW}_{T=0}=-a'\rho^{2}$, where $a'$  parametrizes the dipole-dipole interaction. Applying Eq.(\ref{P}) to Eq.(\ref{FvdW}),
\begin{equation}\label{ladvander}
\mathcal{P}^{vdW}=-\frac{\hbar\omega_{0}}{6\pi}\frac{\alpha_{0}^{2}}{\xi^{3}}\:\rho^{2},
\end{equation}
we identify $a'=\frac{\hbar\omega_{0}}{6\pi}\alpha_{0}^{2}/\xi^{3}$. Should we introduce the higher order terms in $\rho$ of Eq.(\ref{laO1}) we would be performing the virial expansion of $P^{vdW}_{T=0}$ restricted to van-der-Waals short-ranged forces going like $\sim1/r^{7}$.\\
\indent For the sake of completeness we compute in the same approximation $\mathcal{P}_{rad}$. We will restrict ourselves to the simplified case in which the dielectric possesses a unique resonance with not too much loss of generality. Making use of Eq.(\ref{fdiel}) together with Eq.(\ref{laeps}), we obtain
\begin{equation}
\mathcal{F}_{rad}^{diel}\simeq\frac{3}{64\pi^{2}}\hbar\alpha_{0}k_{0}^{2}\rho\:(\Gamma_{0}+\Delta\Gamma)
(k_{0}+\Delta k_{res}+\Delta k_{L}),
\end{equation}
where only finite $\rho$-dependent contributions have been considered in the range of validity of Eq.(\ref{laeps}). Next, using the values of the $\gamma^{[1]}$-factors computed in Section \ref{Sect5A},
\begin{eqnarray}
\Delta\Gamma^{[1]}\simeq\frac{7}{6}\rho\alpha_{0}\Gamma_{0},\qquad\Delta k^{[1]}_{res}\simeq-\frac{k_{0}}{6\pi}\alpha_{0}^{2}\rho/\xi^{3},\qquad\Delta k_{L}=-\frac{1}{3}k_{0}\alpha_{0}\rho,\label{deltas}\\
\textrm{we obtain, }\qquad\qquad\qquad\qquad\qquad\qquad \mathcal{F}_{rad}^{diel}\simeq\frac{c\hbar}{128\pi^{3}}k_{0}^{7}\alpha^{2}_{0}\rho\:
[1+\rho\alpha_{0}(\frac{5}{6}-\frac{\alpha_{0}}{6\pi\xi^{3}})]\label{fdiel1}.
\end{eqnarray}
Applying Eq.(\ref{P}), we get at leading order in $\rho\alpha_{0}$,
\begin{equation}\label{laP1}
\mathcal{P}^{[1]}_{rad}\simeq\frac{c\hbar}{128\pi^{3}}k_{0}^{7}\alpha^{3}_{0}\rho^{2}\:
(\frac{5}{6}-\frac{\alpha_{0}}{6\pi\xi^{3}}).
\end{equation}
For realistic values of $\xi$ and $\alpha_{0}$, $\mathcal{P}_{rad}$ is positive except for extremely long-wavelength resonances of the order of $10^{6}$nm. This means that the contribution of the individual self-energies of the dipoles in $\mathcal{P}^{[1]}_{rad}$ is negligible in comparison to that coming from the additional decay rate, $\Delta\Gamma^{[1]}$, and the Lorentz-shift, $\Delta k_{L}$. The latter being just proper of the effective medium. Because $\Delta k_{L}$ enters with opposite sign w.r.t. that of $\Delta\Gamma^{[1]}$, its net effect is to subtract from $\Delta\Gamma^{[1]}$ the longitudinal contribution of a local field factor at leading order in $\rho\alpha_{0}$, $\Delta\Gamma^{[1]}+\Delta k_{L}=\Delta\Gamma^{[1]}_{\perp}$. The ratio between Eq.(\ref{laP1}) and Eq.(\ref{ladvander}) is,
\begin{equation}
\frac{\mathcal{P}^{[1]}_{rad}}{\mathcal{P}^{vdW}}\simeq\frac{-5}{128\pi^{2}}k_{0}^{6}\alpha_{0}\xi^{3}
\lesssim-10^{-12},
\end{equation}
for realistic values of $\alpha_{0}$. Therefore, we find that $\mathcal{P}_{rad}$ is negligible for a van-der-Waals gas in comparison to $\mathcal{P}_{mat}$.\\
\indent We finalize this Section speculating about possible observational effects of the above forces at cosmological scales. It is known that the magnetic dipole associated to rotating galaxies is strong. Therefore, the interactions between distant galaxies or clusters of these which are not gravitationally bounded might be predominantly electromagnetic. As long as galaxies can be treated as point dipoles, we can model the universe as a cold van-der-Waals gas. Its pressure, disregarding the coupling to Hubble's flow, would be determined by the van-der-Waals-Casimir forces and the vacuum radiative pressure.
However, we must emphasize that this does not imply by any means that the associated energy $\mathcal{F}^{selfE}_{mat}+\mathcal{F}^{diel}_{rad}$ must be of any gravitational relevance. Said this, it results tentative to associate the negative pressure  $\mathcal{P}_{T=0}^{vdW}$ --which however only in flat space and ignoring retardation effects has been proved here to be negative- with a cosmological constant. Nevertheless, $\mathcal{P}_{tot}$ might generate a
 red/blue-shift of similar characteristics to the cosmological one at scales where the distribution
of galaxies is really isotropic and homogeneous. Otherwise, it would yield small variations on top of the
cosmological red-shift, due to the generation of peculiar velocities. Evidently, curvature, causality,
magnetic effects and retardation effects, so far ignored in the present work, should be considered. Also, we
would like to stress that, although we have showed that ordinary QED effects are explained without reference
at all to the ZPE, this does not imply that current investigations on more exotic Casimir-like effects are meaningless. Note that our derivation of the total pressure in Eq.(\ref{P}) considers only variations on the configuration of actual dipoles, which does not affect to either $\mathcal{F}_{0}$ or $\mathcal{E}^{0}$. However, other kinds of variations may be thought of which so do. Hence, in the so-called Casimir effect in compact dimensions and the dynamical Casimir effect in a manifold with moving boundaries,
the resultant pressure is associated to variations on the geometry of the space-time manifold. That gives rise to the aperture/clousure of channels for the normal modes of quantum fields and hence to variations on $\mathcal{E}^{0}$.
 The fact that the 'ordinary' Casimir effect between parallel plates can be explained effectively
in similar terms is just due to the fact that the interaction between perfectly metallic plates can be well
approximated by boundary conditions over the mediating EM field which 'mimic' the effect of a compact space \cite{Jaffetpaper}.
However, those conditions are imposed over $\mathcal{\bar{G}}$ and hence give rise to changes on the
self-energy  of the plates \cite{Milton} and not over $\mathcal{E}^{0}$. In this sense, $\mathcal{P}_{mat}$
for a gas of galaxies is the actual cosmological analog to the ordinary Casimir's force.
\section{Conclusions}\label{TheEnd}
We have studied the phenomenon of electric dipole emission in statistically homogeneous complex media. Based on the computation of the spectrum of sourceless normal modes and the spectrum of the EM fluctuations which yield the self-energy of the dielectric matter, we postulate the distinction between a sourceless vacuum, $|\Omega\rangle^{s.l.}$, and a self-polarization vacuum, $|\Omega\rangle^{s.p.}$. The former is a proper vacuum as it is translation-invariant. The latter is defined upon each dipole. In physical grounds, the distinction bases on the observation that the constituents of a dielectric play a double role. In the first place, they polarize the zero-point vacuum, $|0\rangle$. Complementarily, they renormalize their own polarizability, $\alpha_{0}$, for the multiple-dipole interactions generate a self-energy. The EM energy in $|\Omega\rangle^{s.l.}$ is computed non-perturbatively. The dielectric self-energy enters at second-order of perturbation theory following Fermi's Golden rule.\\
\indent In application of the fluctuation-dissipation theorem, the spectrum of EM fluctuations of
$|\Omega\rangle^{s.l.}$ is determined by he Dyson propagator, $\bar{G}$. The spectrum of fluctuations of the self-polarization field in $|\Omega\rangle^{s.p.}$ is determined by $\bar{\mathcal{G}}$. The latter satisfies a Lippmann-Schwinger equation with a stochastic kernel. In a homogeneous dielectric the kernel is given by
\begin{equation}
\Xi^{VC}_{\perp,\parallel}(k)=-\frac{\rho\tilde{\alpha}}{\chi_{\perp,\parallel}(k)\:G^{(0)}_{\perp,\parallel}(k)}
\Bigl[1\:-\:\frac{\chi_{\perp,\parallel}(k)}{\rho\tilde{\alpha}}\:+\:k_{0}^{2}\chi_{\perp,\parallel}(k)\:G^{(0)}_{\perp,\parallel}(k)\Bigr].\nonumber
\end{equation}
As a result,
\begin{eqnarray}
\mathcal{G}^{VC}_{\perp}(k)&=&\frac{1}{\rho\tilde{\alpha}}\:\chi_{\perp}(k)\:G_{\perp}(k)=\frac{1}{\rho\tilde{\alpha}}\:
\frac{\chi_{\perp}(k)}{k_{0}^{2}[1+\chi_{\perp}(k)]-k^{2}},\nonumber\\
\mathcal{G}^{VC}_{\parallel}(k)&=&\frac{1}{\rho\tilde{\alpha}}\:\chi_{\parallel}(k)\:G_{\parallel}(k)
=\frac{1}{\rho\tilde{\alpha}}\:\frac{\chi_{\parallel}(k)}{k_{0}^{2}[1+\chi_{\parallel}(k)]}.\nonumber
\end{eqnarray}
The above equation is exact in the strictly virtual cavity scenario (VC). For the self-polarization of an impurity, the computation of $\mathcal{G}$ is model-dependent. The model is generically referred to as real-cavity model (RC). It is possible to compute the kernel of the equation for $\mathcal{G}$ in closed form for the case the impurity is placed in a large cavity with $R\gg\xi$,
\begin{equation}
\Xi^{RC}_{\perp,\parallel}(k)=-\frac{1}{\kappa^{1PI}_{\perp,\parallel}(k)\:\chi_{\perp,\parallel}(k)\:G^{(0)}_{\perp,\parallel}(k)}
\Bigl[1\:-\:\kappa^{1PI}_{\perp,\parallel}(k)\:\chi_{\perp,\parallel}(k)\:+\:k_{0}^{2}\chi_{\perp,\parallel}(k)\:G^{(0)}_{\perp,\parallel}(k)\Bigr],\nonumber
\end{equation}
where $\kappa^{1PI}_{\perp,\parallel}(k)$ is derived formally in Section \ref{Sect4B2}. For a medium in which dipoles remain fixed $\mathcal{G}$ can be expressed in function of the transference matrix as,
\begin{equation}
\bar{\mathcal{G}}(\vec{r}_{0},\vec{r}_{0})=\frac{1}{-k_{0}^{2}\tilde{\alpha}}\sum_{\{\vec{r}_{i}\}}
\bar{G}^{(0)}(\vec{r}_{0},\vec{r}_{i})\cdot\bar{t}(\vec{r}_{i},\vec{r}_{0}).\nonumber
\end{equation}
\indent The local density of EM states in $|\Omega\rangle^{s.p.}$ is denoted by LDOS$_{\omega}^{emission}$ for it corresponds to the density of channels accessible to the emitted photons. The power emission, $W_{\omega}$, the decay rate, $\Gamma$ and LDOS$^{emisssion}_{\omega}$ are decomposed according to their transverse/longitudinal and coherent/incoherent nature as,
\begin{eqnarray}
2W^{Coh.}_{\perp}&=&W_{o}\int\frac{\textrm{d}^{3}k}{(2\pi)^{3}}\:2\Re{\{\frac{\chi_{\perp}(k)}{\rho\tilde{\alpha}}\}}\Im{\{G_{\perp}(k)\}},
\nonumber\\
2W^{Incoh.}_{\perp}&=&W_{o}\int\frac{\textrm{d}^{3}k}{(2\pi)^{3}}\:2\Im{\{\frac{\chi_{\perp}(k)}{\rho\tilde{\alpha}}\}}\Re{\{G_{\perp}(k)\}},
\nonumber\\
W^{Coh.}_{\parallel}&=&W_{o}\int\frac{\textrm{d}^{3}k}{(2\pi)^{3}}\:\Re{\{\frac{\chi_{\parallel}(k)}{\rho\tilde{\alpha}}\}}\Im{\{G_{\parallel}(k)\}},
\nonumber\\
W^{Incoh.}_{\parallel}&=&W_{o}\int\frac{\textrm{d}^{3}k}{(2\pi)^{3}}\:\Im{\{\frac{\chi_{\parallel}(k)}{\rho\tilde{\alpha}}\}}\Re{\{G_{\parallel}(k)\}},
\nonumber
\end{eqnarray}
where $W_{o}$ must be substituted by the appropriate constants in the case of $\Gamma$ and LDOS$^{emisssion}_{\omega}$. We found that the vacuum in which coherent emission propagates, $|\Omega\rangle^{Coh}$, is additionally polarized w.r.t. that of bulk normal modes, $|\Omega\rangle^{s.l.}$.  The renormalization function being, $Z_{\perp,\parallel}=\Re{\{\frac{\chi_{\perp,\parallel}}{\rho\tilde{\alpha}}
\}}$. In a Maxwell-Garnett dielectric, it equals a local field factor.\\
\indent We have computed the decay rate of a weakly-polarizable interstitial atom in an MG dielectric. The results are meant to fit the experimental data on the life-time of Ce$^{+3}$ ions. We obtained,
\begin{eqnarray}
\Gamma_{MG}^{Coh.}&=&2\tilde{\Gamma}_{o}\Re{\Bigl\{\frac{\epsilon_{MG}+2}{3}\Bigr\}}\Re{\{\sqrt{\epsilon_{MG}}\}},\nonumber\\
\Gamma_{MG}^{\parallel,Disper.}&=&
2\tilde{\Gamma}_{o}\Re{\{\frac{1}{3}\chi_{MG}
+\frac{1}{3}\chi^{2}_{MG}-0.051\:\chi^{3}_{MG}+0.055\:\chi^{4}_{MG}
-0.015\:\chi^{5}_{MG}+...\}}\nonumber\\
\Gamma_{MG}^{Absorb.}&=&-2\tilde{\Gamma}_{o}\Im{\Bigl\{\frac{\epsilon_{MG}+2}{3}\Bigr\}}\Im{\{\sqrt{\epsilon_{MG}}\}}\nonumber\\
&+&\frac{2\tilde{\Gamma}_{o}}{\zeta}\Im{\{\frac{1}{2}\chi_{MG}
+\frac{1}{2}\chi^{2}_{MG}-0.076\:\chi^{3}_{MG}+0.083\:\chi^{4}_{MG}
-0.022\:\chi^{5}_{MG}+...\}}\nonumber\\
&+&\frac{2\tilde{\Gamma}_{o}}{\zeta^{3}}\Im{\{\chi_{MG}
+\frac{1}{3}\chi^{2}_{MG}-\frac{1}{8}\chi^{3}_{MG}+0.073\:\chi^{4}_{MG}
-0.028\:\chi^{5}_{MG}+...\}},\nonumber
\end{eqnarray}
where $\tilde{\Gamma}_{o}$ contains the renormalization of $\mu$. In the single-scattering approximation, a classification of the decay rate in terms of radiative and non-radiative energy transfer has been given --see Fig.\ref{fig13}.\\
\indent The computation of the dielectric constant of an MG dielectric is a problem of self-consistency due the double polarization role played by point dipoles.\\
\indent We have computed the decay rate of a low-polarizable substantial impurity within a large cavity. The results are meant to fit the experimental data on the life-time of Eu$^{+3}$ ions. Formulae are exact up to order $\mathcal{O}(\chi^{2})$,
\begin{eqnarray}
\Gamma_{RC}^{Coh.}&=&\Gamma_{0}\Bigl[\Re{\{\sqrt{\epsilon}\}}\Re{\Bigl\{\Bigl(\frac{\epsilon+2}{3}\Bigr)^{2}\Bigr\}}
-\frac{1}{3}\Re{\{\epsilon-1\}}\Bigr],\nonumber\\
\Gamma_{RC}^{\parallel,Disper.}&=&\Gamma_{0}\Re{\Bigl\{\frac{\epsilon-(\epsilon-2)^{2}}{9\epsilon}\Bigr\}},\nonumber\\
\Gamma_{RC}^{Absorb.}&=&-\Gamma_{0}\Bigl[\Im{\{\sqrt{\epsilon}\}}\Im{\Bigl\{\Bigl(\frac{\epsilon+2}{3}\Bigr)^{2}\Bigr\}}\nonumber\\
&+&\Bigl(\frac{1}{(k_{0}R)^{3}}+\frac{1}{k_{0}R}\Bigr)\Im{\{\frac{2(\epsilon-1)^{2}}{3\epsilon}-(\epsilon-1)\}}\Bigr].\nonumber
\end{eqnarray}
\indent In comparison to previous works, we have found that the usual formulae for the decay rate in the virtual and the real cavity models are erroneous in physical grounds. The main reason being that the inclusion of one local field factor per field operator entering Fermi's Golden rule is an erroneous assumption. Also, only coherent emission can be accurately computed using macroscopic fields and incoherent radiative emission is longitudinal.\\
\indent The total EM energy density is the sum of the energy of the EM fluctuations in $|\Omega\rangle^{s.l.}$, $\mathcal{F}_{rad}=-\frac{\hbar}{2\pi c^{2}}
\int\omega^{2}\textrm{Tr}\Bigl\{\Im{\{G_{\perp}(\vec{r},\vec{r})\}}\Bigr\}\textrm{d}\omega$, plus the total self-energy stored in the dielectric, $\mathcal{F}_{mat}=\mathcal{F}_{0}+\frac{2\rho}{3\epsilon_{0}}|\mu|^{2}\tilde{k}^{2}
\Re{\{2\gamma_{\perp}^{\tilde{k}}+\gamma_{\parallel}^{\tilde{k}}\}}|_{\tilde{k}=k_{res}}$. It is the variation of the latter w.r.t. external parameters, $\xi$ and $\rho$, that gives rise to generalized van-der-Waals-Casimir forces in a complex media. Neither zero-point EM vacuum energy nor $\mathcal{F}_{0}$ contribute. The variation of $\mathcal{F}_{rad}+\mathcal{F}_{mat}$ w.r.t $\rho$ yields an isotropic pressure. The zero-temperature pressure of a van-der-Waals gas is found to be $\mathcal{P}^{vdW}=-\frac{\hbar\omega_{0}}{6\pi}\alpha_{0}^{2}\:\rho^{2}/\xi^{3}$. Higher order terms of the virial expansion have been found together with additional contributions from long-ranged forces. The radiative pressure at zero temperature is estimated to be at least twelve orders of magnitude weaker at leading order in $\rho$.\\
\indent We have argued on the possibility of testing our formulae for $\epsilon_{eff}$, $\Gamma$ and resonance shifts in experimental setups. Cold atoms and colloidal liquids are suggested to be suitable candidates. In order to distinguish ours from the usual formulae of the virtual cavity and the real cavity models, it has been argued that either high refractive index or optically thick media are needed. Alternatively, measuring the coherent radiation instead of integrating the total one may do.\\
\indent The detection of the zero-temperature radiative pressure, which is estimated inviable in  gasses, might be possible in highly correlated cold fluids which present structural resonances. However, we need a more precise dielectric function than that for the two-level mono-atomic dielectric used in the derivation of Eq.(\ref{laP1}). The dielectric constant should be valid for a much wider range of frequencies. To this respect, Kramers-Kronig relations \cite{KK} might help in performing the integration of $\mathcal{F}^{diel}_{rad}$. More investigation is needed on this point. Finally, predictions at cosmological scales need of the introduction in the present formalism of curvature, retardation and magnetic effects. Work on this matter is in progress.\\\\
\indent We thank S.Albaladejo, L.Froufe, J.J. Saenz, R. Carminati and I. Suarez for
fruitful discussions and suggestions. This work has been supported by the Spanish integrated
project Consolider-NanoLight CSD2007-00046, the EU project
 NanoMagMa EU FP7-NMP-2007-SMALL-1 and the Scholarships Program 'Ciencias de la Naturaleza' of Ramon Areces Foundation.


\begin{thebibliography}{99}
\bibitem{Miltonbook} K.A. Milton, \emph{The Casimir effect, physical manifestations of zero-point energy} World Scientific, New Jersey (2001).
\bibitem{Jaffetpaper} R.L. Jaffe, Phys.Rev. D {\bf 72}, 021301 (2005).
\bibitem{Bordag} M.Bordag, U. Mohideen, V.M. Mostepanenko, e-print arXiv:quant-ph/0106045v1.
\bibitem{Purcell} E.M. Purcell,  Phys. Rev. {\bf 69}, 681 (1946).
\bibitem{JohnI} E. Yablonovitch,  Phys. Rev. Lett. {\bf 58}, 2059
(1987); S. John,  Phys. Rev. Lett. {\bf 58}, 2486 (1987); S. John
and T. Quang,  Phys. Rev. A {\bf 50}, 1764 (1994).
\bibitem{Suhling} K. Suhling, P.M.
W. French and D. Phillips, Photochem. Photobiol. Sci. {\bf 4}, 13
(2005).
\bibitem{Antenas} V.V. Protasencko, A.C. Gallagher, Nano. Lett. {\bf 4}, 1329 (2004);
J.N. Farahani, D.W. Pohl, H.J. Eisler and B. Hecht, Phys. Rev.
Lett. {\bf 95}, 017402 (2005); R. Carminati \emph{et al.}, Opt. Com. {\bf 261} (368).
\bibitem{PendryLiu} Z. Liu \emph{et al.}  Science {\bf 289}, 1734 (2000); D.R. Smith, J.B. Pendry and M.C.K. Wiltshire, Science {\bf 305}, 788 (2004).
\bibitem{Foldy} L.L. Foldy, Phys. Rev. {\bf 67}, 107 (1945).
\bibitem{Lax} M. Lax, Rev. Mod. Phys. {\bf 23}, 287 (1951); Phys. Rev. {\bf 85}, 621 (1952); Rev. Mod. Phys. {\bf 38}, 359 (1966).
\bibitem{Frisch} U. Frisch, Ann. Astroph. {\bf 29}, 645 (1966); {\bf 30}, 565 (1967).
\bibitem{Bullough} R.K. Bullough, J. Phys. A {\bf 1}, 409 (1968); {\bf 2}, 477 (1969); {\bf 3}, 708 (1970);
{\bf 3}, 726 (1970); {\bf 3}, 751 (1970).
\bibitem{BulloughHynne} F. Hynne, R.K. Bullough, Phil. Trans. R. Soc. Lond. A {\bf 312}, 251 (1984);
{\bf 321}, 305 (1987); {\bf 330}, 253 (1990).
\bibitem{BulloughHynne2} F. Hynne, R.K. Bullough J. Phys. A {\bf 5}, 1272 (1972).
\bibitem{Feldoher} B.U. Felderhof, G.W. Ford and E.G.D. Cohen, J. Stat. Phys. {\bf 33}, 241 (1983).
\bibitem{Feldoher2} B.U. Felderhof and B. Chichocki, J. Stat. Phys. {\bf 55}, 1157 (1988);  B.U. Felderhof and R.B Jones, Phys. Rev. A, {\bf 39}, 5669 (1989); B.U. Felderhof and B. Chichocki, J. Chem. Phys. {\bf 92}, 6104 (1990).
\bibitem{Einstein} A. Einstein,  Z. Phys. {\bf 18}, 121 (1917).
\bibitem{Sakurai} J.J. Sakurai, \emph{Advanced Quantum Mechanics} Addison-Wesley (1994).
\bibitem{LaudonJPB} S.M. Barnett, B. Huttner, R. Loudon and R. Matloob, J. Phys. B {\bf 29}
 3763 (1996).
\bibitem{Huang} K. Huang, \emph{Statistical Mechanics} John Wiley and Sons, (1987).
\bibitem{CrenBow} M.E. Crenshaw, C.M. Bowden, Phys.Rev.Lett. {\bf 85}, 1851 (2000).
\bibitem{Crenshaw} M.E. Crenshaw, Phys. Rev. A {\bf 78}, 053827 (2008).
\bibitem{BerMil} P.R. Berman, P.W. Milonni, Phys.Rev.Lett. {\bf 92}, 053601 (2004).
\bibitem{Agerwal}
G.S. Agarwal, Phys. Rev. A {\bf 11}, 253 (1975).
\bibitem{Bloch} C.R. Balian and C. Bloch, Ann. Phys. {\bf 64}, 271 (1971).
\bibitem{Economou} E.N. Economou, \emph{Green's Functions in Quantum Physics} Springer-Verlag, Berlin (1983).
\bibitem{Ranvdal} X. Kong and I. Ravndal, Nuc. Phys. B {\bf526}, 627 (1998).
\bibitem{Andrews} D.L. Andrews and D.S. Bradshaw, Eur. J. Phys. {\bf 25}, 845 (2004).
\bibitem{Peskin} M.E. Peskin and D.V. Schroeder, \emph{An Introduction to Quantum Field Theory} Addison-Wesley  (1995).
\bibitem{Greiner} W. Greiner, \emph{Quantum Mechanics. Special Chapters} Springer-Verlag, Berlin Heidelberg  (1998).
\bibitem{BerBoMil} P.R. Berman, R.W. Boyd and P.W. Milonni, Phys. Rev. A {\bf 74}, 053816 (2006).
\bibitem{BarLou} R. Loudon and S.M. Barnett, J. Phys. B {\bf 39}, S555 (2006).
\bibitem{Birula} I. Bialynicki-Birula and T. Sowinski, Phys. Rev. A {\bf 76}, 062106 (2007).
\bibitem{MilLouBerBar} P.W. Milonni, R. Loudon, P.R. Berman and S.M. Barnett, Phys. Rev. A {\bf 77}, 043835 (2008).
\bibitem{RMPdeVries}
P. de Vries, D.V. van Coevorden, A.Lagendijk,
Rev. Mod. Phys. {\bf 70}, 447 (1998).
\bibitem{Fano} F. Fano, Phys. Rev. {\bf 103}, 1202 (1956).
\bibitem{Hopfield} J. Hopfield, Proc. Roy. Soc. (London) {\bf A68}, 441 (1955); J. Hopfield, Phys. Rev. {\bf 112}, 1555 (1958).
\bibitem{KnollBarnett} S. Scheel, L. Kn\"{o}ll, D.-G. Welsch and S.M. Barnett,
Phys. Rev. A {\bf 60}, 1590 (1999).
\bibitem{Breuer} H.P. Breuer, F. Petruccione \emph{The Theory of Open Quantum Systems} Clarendon Press, Oxford (2006).
\bibitem{Wolf} M. Born and E. Wolf, \emph{Principles of Optics} Pergamon Press, Beijing (1980).
\bibitem{Glauber} R.J. Glauber, Phys. Rev. {\bf 130}, 2529 (1963); {\bf 131}, 2766 (1963).
\bibitem{Citrus} D.S. Citrin, Nano Lett. {\bf 4} (9), 1561 (2004).
\bibitem{Japa} A.D. Yaghjian, Proc. IEEE, {\bf 68}, 248 (1980).
\bibitem{Davydov} A.S. Davydov, \emph{Quantum Mechanics} Pergamon Press, New York (1976).
\bibitem{Band} Y.B. Band, \emph{Light and Matter} Wiley, West Sussex UK (2006).
\bibitem{Milonnibook} P.W. Milonni, \emph{The Quantum Vacuum. An Introduction to Quantum Electrodynamics} Academic Press, Inc. (1994).
\bibitem{Wylie} J.M.Wylie, J.E. Sipe, Phys. Rev. A {\bf 30}, 1185 (1984).
\bibitem{MeI} M.Donaire, e-print arXiv:0811.0323.
\bibitem{RemiMole} R. Carminati and J.J. Saenz, Phys.Rev. Lett. {\bf 102}, 093902 (2009).
\bibitem{Toptygin} D. Toptygin, J. Fluoresc. {\bf 13}, 201 (2003).
\bibitem{Latakia} A. Lakhtakia, Astrophys. J. {\bf 394}, 494 (1992).
\bibitem{Draine}
B. Draine and P. Flatau, J. Opt. Soc. Am. A {\bf 11}, 1491 (1994).
\bibitem{LagvanTig} A.Lagendijk, B.van Tiggelen, Phys.Rep. {\bf 270}, 143 (1996).
\bibitem{LuisRemiMole} L.S. Froufe-Perez, R. Carminati and J.J. Saenz,
Phys. Rev. A {\bf 76}, 013835 (2007).
\bibitem{Lorentz} H.A.Lorentz, Wiedem. Ann. {\bf 9}, 641 (1880);
 L.Lorenz, Wiedem. Ann. {\bf 11}, 70 (1881).
\bibitem{Onsager} L.Onsager, J. Am. Chem. Soc. {\bf 58}, 1486
(1936).
\bibitem{PRLdeVries}
P.de Vries, A.Lagendijk, Phys.Rev.Lett. {\bf 81}, 1381(1998).
\bibitem{Maxwell} J.C. Maxwell-Garnett, Phil. Trans. R. Soc. Lond. A {\bf 203}, 385 (1904).
\bibitem{vanTigg} A. Lagendijk, B. Nienhuis, B. van Tiggelen and P. de Vries, Phys. Rev. Lett. {\bf 79}, 657 (1997).
\bibitem{Andrews2} D.L. Andrews, Chem. Phys. {\bf 135}, 195 (1989).
\bibitem{Foster} T. F\"{o}rster, Discussion Faraday Soc. {\bf 27}, 7 (1959).
\bibitem{Chew} H. Chew, Phys. Rev. A {\bf 38}, 3410 (1988).
\bibitem{Ruskies} K.K. Pukhov, T.T. Basiev, Yu.V. Orlovskii, JETP Lett. {\bf 88}, 12 (2008).
\bibitem{Duan2} C.K. Duan, M.F. Reid, Curr. App. Phys. {\bf 6}, 348 (2006).
\bibitem{Duan1} C.K. Duan, M.F. Reid and Z. Wang, Phys. Lett. A {\bf 343}, 474 (2005).
\bibitem{PRLMakietal} J.J. Maki, M.S. Malcuit, J.E. Sipe and R.W. Boyd,
Phys. Rev. Lett. {\bf 67}, 972 (1991).
\bibitem{Hilborn} R.C. Hilborn, Am. J. Phys. {\bf 50}, 982 (1982).
\bibitem{tobepublished} M. Donaire, In preparation.
\bibitem{Potter} M.R. Potter and G.W. Green, J. Phys. F {\bf 5}, 1426 (1975).
\bibitem{Cold} L. Froufe-Perez, W. Guerin, R. Caminati and R. Kaiser, Phys. Rev. Lett. {\bf 102}
 173903 (2009).
\bibitem{PRLRojasFranck} L.F. Rojas-Ochoa, J.M. Méndez-Alcaraz, J.J. Sáenz, P. Schurtenberger
and F. Scheffold, Phys. Rev. Lett. {\bf 93}, 073903 (2004).
\bibitem{SchuEu} J.P. Schuurmans \emph{et al.}, Phys. Rev. Lett. {\bf 80}
 5077 (1998).
\bibitem{MeII} M.Donaire, e-print arXiv:0902.1783.
\bibitem{Jackson} J.D. Jackson, \emph{Classical Electrodynamics} John Wiley and Sons, (1962).
\bibitem{Lavallard} P. Lavallard, M. Rosenbauer and T. Gacoin,
Phys. Rev. A {\bf 54}, 5450 (1996).
\bibitem{Schuurmans1} F.J.P. Schuurmans, A. Lagendijk,
J. Chem. Phys. {\bf 113}, 3310 (2000).
\bibitem{Schuurmans2} F.J.P. Schuurmans, P. de Vries, A. Lagendijk,
Phys. Lett. A {\bf 264}, 472 (2000).
\bibitem{LaudonPRL} S.M. Barnett, B. Huttner, R. Loudon and R. Matloob, Phys. Rev. Lett. {\bf 68}
 3698 (1992).
\bibitem{Nonhenhaimen} G. Nienhuis and C.T.J. Alkemade, Physica B,C {\bf 81}, 181 (1976).
\bibitem{Tip} C.A. Gu\'{e}rin, B. Gralak and A. Tip,  Phys. Rev. E
{\bf 75}, 056601 (2007).
\bibitem{Juzel} G. Juzeli\={u}nas, J. Lumin. {\bf 76,77}, 666 (1998).
\bibitem{Welsh} S. Scheel, L. Kn\"{o}ll and D.-G. Welsch,
Phys. Rev. A {\bf 60}, 4094 (1999).
\bibitem{Weinberg} S. Weinberg, Rev. Mod. Phys. {\bf 61}, 1 (1989).
\bibitem{Peeble} P.J.E. Peebles, Nucl. Phys. B {\bf 138}, 5 (2005).
\bibitem{Straumann1} N. Straumann, Mod. Phys. Lett. A {\bf 21}, 1083 (2006).
\bibitem{PeebleRatra} P.J.E. Peebles, B. Ratra, Rev. Mod. Phys. {\bf 75}, 559 (2003).
\bibitem{Brodsky} S.J. Brodsky, R. Shrock, Pre-print:arXiv:0803.2554v1 (2008).
\bibitem{Beck} C. Beck, M.C. Mackey, Phys. Lett. B {\bf 605}, 295 (2005).
\bibitem{JetStrau1} P. Jetzer, N. Straumann, Phys. Lett. B {\bf 606}, 77 (2005).
\bibitem{JetStrau2} P. Jetzer, N. Straumann, Phys. Lett. B {\bf 639}, 57 (2006).
\bibitem{Milton} K.A. Milton, P. Parashar, K.V. Shajesh and J. Wagner, J. Phys. A {\bf 40}, 10935 (2007).
\bibitem{Lifschitz} E.M. Lifschitz, Soviet. Phys. JTEP {\bf 2}, 73 (1956).
\bibitem{Milton2} K.A. Milton, J. Phys. A {\bf 37}, R209 (2004).
\bibitem{MiltonGrav} K.A. Milton \emph{et al.}, J. Phys. A {\bf 41}, 164052 (2008).
\bibitem{KK} J.S. Toll, Phys. Rev. {\bf 104}, 1760 (1956).
\end{thebibliography}
\end{document}